\documentstyle[epsfig,natbib2,natbibmnfix]{mn}

\newcommand{\be}{\begin{equation}}
\newcommand{\ee}{\end{equation}}
\newcommand{\bea}{\begin{eqnarray}}
\newcommand{\eea}{\end{eqnarray}}
\newcommand{\bc}{\begin{center}}
\newcommand{\ec}{\end{center}}

\renewcommand{\thefootnote}{\fnsymbol{footnote}}

\title{Hydrodynamical simulations of cluster formation with central
  AGN heating}

\author[D.~Sijacki et al.]
       {\parbox{18cm}{D.~Sijacki\footnotemark[1] and
       V.~Springel}\vspace{0.3cm}\\ 
       Max-Planck-Institut f\"{u}r Astrophysik,
       Karl-Schwarzschild-Stra\ss{}e 1, 85740 Garching bei M\"{u}nchen, Germany}

\begin{document}

\maketitle
\begin{abstract}
  We analyse a hydrodynamical simulation model for the recurrent
  heating of the central intracluster medium (ICM) by active galactic
  nuclei (AGN).  Besides the self-gravity of the dark matter and gas
  components, our approach includes the radiative cooling and
  photoheating of the gas, as well as a subresolution multiphase model
  for star formation and supernova feedback. Additionally, we
  incorporate a periodic heating mechanism in the form of hot, buoyant
  bubbles, injected into the intragalactic medium (IGM) during the
  active phases of the accreting central AGN. We use simulations of
  isolated cluster halos of  different masses to study the bubble
  dynamics and the heat transport into the IGM.  We also apply our
  model to self-consistent cosmological simulations of the formation
  of galaxy clusters with a range of masses.  Our numerical schemes
  explore a variety of different assumptions for the spatial
  configuration of AGN-driven bubbles, for their duty cycles and for
  the energy injection mechanism, in order to obtain better
  constraints on the underlying physical picture.  We argue that AGN
  heating can substantially affect the properties of both the stellar
  and gaseous components of clusters of galaxies. Most importantly, it
  alters the properties of the central dominant (cD) galaxy by
  reducing the mass deposition rate of freshly cooled gas out of the
  ICM, thereby offering an energetically plausible solution to the
  cooling flow problem. At the same time, this leads to reduced or
  eliminated star formation in the central cD galaxy, giving it red
  stellar colours as observed.
\end{abstract}

\section{Introduction}

\renewcommand{\thefootnote}{\fnsymbol{footnote}}
\footnotetext[1]{E-mail: deboras@mpa-garching.mpg.de }

Clusters of galaxies are the largest virialised objects in the Universe and
are thought to contain a representative fraction of baryons \citep[]{White93}.
Most of these baryons can be found in the diffuse gas of the intracluster
medium (ICM), which is directly observable in X-rays, making clusters of
galaxies an almost ideal laboratory for studying the physical processes that
shape galaxies and halos in the Universe. Clusters of galaxies are also a
useful cosmological probe \citep[for a recent review see][]{Voit2004}, and
therefore have been a prime target for theoretical modelling early on, both
numerically and analytically.

A first order approximation for the ICM is to represent it as an ideal,
non-radiative gas. This leads to the predications of scale invariant relations
between X-ray luminosity, mass and temperature \citep[]{Kaiser86}.  However,
it has long been established that the observed relations do not agree in
detail with these assumptions, e.g.~the observed $L_{\rm X}$--$T$ relation is
much steeper than expected based on this simple model. In addition, recent
observations with radio and X-ray telescopes have revealed a stunning
complexity of the ICM physics, including phenomena such as cold fronts, radio
ghosts, cluster turbulence, and apparently nearly uniform enrichment to high
metallicity.

However, possibly the most puzzling observational fact is the ``cooling-flow''
problem.  Since the cooling time in the central regions of galaxy clusters is
smaller than the age of the clusters themselves, a central inflow of cool gas
is expected to occur \citep[e.g.][]{FabianNulsen77, Cowie77, Fabian94}. The
rate of gas cooling can be estimated by energetic arguments if one assumes
that X-ray cooling radiation is fed by the thermal reservoir of the hot
cluster plasma.  Based on this, the estimated rate of accretion onto the
central galaxy is rather high in many cases
\citep[e.g.][]{Fabian84,White94,WJF97, Allen00, Allen2001}, implying that a
significant amount of gas cooler than $1-2\,{\rm keV}$ should be present in the
centre.  However, up to the present time, optical and X--ray observations have
failed to detect the required amount of this cool gas, suggesting
that it is simply not there \citep[e.g.][]{McNamara00, Peterson01,Tamura01,
  Balogh01, Kaastra01, Edge01, Edge02, Edge03, Fabian01, Boehringer2002,Peterson03,Salome2004}. The
low current star formation rates of central galaxies
\citep[e.g.][]{O'Connell89, Johnstone87,Allen95} provide additional support
for the absence of strong cooling flows. Apparently, there must be a physical
process that offsets the radiative cooling in the centre, preventing the gas
from falling out of the ICM in a cooling flow.

Theoretical studies have therefore often invoked some sort of
non--gravitational heating to explain the cluster scaling relations
\cite[e.g.][]{Kaiser91, Navarro95, Bower97, TozziN01, Borgani01, VoitBryan01,
  Babul02, Voit02, Voit03, OhBenson03, Tornatore03, Borgani04}. The main
unsolved issue in these models remains the origin and nature of the physical
sources that cause the extra--heating of the ICM. Perhaps the most obvious
heat source is supernovae associated with star formation, but it seems
questionable that they are able to supply the required amount of feedback
energy.  Curiously, radiative cooling alone may also account for the steepness
of the $L_{\rm X}$--$T$ relation by eliminating gas more efficiently in
low-mass systems \citep[e.g.][]{Lewis00, VoitBryan01, Muanwong01, Yoshida02,
  WuXue02, Voit02, McCarthy04}, but this produces a drastic overprediction of
the amount of cold gas (apart from a problem with the $L_{\rm X}$--$T$
zero-point) and is therefore disfavoured. Models that self-consistently
incorporate SNe heating and radiative cooling processes are also only found to
have limited success \citep[e.g.][]{Borgani04}. The over--cooling problem
therefore remains unsolved. Another problem is posed by simulated temperature
profiles, which typically exhibit a trend to increase towards the cluster
centre, in disagreement with observational inferences \cite[e.g.][]{ Allen01,
  DeGrandiM02}.

An array of different physical hypothesis have been proposed to solve the
cooling-flow paradox, including thermal conduction, magnetic fields, cosmic
rays, and hot buoyant bubbles from AGN jets.  Thermal conduction may in
principle offset central cooling losses by inducing a heat current from outer,
hotter regions of clusters \citep[e.g.][]{Narayan01}, provided the
conductivity is not strongly suppressed by tangled magnetic fields. Analysis
of static cluster models with conduction have been able to provide good
matches to observed temperature profiles in some cases
\citep[e.g][]{Voigt02, Zakamska03, Voigt03} but detailed
self-consistent numerical simulations which followed conduction still
encountered the cooling flow problem \citep[e.g.][]{Jubelgas04, Dolag04},
making it questionable whether this can be the real solution.

The more widely favoured hypothesis is instead that the central AGN may supply
the required amount of energy. Accretion onto supermassive black holes is
thought to liberate of order $\sim 10\%$ of the accreted rest mass energy,
implying that even for low accretion rates onto a supermassive black hole,
offsetting the cooling flows is energetically quite possible. In fact, such
accretion powers high-redshift quasars, the most luminous sources in the
universe. Quasar activity is likely to be triggered by mergers of galaxies,
where cold gas is forced to the nuclei by gravitational tidal forces. This
accretion and the associated quasar feedback has recently been incorporated
into simulations, and shown to play a potentially important role in shaping
the properties of elliptical galaxies \citep[]{Springel2005}.

In clusters of galaxies, however, it seems clear that the central AGN activity
that causes bubbles is of a different nature, and needs not be triggered by
galaxy mergers.  Observationally, many clusters of galaxies show evidence for
X-ray cavities filled with radio plasma \citep[e.g.][]{Owen2000,Blanton01}, which are
thought to be inflated by relativistic jets from the AGN. Theoretically, it has
been shown that these bubbles may rise buoyantly and raise some of the
central cool gas \citep[e.g.][]{Churazov01}, allowing it to mix with
the hotter gas further out. Together with the accompanying mechanical and
possibly viscous heating, this can then constitute an efficient feedback
mechanism.

In this paper, we focus on the phenomenology of this bubble feedback, without
addressing the small scale physics of the accretion onto the black hole.  This
extends earlier simulation studies which all employed hydrodynamical mesh
codes, but which focused exclusively on highly idealised cluster models
\citep[e.g.][]{Churazov01, Quilis01, Ruszkowski02, Churazov02,
  Brueggen02,Brueggen02b,Brueggen03,Nulsen03, DVecchia04, Hoeft04}. A first
goal of our work is to demonstrate that such simulations are also possible with the
smoothed particle hydrodynamics (SPH) technique, and give results
consistent with earlier studies. This is important because
the Lagrangian nature of SPH is ideal for cosmological simulations of
structure formation, and if applicable for bubble feedback, will allow us to
carry out the first self-consistent cosmological simulations with
AGN-driven bubble heating.  An equally important goal of our simulations is to gain new
insights into the efficiency of bubble feedback associated with AGN for
modifying the thermodynamic state of the ICM and the properties of cluster
galaxies over the course of cosmic history. Our modelling can hence inform
semi-analytic models of galaxy formation that have just begun to include AGN
feedback \citep[e.g.][]{Croton05}, and provide crucial input for
future hydrodynamic simulations that try to incorporate the growth of
supermassive black holes both from the quasar- and the radio-mode.

The outline of this paper is as follows. In Section~\ref{MET}, we describe the
characteristics of our simulation code and the numerical method adopted to
introduce bubble heating. In Section~\ref{ISO} we analyse the AGN heating in
isolated galaxy halos, spanning a wide range in mass, and we present some
Chandra-like photon images of simulated bubbles. The effects of AGN
heating in cosmological simulations of galaxy cluster 
formation is discussed in Section \ref{AGN_cosmo}. Finally, in Section
\ref{DIS} we discuss successes and limitations of our model, and we present
our conclusions.

\section{Methodology} \label{MET}

\subsection{Basic code properties}

Our simulations have been performed with the parallel TreeSPH-code {\small
  GADGET-2} \citep{Gadget2,SYW01b}. We use the `entropy formulation' for SPH
suggested by \citet{SH02}, which manifestly conserves both energy and entropy
when adaptive smoothing lengths are used.  Besides gravitational and
hydrodynamical processes, we include radiative cooling of the gas component,
together with heating by a spatially uniform, time dependent UV background
modelled as in \citet{KWH96}. The gas consists of an optically thin primordial
plasma of hydrogen and helium. In addition, a multiphase subresolution model
for the treatment of star formation and associated feedback mechanisms has
been adopted \citep{SH03}. In this model, stars form from dense cold gas
clouds assumed to reside at pressure equilibrium in a surrounding hot phase of
the interstellar medium. Supernova explosions heat the hot medium and
evaporate cold clouds, thereby providing a self-regulation cycle for star
formation, and a net pressurisation for the highly overdense ISM.
Additionally, we use a simple prescription for metal enrichment, which assumes
that each star-forming gas element can be locally approximated by a closed box
model in which perfect and instantaneous mixing of metals between cold clouds
and ambient gas occurs, as explained in detail in \cite{SH03}.

\subsection{Phenomenological description of AGN heating in clusters}

Besides considering the physical processes already implemented in {\small
  GADGET-2}, we have implemented for this study a new model that accounts for
heating by the central AGN in clusters of galaxies.  This model does not
attempt to provide a fully self-consistent ab initio treatment of the complex
physical processes related to accretion onto supermassive black holes in
clusters and the associated AGN activity. Rather, we try to mimic the observed
phenomenology of hot bubbles in clusters directly in our simulations, without
addressing the jet physics that presumably inflates the bubbles in the first
place. We therefore assume as a starting point that such bubbles are generated
during phases in which an AGN is ``switched on'', and introduce them into the
IGM in a phenomenological fashion.  This allows us to study how the bubbles
affect the properties of the central ICM as a function of their
characteristics, in particular with respect to distributing their energy
content to the surrounding cooler gas.

For definiteness, we assume in our model that a certain amount of thermal
energy is injected in the form of centrally concentrated bubbles spaced in
uniform time intervals. We parameterise this scheme in terms of the AGN duty
cycle, the amount of energy $E_{\rm bub}$ injected, and by the radius $R_{\rm
  bub}$ and distance $d_{\rm bub}$ of the buoyant bubbles from the cluster
centre, respectively.

We first test our scheme for AGN-heating on isolated, axisymmetric halo
models. These systems are clean laboratories which permit us to compare
directly with analogous modelling in the literature \citep[e.g.][]{Churazov01,
  Quilis01, DVecchia04}, and hence to evaluate whether SPH is suitable for
such simulations.  Moreover, these simplified models give us the possibility
to explore straightforwardly and with comparatively low computational cost a
large number of cases. In this way we can investigate the importance of
different physical parameters of the bubbles, thus constraining their
dynamical evolution and the heat transport into the ICM.

As a second step, we apply the model for bubble heating to fully
self-consistent cosmological simulations of galaxy cluster formation. Here, we
also investigate different redshift-dependent energy injection schemes,
allowing us to gain some insight in how the AGN activity influences the
hierarchical galaxy cluster growth and the characteristics of the central
cluster galaxy, and to elucidate the relative importance of AGN heating with
respect to the other physics included.  We consider a set a galaxy clusters
spanning a range in mass because we expect the efficiency of bubble heating
to have a significant mass dependence.

Both for isolated halos and in cosmological simulations, we explored
two different schemes for spatially placing the bubbles around the 
cluster centres.  In the first scheme, the bubbles are introduced randomly
within a sphere with a radius given by $d_{\rm bub}$ around the centre, while in the second
approach, two symmetric bubbles are placed randomly along a fixed axis
of length $2\times d_{\rm bub}$, which has an orientation preserved
in time during subsequent bubble events.  The latter hence mimics a situation
where the AGN jet that inflates the bubbles has directional stability over
time, which could arise due to some coupling with the host galaxy's angular
momentum, for example.  At the present time there is no clear evidence either
way concerning what is the preferred scenario, therefore our main aim is to
investigate the possible differences in the ICM properties between these two
bracketing scenarios.

\subsection{Constraining the model parameters}

Our choice for the values of $R_{\rm bub}$ and $d_{\rm bub}$ has been guided
by observational constraints on X--ray cavities in clusters, and also by the
values typically adopted in previous numerical works, for easier comparison.
For simplicity, we restricted most of our simulations to the case where the
values of $R_{\rm bub}$ and $d_{\rm bub}$ depend only on the mass of the halo
under consideration, and on the redshift in the case of cosmological
simulations. Specifically, we adopted \be
\label{Rbub(M_200)} 
R_{\rm bub} \, \propto \, M_{200}(z)^{1/3}\times
\frac{1+z}{(\Omega_{0m}(1+z)^3 + \Omega_{0\Lambda})^{1/3}}\,, \ee where
$M_{200}(z)$ is the virial mass of the host galaxy cluster at given redshift
of AGN activity, and the same scaling has been adopted for $d_{\rm bub}$. For
the simulations of isolated halos, we used the same dependence of $R_{\rm
  bub}$ and $d_{\rm bub}$ on cluster mass, setting $z=0$.

We study multiple bubble injection events in order to analyze how AGN heating
couples with radiative cooling losses over a sufficiently long time interval.
Thus, our modeling requires prescriptions both for the AGN duty cycle and for
the time evolution of the energy content stored in the bubbles. However, most
of the observed AGN-driven bubbles are found at low redshifts
\citep[e.g.][]{Birzan04}, and only recently some observational evidence for
X--ray cavities in more distant galaxy clusters has been found
\citep{McNamara05}. Therefore, the properties and presence of radio-bubbles at
higher redshifts, and their evolution with time, are observationally rather
unconstrained.  We hence limit ourselves in this work to simple parametric
prescriptions for the evolution of $E_{\rm bub}$, derived from basic
theoretical considerations and empirical laws, which hopefully bracket
reality. Typically, we started injecting bubbles at redshift $z=3$, which is
the epoch that roughly corresponds to the peak of the comoving quasar space
density, but we also tested an earlier epoch given by $z=6$ for the start of
the bubble activity. For our modeling of the evolution of $E_{\rm bub}$ with
time, we adopted two scenarios with rather different behaviour. In the first
one, most of the energy is released at late epochs, while in the second one,
the bubble energy is coupled more closely to an assumed BH accretion rate
(BHAR) model for the growth of the black hole population as a whole, such that
the energy release is more pronounced at high redshifts.  

More specifically, our first model is loosely motivated by the
Magorrian relationship, which implies $M_{\rm BH} \propto
\sigma^4$. A relation between the bubble mechanical luminosity
and the black hole accretion rate, $\dot M_{\rm BH}$, can be derived
by assuming that only a small fraction of the total bolometric
luminosity thermally couples with the ICM. Hence, $L_{\rm bub} = f
\times L_{\rm bol} = f \times \epsilon \dot M_{\rm BH}c^2$. The factor
$f$ sets the efficiency of thermal coupling with the ICM, and is
typically assumed to lie in the range of 1-5\%, while $\epsilon$ is
the radiative efficiency factor. Assuming that the mechanical
luminosity for Eddington-limited accretion is directly proportional to
the black hole mass, the energy content $E_{\rm bub}$ of the bubbles
is then proportional to $M_{200}(z)^{4/3}$, provided the mass of the
central cluster galaxy scales self-similarly with the cluster mass.
Hence, it follows that in this model the bubble energy content is
determined by the mass assembly of the host galaxy cluster with time.

In our second scenario, we instead relate the amount of bubble energy to the
average growth rate of supermassive central black holes. To describe the
latter, we employ an estimate of the BHAR by \cite{DiMatteo03}, who give an
analytic fit \be \dot \rho(z) \,= \, \epsilon_{\rm BH} \,\frac{b \,
  \rm{exp}\it[a(z-z_m)]}{b - a + a\,\rm{exp} \it[b(z-z_m)]} \, , \ee for their
numerical results, with the parameters $a=5/4$, $b=3/2$, $z_m=4.8$, and $
\epsilon_{\rm BH} = 3 \times 10^{-4} {\rm M}_\odot \, {\rm yr}^{-1} \, {\rm
  Mpc}^{-3}$.  Thus, for every duty cycle of AGN activity we can directly
relate $\dot M_{\rm BH}$ with $E_{\rm bub}$ in the following way, \be
\label{Ebub_BHAR_eq} \frac{E_{\rm bub}}{E_{\rm norm}} = f \times \epsilon \times c^2
\int_{z_1}^{z_2} \dot \rho(z)\, {\rm d}z \, , \ee where a normalisation
factor, $E_{\rm norm}$, has been introduced which we set such that the total
energy injected over all duty cycles is the same in our two schemes.   We note that the different temporal evolution of the BH mass in this
approach implies a significantly reduced energy content of the bubbles at low
redshifts.  Finally, there is still one free constant of integration which we
choose by requiring that the assumed mass of the black hole is the same at
$z=0$ in both scenarious.

A number of observational and theoretical works
\citep[e.g.][]{Birzan04,Sanderson2005,McNamara05,Nulsen05,Nulsen2005,
  Donahue2005,Voit2005} have constrained the plausible time interval between
two successive bubble injection episodes to be of order of $\Delta t_{\rm
  bub}\sim 10^8$yrs. Clearly, $\Delta t_{\rm bub}$ could vary both for
clusters of different mass and also in time, especially if the bubble activity
is triggered by a self-regulated mechanism that operates between AGN feedback
and the cooling flow.  Nevertheless, given our simple phenomenological
approach and lack of any better observational constraints, we adopt the value
of $\Delta t_{\rm bub}\,=\, 10^8$yrs for all of our cluster simulations,
independent of the cosmological epoch.

While some of our prescriptions for bubble parameters are motivated by
``quasar-like'' phenomena, our models are really meant to reflect a mode of
feedback by supermassive black holes different from that of ordinary quasars.
Instead of being triggered by mergers and being fueled with dense and cold ISM
gas, the bubbles are a model for the radio activity observed in clusters.
Note that there are also newly emerging theoretical models
\citep[e.g.][]{Croton05,Churazov2005} on how both quasar activity at higher
redshifts and AGN-driven radio bubbles at lower redshifts can be described
within a common unified framework. We will discuss this possibility in more
detail in our conclusions.

\section{AGN heating of isolated galaxy clusters} \label{ISO}

We here analyse simulations of isolated halos, consisting of a static NFW dark
matter halo \citep{NFW96, NFW97} with a gaseous component that is initially in
hydrostatic equilibrium and chosen to follow a density distribution similar to
the NFW dark matter profile, but slightly softened at the centre according to
\be \rho_g(r) \, = \,\frac{f_b \, \delta_0 \, \rho_{\rm
    crit}}{(r+r_0)/r_s(1+r/r_s)^2}, \ee where $r_0$ is a parameter introduced
to mimic a gas core radius.  The baryonic fraction is given by $f_b$, while
$\rho_{\rm crit}$ is the critical density, $\delta_0$ is the characteristic
overdensity and $r_s$ is the scale radius.  The gas follows the equation of
state of an ideal monoatomic gas with adiabatic index $\gamma = 5/3$.
Besides, a certain amount of angular momentum has been imposed that can be
quantified by the dimensionless spin parameter of a halo, \be \lambda \, = \,
\frac{J|E|^{1/2}}{GM^{5/2}}, \ee where $J$ represents the angular momentum,
$M$ is the halo mass, and $E$ its total energy.

The boundary conditions were chosen to be vacuum, i.e. both density and
pressure are initially zero outside the virial radius (defined here as the
radius enclosing a mean density equal to $200\,\rho_{\rm crit}$). We have
simulated halos with a wide range of masses, with virial radii and
concentration parameters as listed in Table \ref{tab_simpar_iso}. The baryonic
fraction, $f_b=0.12$, the spin parameter, $\lambda=0.05$, and $r_0=0.02
R_{200}$ were kept fixed for all the halos. When evolved without radiative
cooling, these initial models are perfectly stable for more than $1/4$ of the
Hubble time, as we explicitly checked. This is the timespan we subsequently
consider in all our non-trivial simulations, both for the case with cooling
and star formation only, and also for the case with additional AGN heating.

\begin{table*}
\bc
\begin{tabular}{cccccccc}
\hline
\hline
$M_{200}$ [$\,h^{-1}{\rm M}_\odot\,$] & $R_{200}$ [$\,h^{-1}{\rm kpc}$] & $c$ & $N_{\rm gas}$ & $m_{\rm gas}$ [$\,h^{-1}{\rm M}_\odot\,$] & $\epsilon$ [$\,h^{-1}{\rm kpc}\,$] \\
\hline
$ 10^{12}$ &  206 & 12.0  & $3 \times 10^{5}$ & $4.0 \times 10^5$ &  1.0   \\
$ 10^{13}$ &  444 &  6.5  & $3 \times 10^{5}$ & $4.0 \times 10^6$ &  2.0   \\
$ 10^{14}$ &  957 &  8.0  & $3 \times 10^{5}$ & $4.0 \times 10^7$ &  5.0   \\
$ 10^{15}$ & 2063 &  5.0  & $3 \times 10^{5}$ & $4.0 \times 10^8$ & 10.0   \\
$ 10^{15}$ & 2063 &  5.0  & $1 \times 10^{6}$ & $1.2 \times 10^8$ &  6.5 \\
\hline
\hline
\end{tabular}
\caption{Numerical parameters of the isolated galaxy clusters. The
  first two columns give the virial mass and radius of the halos,
  evaluated at $200\, \rho_{\rm crit}$. The assumed values for the
  concentration parameter are in the third column, while the initial
  number and the mass of the gas particles is shown in the fourth and
  the fifth columns, respectively. The mass of the star particles is
  half that of the gas particles, because we set the number of
  generations of star particles that a gas particle may produce to
  two. Note that there are no parameters for the dark matter particles
  in these run, because we modelled the dark halo with a static NFW potential.
  Finally, in the last column, the gravitational softening length
  $\epsilon$ for the gas and star particles is given.
\label{tab_simpar_iso}}
\ec
\end{table*}

For the $10^{15} h^{-1}{\rm M}_{\odot}$ isolated cluster, our fiducial set of
AGN heating parameters is (if not explicitly stated otherwise) $E_{\rm bub} =
5 \times 10^{60}\,{\rm erg}$, $R_{\rm bub} = 30\,h^{-1}{\rm kpc}$, $d_{\rm
  bub} = 50\,h^{-1}{\rm kpc}$, and a duty cycle of $\Delta t_{\rm bub} =
10^8{\rm yrs}$, all kept fix in time. The thermal energy injected in the form
of bubbles has been estimated using the simple relations given in Section
\ref{MET}, assuming a $\sim (5 \times 10^8 - 3 \times 10^9) {\rm M}_{\odot}$
black hole in the cluster centre (depending on the thermal-coupling efficiency
factor $f$). For the halos of lower mass, we assumed that the thermal content
of the bubble is proportional to $M_{200}^{4/3}$, in analogy to our first
scenario for scaling the bubble energy in a cosmological setting.  This energy
scaling is motivated by the well established observational relation between
the black hole mass and the velocity dispersion of the stellar component of
the bulge \citep[e.g.][]{Tremaine02}, given by \be M_{\rm BH} = (1.5 \pm 0.2)
\times 10^8\, {\rm M}_{\odot} \ \bigg(\frac{\sigma}{200 \ {\rm km \ 
    s^{-1}}}\bigg)^{4.02 \pm 0.32}, \, \ee and on the hypothesis that the
central cluster galaxy scales self-similarly with the mass of the cluster
itself. Even though these assumptions are certainly very restrictive, they
provide us with a definite model that allows a straightforward interpretation
of the trends with mass, and supply some guidance for what to expect in full
cosmological simulations.  We also note that the recent numerical work of
\cite{DiMatteo03} suggests that, once the $M_{\rm BH}-\sigma$ relation is
established with time, the black hole mass is proportional to $M_{\rm
  DM}^{4/3}$ of the host galaxy.

\subsection{AGN heating of a massive galaxy cluster}

\begin{figure*}
\centerline{
\hbox{
\psfig{file=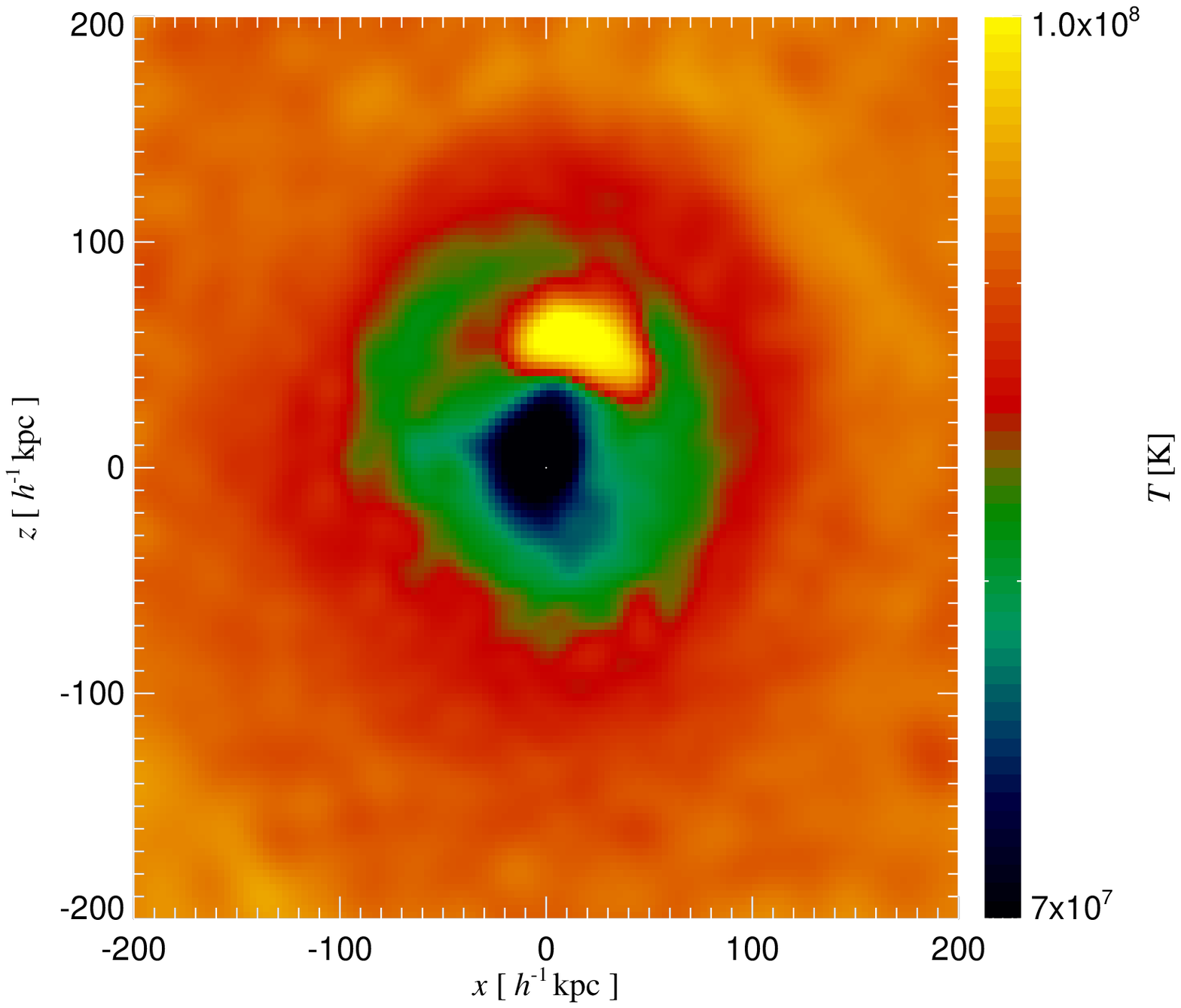,width=9truecm,height=8truecm}
\hspace{0.3truecm}
\psfig{file=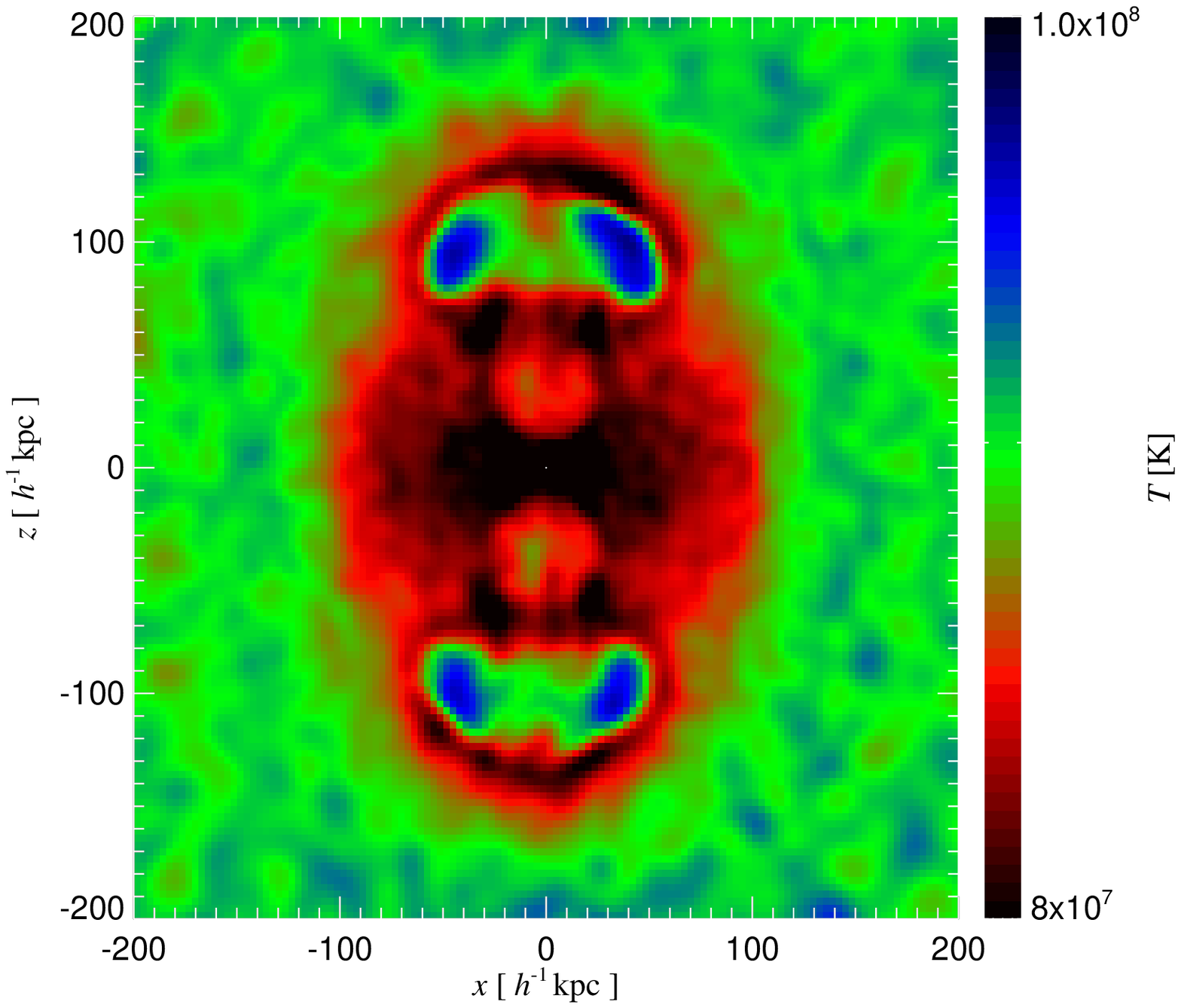,width=9truecm,height=8truecm}
}}
\caption{Projected mass-weighted temperature maps of the central
  regions of an isolated galaxy cluster of mass $10^{15} h^{-1}{\rm
    M}_{\odot}$. In the left panel, bubbles have been introduced with a random
  placement inside a spherical region, while in the right panel, a
  ``jet-like'' injection of bubbles is shown where two bubbles are placed
  opposite of each other, and subsequent generations of bubbles are injected
  along the same spatial axis. $E_{\rm bub}$, $R_{\rm bub}$ and $d_{\rm bub}$
  in both cases are the same, and given by $5 \times 10^{60} {\rm erg}$,
  $30\,h^{-1}{\rm kpc}$ and $50\,h^{-1}{\rm kpc}$, respectively. The maps have
  been constructed for times of $\sim 1.4\, {\rm Gyr}$ and $\sim 0.8\, {\rm
    Gyr}$ after the beginning of the runs.}
\label{Tmap_iso}
\end{figure*}

\begin{figure*}
\centerline{
\hbox{
\psfig{file=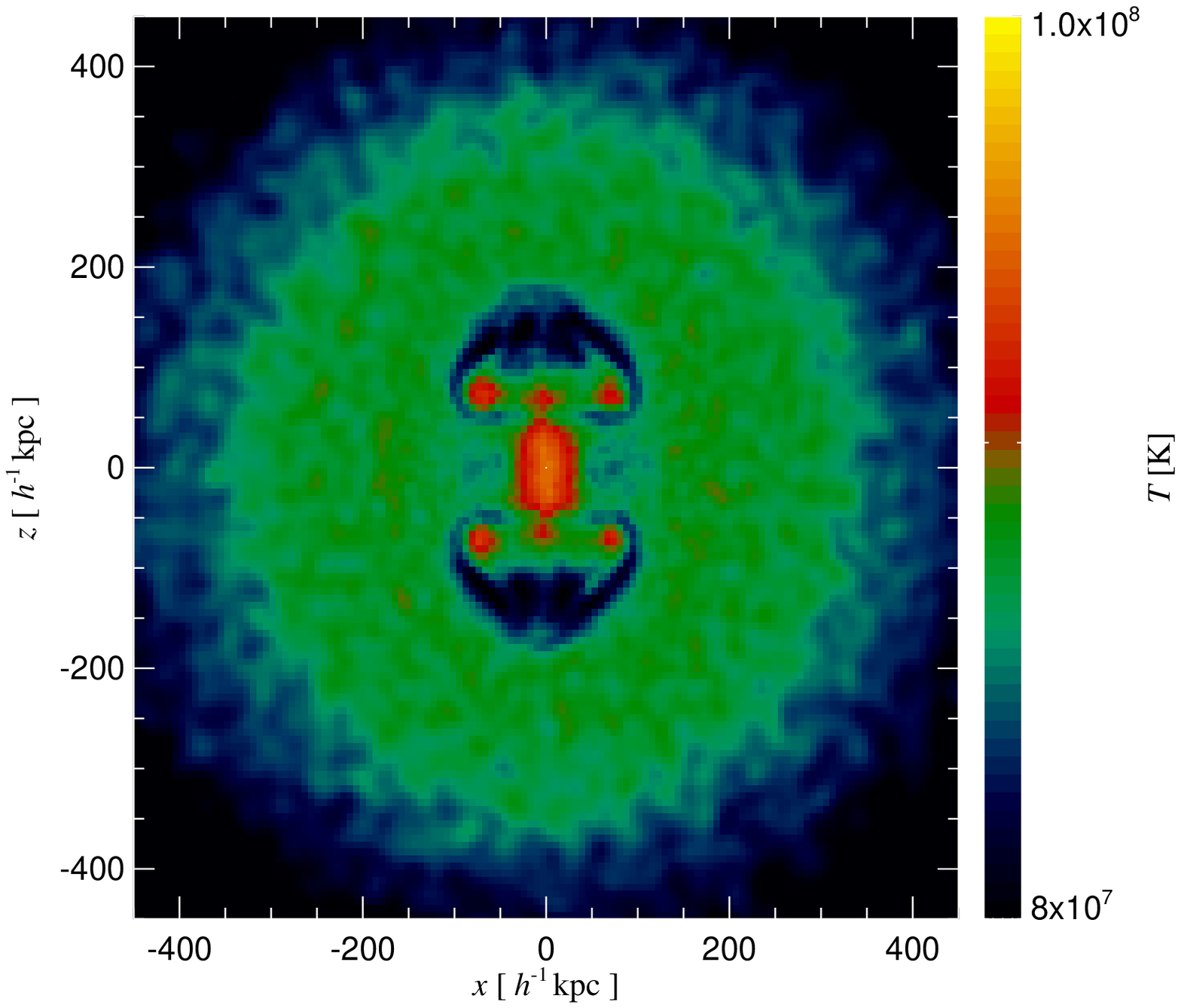,width=9truecm,height=8truecm}
\hspace{0.3truecm}
\psfig{file=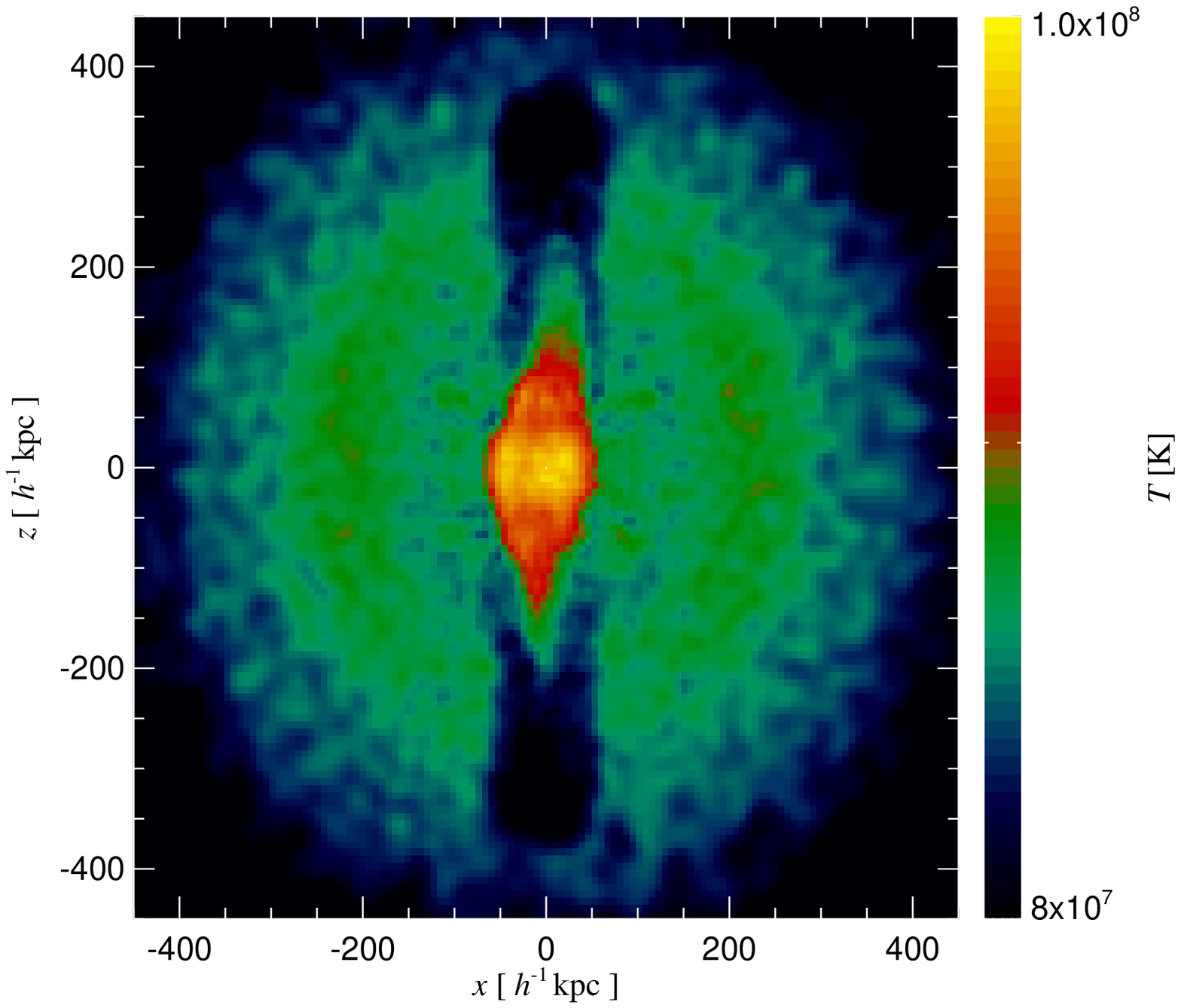,width=9truecm,height=8truecm}
}}
\caption{Time evolution of the isolated $10^{15} h^{-1}{\rm M}_{\odot}$
  galaxy cluster with a jet-like AGN heating. The models are the same as in
  Figure~\ref{Tmap_iso}, only the bubbles have two times bigger radii,
  namely $60\,h^{-1}{\rm kpc}$. It can be noticed how the morphology in the
  central cluster region changes due to bubble-induced motions, from $\sim
  1.8\,{\rm Gyr}$ (left panel) to $\sim 3.3\,{\rm Gyr}$, when a well-defined
  bipolar outflow is clearly visible (right panel).}
\label{Tmap_isobig}
\end{figure*}

In this section we concentrate on the effects of bubble heating on an isolated
galaxy cluster of mass $10^{15} h^{-1}{\rm M}_{\odot}$, while we will discuss
the relative importance of AGN heating as a function of mass in the next
section.  In Figures~\ref{Tmap_iso} and \ref{Tmap_isobig}, we show maps of the
projected mass-weighted temperature of the $10^{15} h^{-1}{\rm M}_{\odot}$
galaxy cluster, focusing on the central regions in order to highlight the
morphology of the bubbles with different injection schemes and at various
evolutionary stages. Both figures were obtained for simulations with cooling
and star formation, and with AGN feedback of the same strength. However, in
the left panel of Figure~\ref{Tmap_iso}, the bubbles were placed randomly
within a sphere of radius $d_{\rm bub}$, while the remaining three panels
illustrate the case of two symmetrical bubbles injected simultaneously,
containing half of the energy each, and with the injection axis preserved with
time for different bubble cycles.

In Figure \ref{Tmap_isobig}, we show results for simulations with the same
feedback energy as in Figure~\ref{Tmap_iso}, but this time the initial radius
of the bubbles was two times larger and equal to $60\,h^{-1}{\rm kpc}$.  After
being injected, the bubbles rise due to buoyancy and start to assume more
elongated, ``pancake-like'' shapes, as clearly visible in the left panel of
Figure~\ref{Tmap_iso}. They continue to rise until the surrounding gas entropy
becomes comparable to their entropy content, at which point they settle into
an equilibrium and dissolve slowly with time. While rising, they push the
intracluster gas above them, and also entrain some of the cooler central gas,
transporting it to larger radii.

A closer comparison of Figures~\ref{Tmap_iso} and \ref{Tmap_isobig} makes it
clear that the smaller bubbles with their significantly higher energy per
particle result in more pronounced mushroom-like structures.  Nevertheless,
they do not shock the surrounding gas which, on the very top of the bubbles,
forms cold rims. At late evolutionary stages, corresponding roughly to a
quarter of the Hubble time, (see the right panel of Figure \ref{Tmap_isobig}),
a characteristic bipolar outflow is visible as a result.

\begin{figure*}
\centerline{\vbox{
\hbox{
\psfig{file=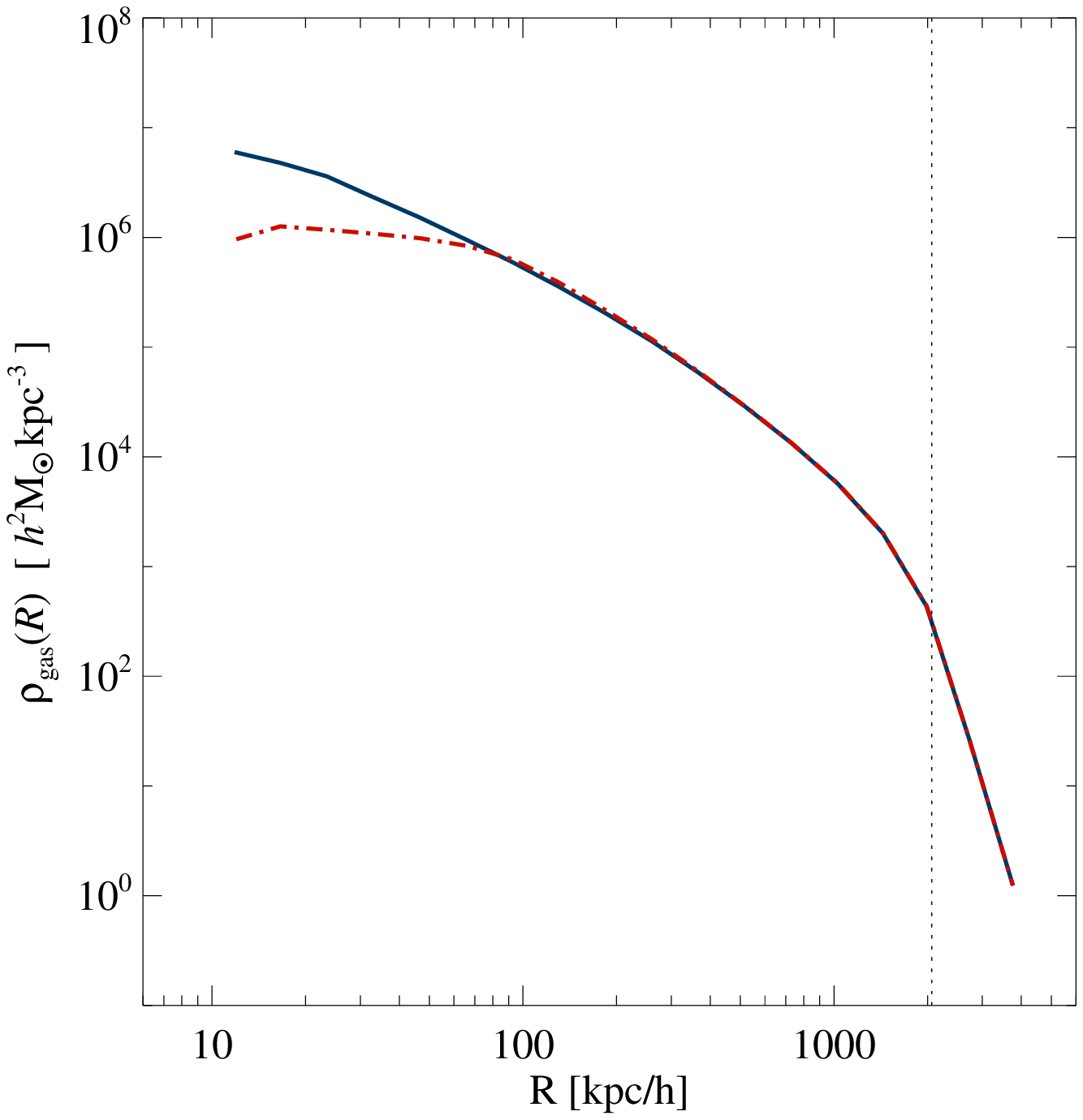,width=8.truecm,height=7.5truecm}
\hspace{1.0truecm}
\psfig{file=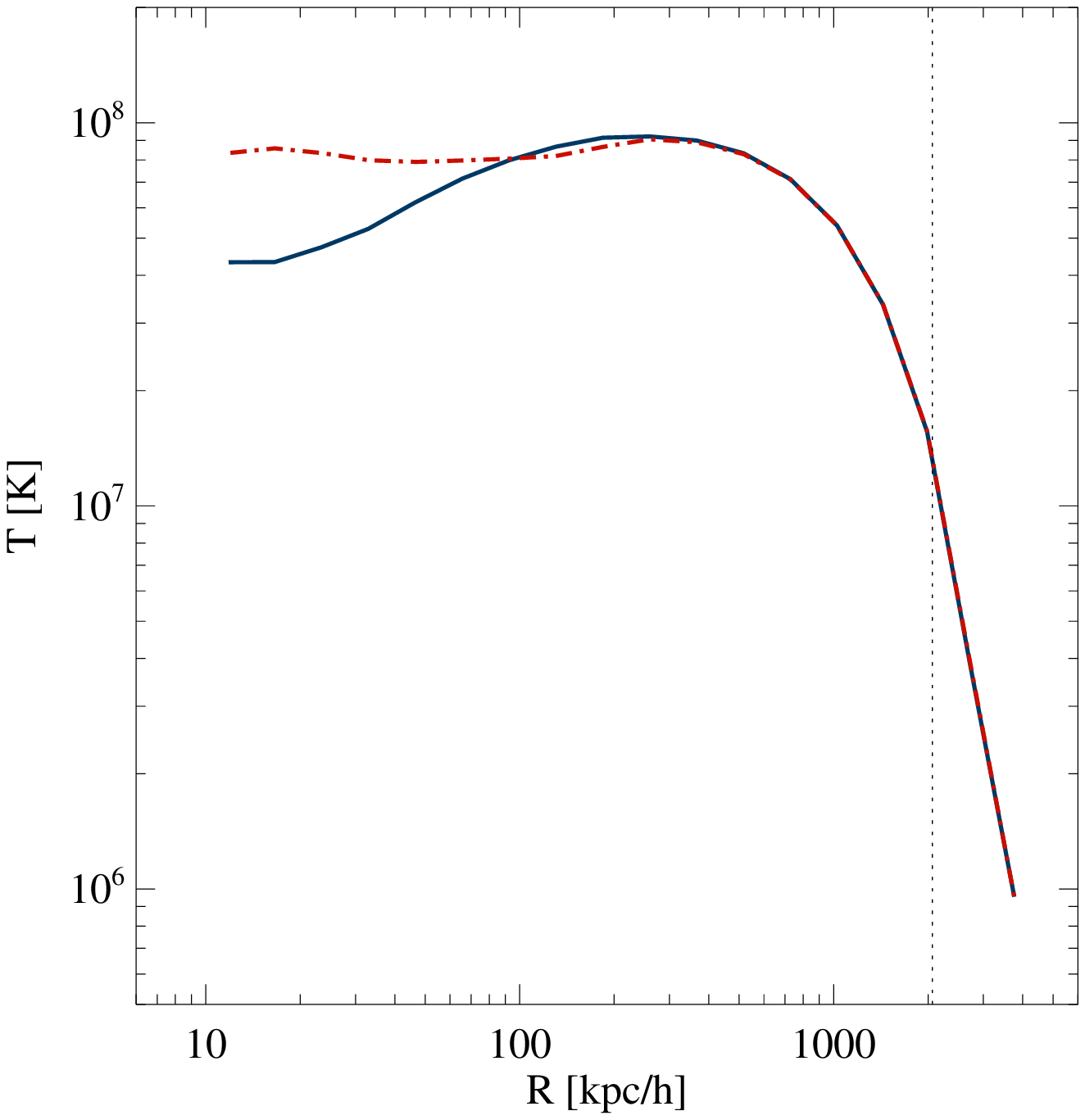,width=8.truecm,height=7.5truecm}
}
\hbox{
\psfig{file=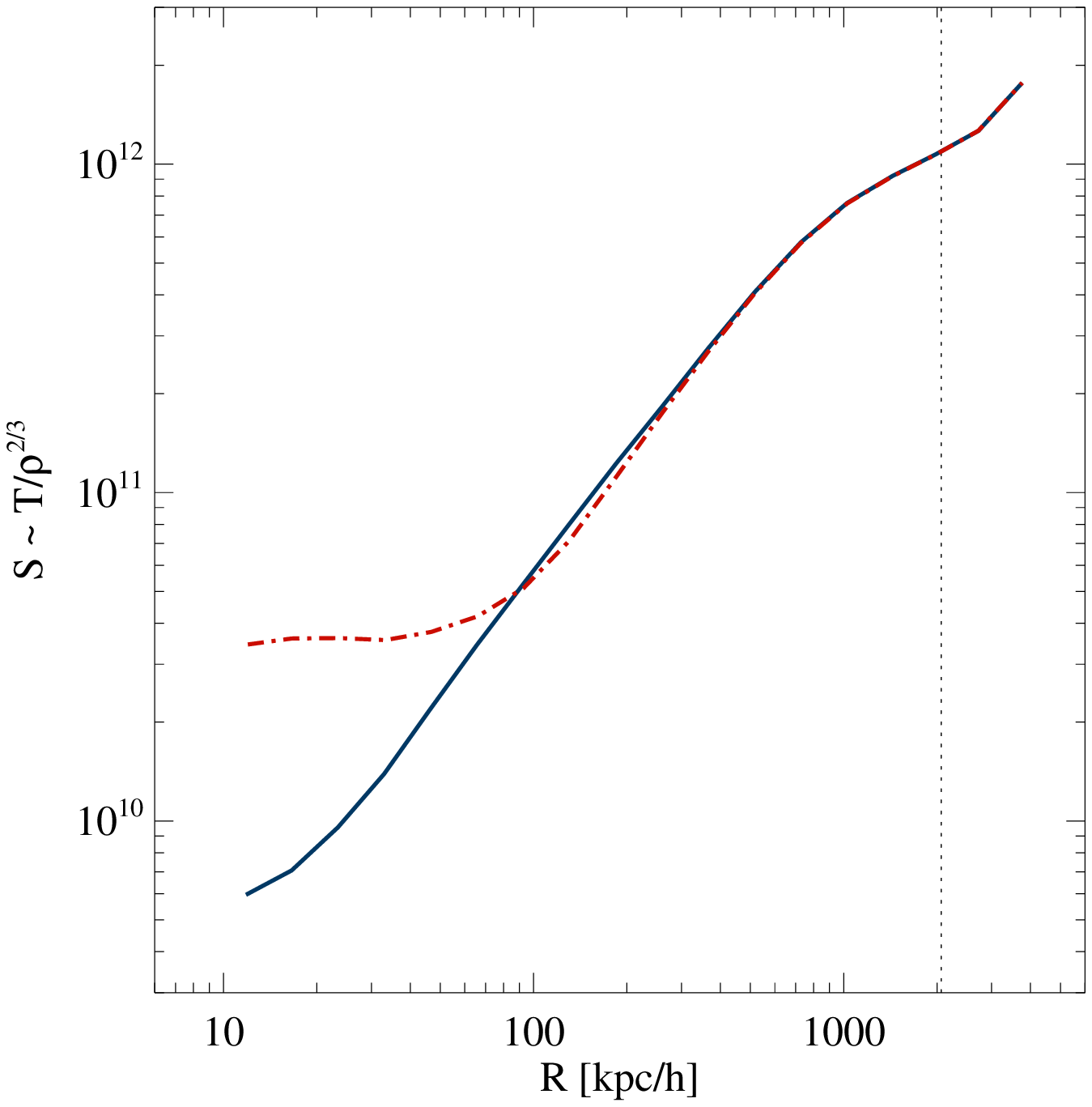,width=8.truecm,height=7.5truecm}
\hspace{1.0truecm}
\psfig{file=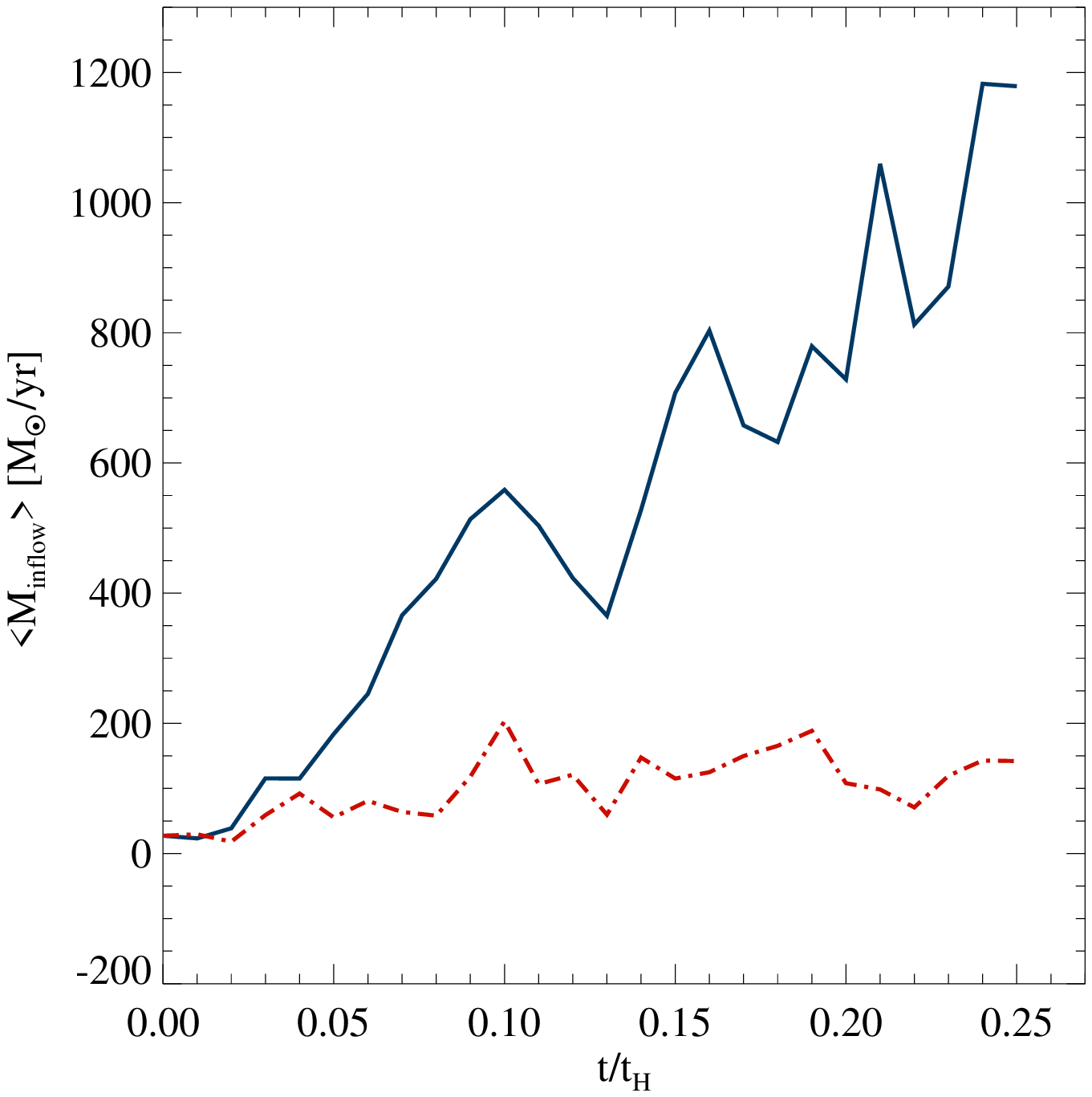,width=8.truecm,height=7.5truecm}
}
}}
\caption{Radial profiles of gas density (upper left panel),
  temperature (upper right panel) and entropy (lower left panel) of
  the isolated $10^{15} h^{-1}{\rm M}_{\odot}$ halo. Blue lines: run
  with cooling and star formation. Red lines: AGN feedback mechanism
  also included (random placement of bubbles). The vertical dotted
  lines denote $R_{200}$. Lower right panel: Mean mass inflow rate in
  the central $30\,h^{-1}{\rm kpc}$ as a function of time, normalised
  to the Hubble time. It can be seen that the mass deposition rate onto
  the central galaxy is substantially reduced with AGN heating and
  stabilised at $\sim 150\,{\rm M}_{\odot}{\rm yr}^{-1}$ after a
  relatively brief period of time.}
\label{Prof_iso}
\end{figure*}

In Figure \ref{Prof_iso}, we analyse the global gas properties of the
cluster in terms of radial profiles of density, temperature and
entropy of the hot gas component, i.e.~gas of the cold interstellar
medium is not included in the plots\footnote{The `cold' gas
component has been defined here as all gas cooler than $1\, \rm{keV}$
and with a density higher than the star formation density
threshold.}.  Bubble injection modifies the inner $100\,h^{-1}{\rm
kpc}$ substantially, reducing the density and increasing the
temperature profile.  Accordingly, the entropy of the central gas
particles is changed as well, and an entropy floor is formed.  The
lower right panel of Figure~\ref{Prof_iso} shows the mean gas inflow
rate in the central $30\,h^{-1}{\rm kpc}$. After a relatively brief
period of time, AGN heating regulates, in a stable fashion, the flow
of gas towards the centre, preventing the unrealistically high mass
deposition rates of a fully developed cooling flow, which can reach up
to $1200\,{\rm M}_{\odot}{\rm yr}^{-1}$ in the case without bubble
heating. Even though a repeated injection of bubbles along the same
spatial axis (``jet-like'') is somewhat more efficient than a random
placement within a sphere, the gas profiles have very similar trends
in both cases, indicating the robustness of the results with respect
to these details of the bubble injection scheme, at least in
situations free of secondary effects due to infalling structures and
mergers.

\begin{figure}
\bc
\centerline{\includegraphics[width=9.5truecm,height=8.5truecm]{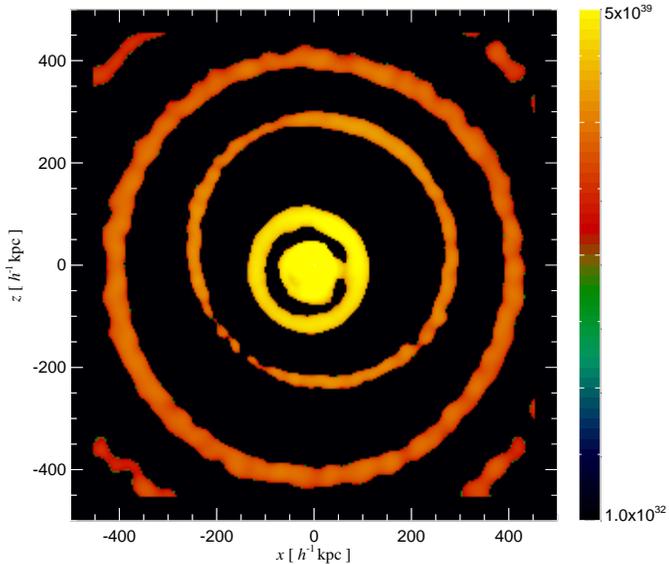}}
\caption{Unsharped masked map of the X--ray luminosity in the central
region of the $10^{15}\, h^{-1} {\rm M}_{\odot}$ isolated halo, at
time $\sim 2.2\,{\rm Gyr}$.  The unsharped masking has been performed
by subtracting a smoothed map from the original projected X-ray
emissivity map, with the smoothing scale set to $ 100\,h^{-1}{\rm
kpc}$.  It can be seen that the AGN bubble heating generates a number
of sound waves, which could gradually release their energy to the ICM
if they are viscously damped on their way to the cluster outskirts.}
\label{Soundwave}
\ec
\end{figure}

In Figure \ref{Soundwave}, we show unsharped masked maps of the X--ray
emissivity of one of our cluster models. The X-ray emission has been estimated
using the bremsstrahlung approximation, \be \label{Lx_eq} L_{\rm X} = 1.2 \times
10^{-24} \frac{1} {\mu^2 m_p^2} \sum_{i=0}^{N_{\rm gas}} m_{{\rm gas},i} \ 
\rho_{i} \ T_i^{1/2} \quad [{\rm erg\,s^{-1}}].  \ee The unsharp-masking has
been performed by subtracting from the original projected $L_{\rm X}$-map the same
map smoothed on a $100\,h^{-1}{\rm kpc}$ scale. A large number of
centrally concentrated ripples are clearly visible in the result.  These
ripples in the X--ray emissivity are sound waves generated by the expansion of
the bubble after the thermal energy is injected.  The sound waves
travel through 
the cluster, and if the IGM has a residual viscosity they can be dissipated,
providing a nonlocal heating of the central cluster volume.  We note that we
have explored different scales over which the smoothing in the unsharped
masked technique is performed, obtaining the sharpest and most prominent
features for smoothing scales corresponding to approximately $100\,h^{-1}{\rm
  kpc}$, which by order of magnitude agrees with the dimension of the ripples
themselves.

We find that the ripples reach distances of $\sim 800\,h^{-1}{\rm kpc}$ after
$1\,{\rm Gyr}$, translating to a velocity of order $\sim 10^3\,{\rm
  km\,s^{-1}}$, which matches the expected sound speed in the ICM of this
cluster. At larger radii, the sound waves are not detectable any more. Note
that we also expect that their velocity drops strongly in the outskirts of the
cluster, where the temperature and hence the sound speed decline.

Upon closer inspection, it can be seen that the ripples are actually slightly
offset from the cluster centre, with their midpoint directly matching the
initial coordinates of the injected bubble. Moreover, the ripples
progressively lose their intensity at larger radii, both due to a $1/r^2$
dilution of their intensity, and to a lesser extent, due to a damping caused
by the residual viscosity of our SPH scheme.  Note that some level of
numerical viscosity is instrinsic to all SPH schemes, even though we are
modelling an ideal gas.  Quantifying the exact magnitude of the resulting
effective viscosity is not trivial, also because it depends on the spatial
resolution achieved in the simulations. However, recent observations of
optical H$\alpha$-filaments \citep{Fabian03} suggest that the gas in the
central region of the cool-core cluster Perseus might be quite viscous, rather
then turbulent \cite[but see][for estimates of gas turbulence on smaller
scales]{Ensslin2005}. If ICM viscosity is relevant, it would imply a high rate
of dissipation of the energy contained in the sound waves at small radii. This
physical viscosity could be easily higher than the numerical viscosity we have
in our simulations.  Naturally, it is then desirable to treat the dissipation
process accurately, which requires an SPH discretization of the Navier-Stokes
equation combined with an assumed level of physical Spitzer-viscosity.
Recently, first mesh-based studies of isolated clusters with viscosity and
bubble heating have appeared
\citep[e.g.][]{Ruszkowski04,Reynolds05,Brueggen05}.  We plan to investigate
this theoretical issue in a forthcoming study.

\begin{figure}
\bc
\centerline{\includegraphics[width=8.6truecm,height=8.3truecm]{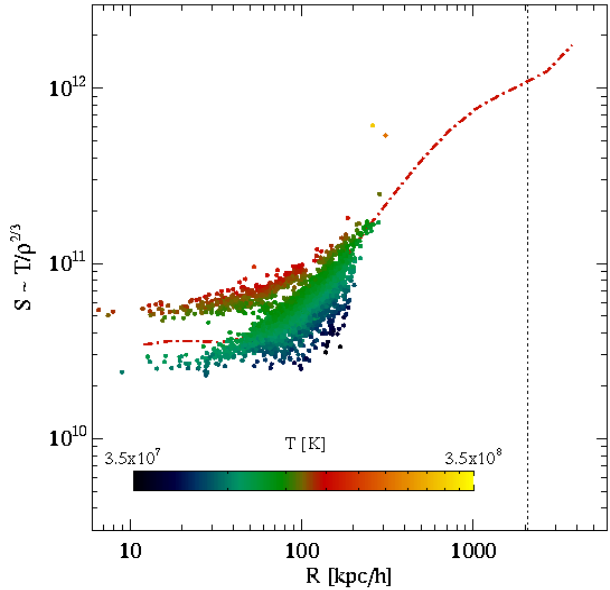}}
\caption{Mean radial entropy profile of the $10^{15}\, h^{-1}{\rm
  M}_{\odot}$ isolated halo. The result for the simulation with AGN
  feedback is given by the red dot-dashed line. The dots show the
  positions of the bubble particles, and they are colour-coded
  according to their temperature.}
\label{BubbleEntropy}
\ec
\end{figure}

In Figure~\ref{BubbleEntropy}, we show the locus of selected particles in the
$\log S - \log R$ plane, at a time equal to one quarter of the Hubble time
(which marks the end of most of our isolated simulations).  We selected only
particles that at least once belonged to one of the injected bubbles.  For
easier comparison, the mean radial entropy profile of the AGN heated cluster
(the same as in Figure~\ref{Prof_iso}) is plotted as a red dot-dashed line.
The dots shown for each individual bubble particle have been colour-coded
according to their relative temperature.  The particles of a recently injected
bubble have the highest temperature values. As expected, their entropy values
lie substantially above the average entropy of the cluster for the same range
of radii, implying that the bubble will rise due to buoyancy.  We thus expect
these bubble particles to reach larger radii and to lose some of their thermal
energy content during the rise, and this expectation is borne out by the
cooler particles from older bubbles.

 The radius at which the bubble particles attain the mean cluster entropy level
is set by their initial density and thermal content after injection, by their
capacity to shed some of the energy to the surrounding ICM, by their
radiative cooling efficiency, and by the amount of mixing. Given
that the different bubbles at various 
epochs have very similar initial temperature and mass,
Figure~\ref{BubbleEntropy} implies that the bubbles reduce their temperature
almost by one order of magnitude, from the injection instant to the final
equilibrium position. If we sum up the total energy injected over the entire
simulated time, and assume that it gets thermalized over the whole cluster, we
obtain that the gas particle temperature is increased by $\sim 0.2\,{\rm
  keV}$, which roughly corresponds to the bubble temperature decrement
mentioned above when the mass fraction of the bubbles is taken into account.
Thus, it appears that the radiative cooling is not severe inside the bubbles,
even though in the cluster as a whole it approximately balances the AGN
feedback mechanism.

We extended our investigation by considering `jet-like' injection of bubbles
and also a scenario in which the bubbles are inflated in a continuous fashion
over some time interval $t_{\rm inj}$. We tried values from $t_{\rm
  inj}=5\times10^7{\rm yrs}$ to $t_{\rm inj}=5\times10^8{\rm yrs}$, which is
significantly longer than the sound crossing time over the scale $R_{\rm bub}$
of the bubble. The maximum radius reached by the bubbles is essentially
invariant in all of these cases, yielding $\sim 250-300\,h^{-1}{\rm kpc}$ at
the final simulated epoch.  Also, the heating efficiency of buoyant bubbles
remains very similar, although the bubbles are somewhat more energetic in the
continuous injection scheme, presumably because cooling losses are reduced
here due to the expansion of the bubble before the bulk of the energy us
released.  This is directly reflected in an even lower mass deposition rate
onto the central object, which always occurs in a stable fashion with time,
where the gas cooling inflow is balanced by the AGN heating rate.
  
It is important to point out that the entropy content of the bubbles and the
maximum distance they can reach from the cluster centre depend upon the
equation of state assumed for the gas belonging to the bubbles. In all our
models bubbles have been simulated assuming the equation of state of an ideal
gas. However, radio observations indicate that AGN-driven bubbles contain
relativistic particles which possibly dominate over the thermal pressure
component, implying a softer equation of state.  Moreover, the energy contrast
of the individual bubbles is not very high in our approach, resulting in a
gentle ICM heating, without presence of significant shocks. In fact, most of
the observations of AGN-heated clusters point out that strong bubble-induced
shocks appear to be absent, although recently a few clusters with moderate
shocks in connection with AGN activity have been discovered
\citep{Fabian2003,Nulsen05,McNamara05}.  Therefore, due to the assumptions of
our model, the maximum possible distance reached by the buoyant bubbles in the
cluster atmosphere may be underestimated with respect to the case where the
relativistic particle component is modelled as well.
       
An important question is whether the bubbles are capable of raising cold gas
from the cluster centre and mixing it with higher entropy gas at larger radii.
Note that the X--ray observations of central metal abundance gradients put a
constraint on the amount of gas mixing \citep{Boehringer04} in the centre. In
order to address this issue, we analysed the gas metallicity distribution in
the clusters. Without AGN feedback mechanism, all metals produced in the
$10^{15}\, h^{-1} {\rm M}_{\odot}$ isolated halo are enclosed in the star
forming region and are confined to the very centre, where the density is
sufficiently high to allow star formation.  The bubble heating instead
produces both a reduction of the star formation rate by heating some of the
gas that would otherwise end-up in the cold interstellar medium, and a
spreading of metals away from the cluster centre.  Moreover, in our simple
model, the metals produced by the central stars are partly entrained and
transported along with the bubbles to larger radii. The spatial distribution
of the stars themselves is unaffected by the bubbles, however. It is important
to note that the metal mixing in our model due to the bubbles represents a
lower limit on the induced additional mixing, because fluid dynamical
processes that produce small-scale mixing tend to be underresolved in
cosmological simulations.

\subsection{Efficiency of bubble heating in halos of different mass}

The radiative cooling times of halos depend on their mass, both because of
their different virial temperatures, and because of the temperature dependence
of the cooling rate. Given that the masses of supermassive black holes, and
hence our assumed bubble feedback, have a mass dependence as well, we expect a
complex interplay between the different heating and cooling processes, and as
a result a mass-dependent efficiency of the feedback. This non-linear dynamics
can be best studied with detailed numerical analysis.  To this end, we
simulated isolated halos for a range of masses, starting from $10^{12}\,
h^{-1}{\rm M}_{\odot}$ and reaching up to $10^{15}\, h^{-1}{\rm M}_{\odot}$,
with the characteristics listed in Table~\ref{tab_simpar_iso}. The case of the
most massive halo has been discussed in some detail in the previous section,
so that we can restrict ourselves here to highlight the differences that occur
when the mass of the systems is lowered. The study of smaller halos gives also
direct insight into the question of how bubble feedback may affect the
hierarchical assembly of present-day massive clusters.

In our numerical simulations, the coupled dynamics resulting from
cooling, subsequent star formation and a given heating mechanism is
quite complex.  The introduction of a certain amount of heating can in
special situations even trigger an increase of the net cooling
rate. For example, let us consider star-forming gas with a density
higher than the density threshold set for star formation. If this cold
gas component receives an amount of thermal energy from bubble heating
that is insufficient to bring it back into the hot phase (where most
of the intracluster gas resides), then a counterintuitive process may
occur. In this case, the thermal energy injection prevents the cold
gas component from forming stars, but the local gas density will
remain comparatively high, which in turn stimulates even larger
radiative cooling losses. Thus, such a gentle heating, especially in
low mass systems where radiative cooling is more pronounced, can in
extreme cases even stimulate an increase of the cold gas component.

\begin{figure}
\bc
\centerline{\includegraphics[width=8.6truecm,height=8.3truecm]{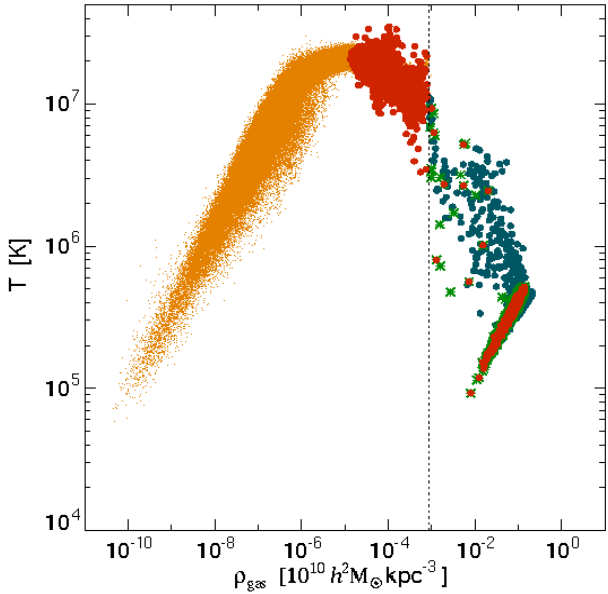}}
\caption{Phase-space diagram of gas temperature versus gas density for
the $10^{14}\, h^{-1}{\rm M}_{\odot}$ galaxy cluster. Small orange
dots are the particles outside the cold star-forming region, while big
blue dots denote the gas particles in the run with no feedback, at
high overdensities and with temperatures below $1\,{\rm keV}$. Green
star symbols are the particles satisfying the same criteria, but when
the AGN heating is included. Finally, the position of the bubble
particles is given by red dots.}
\label{Diagnostic_10^14}
\ec
\end{figure}

Analysing diagnostic phase-space diagrams like the $\log \rho - \log T$ plane,
one can notice that for the $10^{15}\,h^{-1} {\rm M}_{\odot}$ halo the
relative quantity of central cool gas is low, and it is promptly heated once
bubble injection is switched on.  Moreover, the energy per bubble particle is
sufficiently high to push the gas into the ``hot-phase'' (upwards and to the
left in the diagnostic diagram, as illustrated for the $10^{14}\,h^{-1} {\rm
  M}_{\odot}$ halo in Figure~\ref{Diagnostic_10^14}) and the star formation at
late simulated epochs is completely quenched.

Examining the $10^{14}\,h^{-1}{\rm M}_{\odot}$ isolated halo, we find that
bubbles are still very efficient in reducing the star formation rate,
e.g.~after the time $t_{\rm H}/4$, only 14\% of stars are formed with respect to the
run without AGN heating, but the total amount of cold gas, both in the
central regions and out to $R_{200}$, remains very similar. The diagnostic
diagram for this cluster is shown in Figure~\ref{Diagnostic_10^14}. The small
orange dots are the gas particles for the run without AGN feedback, and they
maintain practically the same position even when bubble heating is included.
The big blue dots are those gas particles that belong to the `cold phase',
here defined as having temperatures less than $1\,{\rm keV}$ and densities
higher than the density threshold set for star formation (as indicated by the
vertical dotted line).  With AGN heating included, the red dots denote the
bubble particles, while the green star symbols give the locations of cold
phase gas particles. Two different features due to the presence of bubble
feedback are readily apparent. First, the cold gas fraction for the
intermediate temperature range, from $5 \times 10^{5}\,{\rm K}$ to $10^7\,{\rm
  K}$, is substantially reduced in the case with feedback, because for most of
these particles it is possible to transport them back to the hot phase. In
contrast, along the line in the lower-right part of the diagram, there are
more particles when bubble heating is included. This can be explained by the
fact that this line is determined by the multiphase structure of the ISM,
given here in terms of an effective mass-weighted mean temperature, which is a
combination of the temperature of cold gas clouds and the one of the hot ISM
component \citep[see][]{SH03}. Hence, while the feedback mechanism reduces the
number of stars formed, it produces a higher amount of interstellar gas with
very low temperatures. Note that a large fraction of these cold gas particles
have been a part of a bubble at some earlier epoch.

In the lower mass system of mass $10^{13}h^{-1} {\rm M}_{\odot}$, the final
number of stars is also reduced, this time by only 13\%, and, as well as in
the previous case, the cold gas fraction is essentially unchanged. In the most
extreme case of the $10^{12}\,h^{-1} {\rm M}_{\odot}$ halo, the bubble heating
of the ICM is radiated away on a very short timescale without producing any
substantial modification.

We considered different injection assumptions in order to test whether this
effect can be alleviated and the bubble heating efficiency can be increased.
Focusing on the most difficult case of the $10^{12}\,h^{-1}{\rm M}_{\odot}$
halo, we explored a range of injection energies, from $5 \times 10^{56}\,{\rm
  erg}$ per bubble, to $3 \times 10^{57}\,{\rm erg}$.  Also, we injected
bubbles according to different schemes: in a spatially correlated `jet-like'
fashion, or by inflating the bubbles gradually for $t_{\rm
  inj}=5\times10^7 {\rm yrs}$, i.e.~by
releasing the energy slowly instead of instantaneously.  Finally, we also
tried a model where we artificially prevented bubble particles from cooling
for $10^7-10^8 {\rm yrs}$, motivated by some observational evidences
that the bubbles contain a non-thermal component, which would then be able to
maintain its pressure longer. From these experiments we conclude that relating
the bubble energetics to the underlying dark matter potential by invoking an
assumed relation with its black hole mass, does not in general provide stable
solutions for an efficient elimination of cooling flows in low mass systems.
Unrealistically high bubble energies are required in order to offset the
cooling flow, and also their thermal content has to be fine-tuned to a
restricted range of values, otherwise AGN heating may easily become so strong
that bubbles blow out substantial amounts of mass from the halo or cluster
potential well.
  
This clearly indicates an important deficiency of bubble models like the one
studied here. Since there is no internal trigger for bubble activity, a
self-regulation loop is missing, but this will be required for stability in
more general situations. Therefore, a more detailed physical scenario for the
triggering of bubble activity is needed  which couples the local physics of
the cooling flow with the activity and  energetics of AGN feedback.
We will discuss a number of possibilities for this in Section \ref{DIS}.

\subsection{Observational X--ray features of simulated bubbles}

In the last few years, a growing number of observations performed with
the Chandra X-ray telescope have found the presence of
so called X--ray cavities in central galaxy cluster regions
\citep[e.g.][]{McNamara00, Fabian00, Sanders02, Mazzotta02,
  Birzan04}. These observations support a scenario where relaxed 
cooling-flow/cooling-core clusters are heated due to the presence of a
supermassive black hole at their centres. Hence, it is interesting to
see whether the simulated bubbles produced by our model have
morphologies comparable to the observed ones. 

In order to perform this comparison accurately, it is necessary to produce
artificial emissivity maps which are as similar as possible to realistic
observations based on a finite exposure time on an X--ray
telescope. Recently, a similar analysis has also been performed by
  \cite{Brueggen05}.  To this 
end, we processed selected simulation outputs with the {\small X-MAS} software
package. A final result of this code is a photon event file quite similar to
the one an observer would acquire with the Chandra telescope in ACIS-S3 mode.
The instrument background has been included in our images by taking into
account an appropriate blank-sky background file \citep[]{Markevitch01}. A
detailed description of the {\small X-MAS} simulator can be found elsewhere
\citep[e.g.][]{Gardini04}; here we limit the description to the subsequent
analysis steps performed and the significance of the maps obtained.

\begin{figure*}
\centerline{
\hbox{
\psfig{file=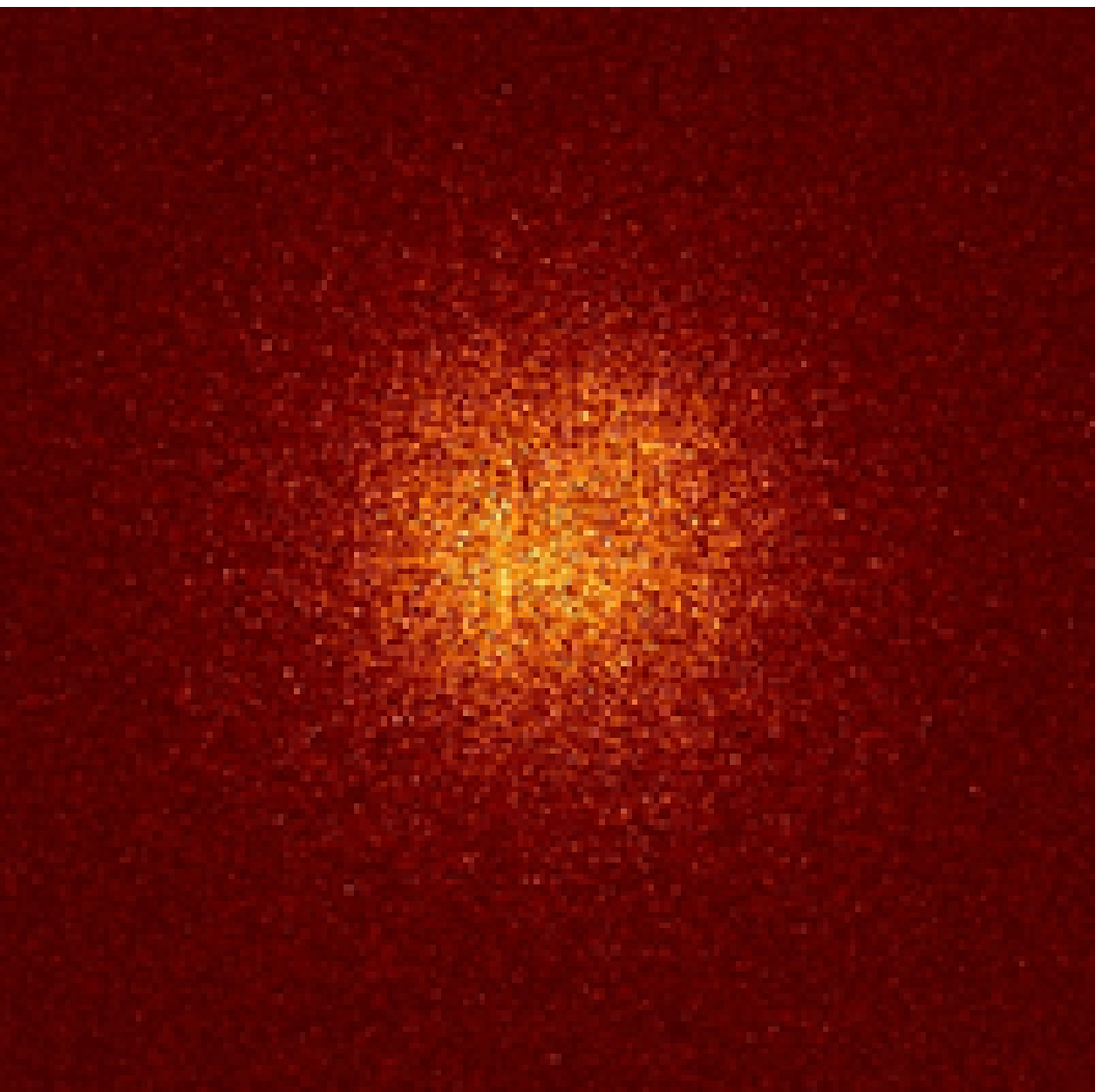,width=4truecm,height=4truecm}
\hspace{0.3truecm}
\psfig{file=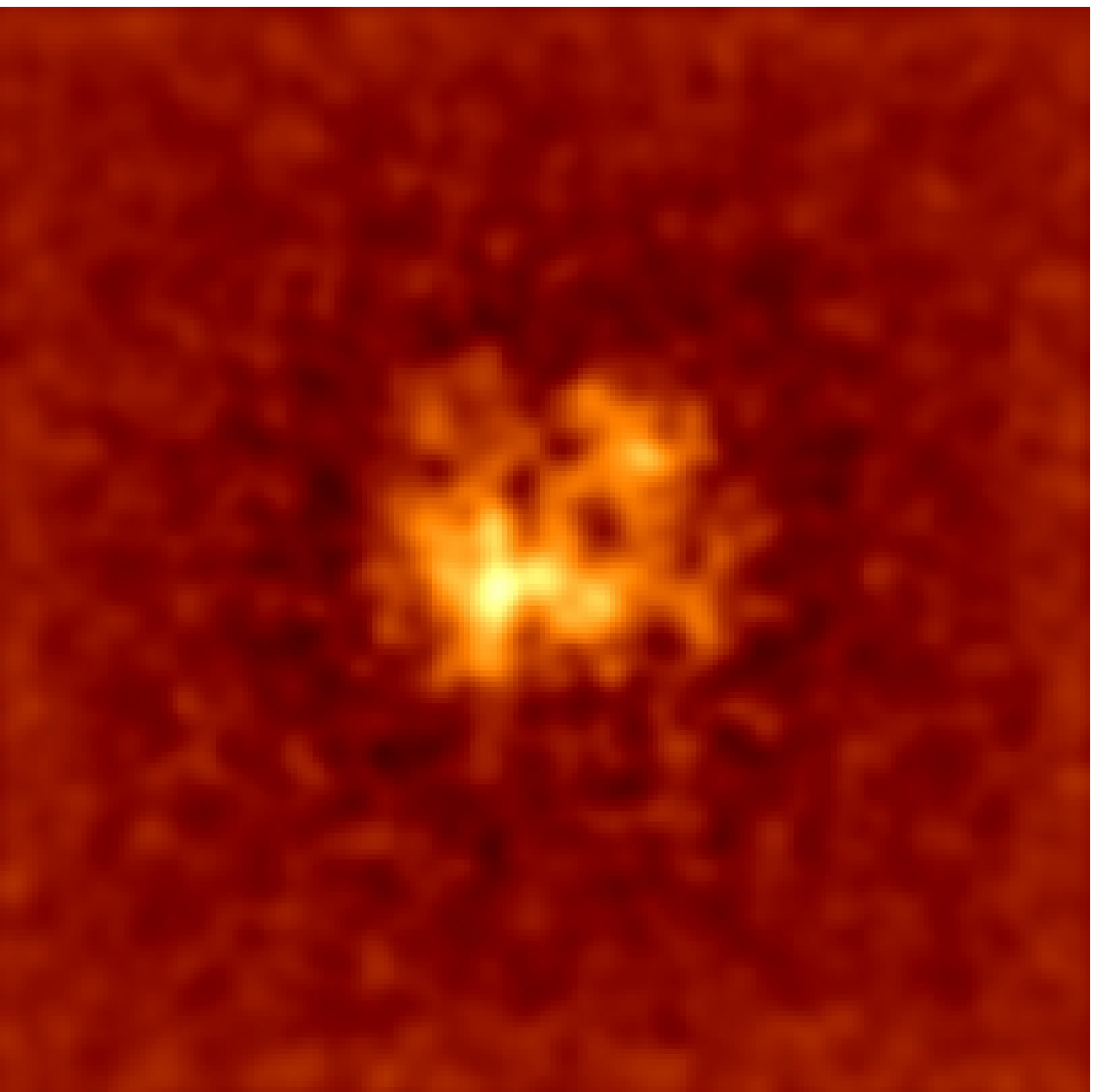,width=4truecm,height=4truecm}
\hspace{0.3truecm}
\psfig{file=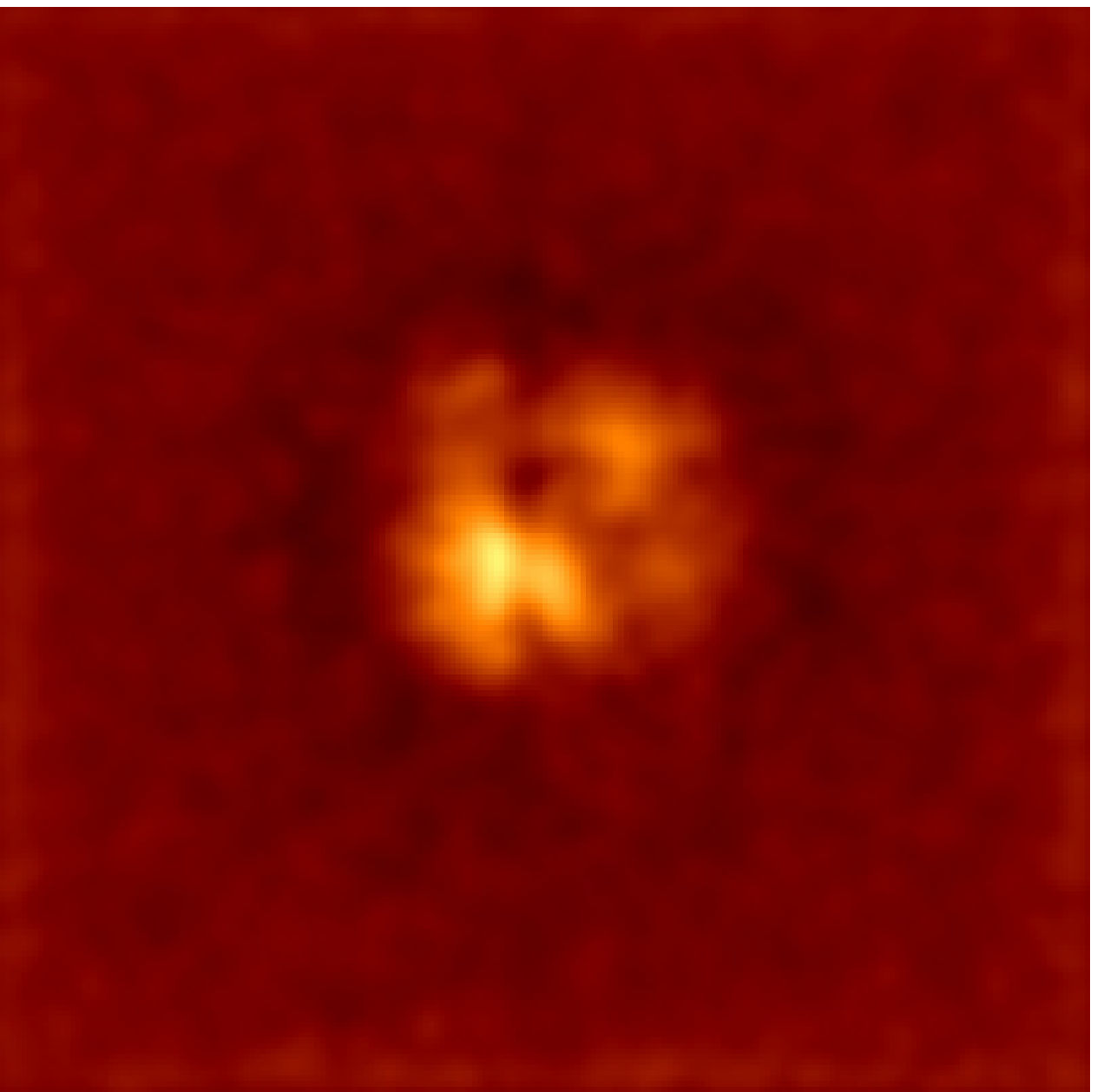,width=4truecm,height=4truecm}
\hspace{0.3truecm}
\psfig{file=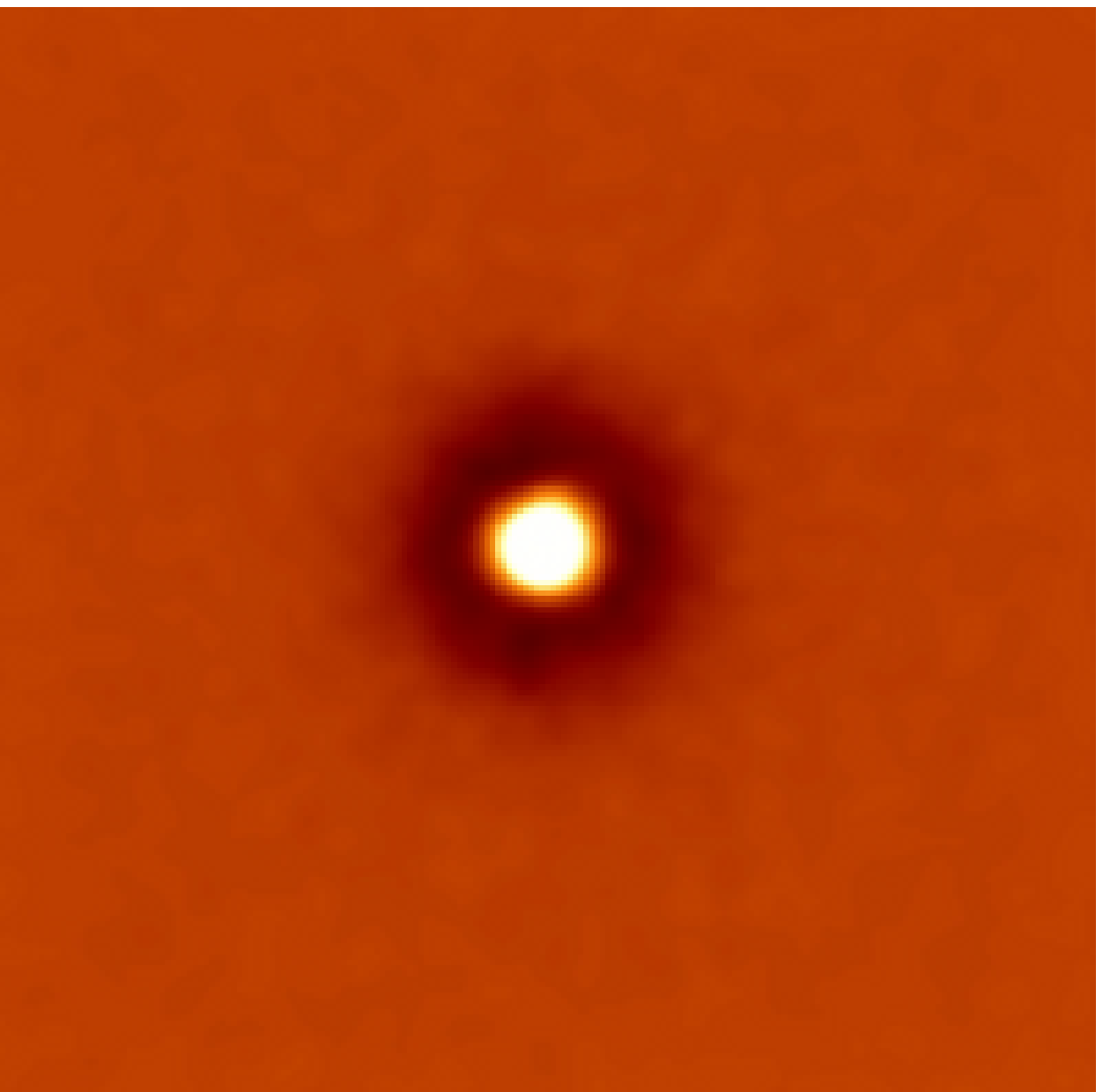,width=4truecm,height=4truecm}
}}
\caption{Artificial photon images for the $10^{15}\, h^{-1}{\rm
  M}_{\odot}$ isolated galaxy cluster obtained with the {\small X-MAS}
  software package. In all the panels, photons have been selected in
  the energy band $\Delta E = [0.3, 1.5]\,{\rm keV}$, maps were binned
  in 4pix/bin, and the physical scale of the maps is $\sim
  670\,{\rm kpc}$. The cluster emissivity map before applying any
  smoothing is illustrated in the first panel, while all the other
  panels have been obtained by unsharp masking previously smoothed
  images, as explained in the text. The second and the third panels
  show qualitatively very similar features. They are for the run with
  AGN feedback and differ only in the exposure time. Finally, the
  fourth panel shows how the same cluster appears when the bubble
  heating is absent.}
\label{XMAS_maps}
\end{figure*}

We have generated event files for different exposure times, ranging from
$10\,{\rm ks}$ to $1\,{\rm Ms}$, both for the runs with and without AGN
feedback.  We then selected a number of different energy bands, performed a
Gaussian smoothing on a range of scales, and finally produced unsharp masked
images to search for evidence of systematic departures of the flux from the
mean. In Figure~\ref{XMAS_maps}, we show photon images of the central region
of the $10^{15}\,h^{-1} {\rm M}_{\odot}$ isolated halo\footnote{We decided to
  perform this analysis on an isolated cluster in order to minimise other
  features in the X--ray emissivity that would have been imprinted by
  possible substructures or merger events. Further discussion of this issue in
  the cosmological framework is given in Section~\ref{S2_24}.}, both in a case
with and without additional AGN heating. The physical scale of the maps
corresponds to $\sim 670\,{\rm kpc}$ (2048pix), the energy band has been
chosen to be $\Delta E = [0.3,1.5]\,{\rm keV}$, and the maps have been
smoothed by summing the pixel fluxes in bins of 4 pixels. For this $\Delta E$,
the instrument background is minimised and the features due to the presence of
the bubbles are more evident.

The first panel of Figure~\ref{XMAS_maps} shows a photon image of the
AGN-heated cluster after $100\,{\rm ks}$ of exposure time, before
applying any smoothing. The rest of the plots have been created by
Gaussian smoothing them first on a small scale (3pix), then
re-smoothing the obtained image on a bigger scale (15pix), and finally
unsharp masking the two smoothed images (i.e. subtracting off the
15pix smoothed version). The smoothing scales have been selected to
maximise flux departures from the mean. The second and the third panel
illustrate how the bubbles introduce emissivity irregularities for two
different exposure times, $100\,{\rm ks}$ and $1\,{\rm Ms}$,
respectively. Finally, the fourth panel presents the cluster photon
image after an exposure time of $100\,{\rm ks}$ and with no AGN
feedback. 

It is clear that the bubbles generate characteristic fluctuations in
the photon counts, both creating bright features and X--ray
depressions. The typical dimension of these irregularities is $\sim
50\,{\rm kpc}$, very similar to the size of the bubbles
themselves. The hot spots can be associated with the most recent
bubble events, containing particles still significantly hotter than
the surrounding ICM (as can be also seen from
Figure~\ref{BubbleEntropy}), whereas the depressions in photon counts
can be explained with previous bubble episodes. These peculiarities in
emissivity are completely absent in the galaxy cluster without AGN
feedback, indicating that they are real features and not artifacts
produced by counting statistics or our analysis. The only feature that
is present in the fourth panel of Figure~\ref{XMAS_maps} is the
central excess due to the prominent cooling flow of $ \sim 100\,{\rm
kpc}$ size in diameter. Based on these results we conclude that in a
relaxed galaxy cluster, departures from the mean flux stemming from
bubbles with characteristics as given by our model can be detected,
provided the exposure times are long enough, and provided that other
sources of photon count deviations are absent or negligible.

\section{Effect of AGN bubble heating in cosmological simulations} \label{AGN_cosmo}

\subsection{Simulation characteristics}
As a next step in our analysis, we consider the importance of AGN feedback in
full cosmological simulations of cluster formation. To this end, we selected a
number of galaxy clusters with a wide range of masses from a parent dark
matter simulation, and resimulated them with higher resolution including gas
dynamics.  Our hydrodynamical simulations account for cooling and star
formation, and additional AGN heating as well. In a subset of our runs, we also
included galactic winds powered by star formation, as implemented by
\cite{SH03}. 
Models with winds provide a better description of some galaxy cluster
properties, e.g.~the distribution of metals, but in order to be able to
cleanly identify the effects of bubble heating, we focus most of our
analysis on simulations without winds. Where appropriate, we will
however briefly discuss any changes of our results when winds are also
included.

Our primary series of simulations consist of resimulations of a cluster
extracted from the GIF $\Lambda$CDM simulation \citep{K99}.  We selected the
second most massive galaxy cluster from this simulation and constructed higher
resolution initial conditions for it using the ``Zoomed Initial
Conditions'' technique \citep{T97}. We carried out two runs with different
resolution in order to test numerical convergence of our results. These
clusters are equivalent to the ones used by \citet{Springel01}, but with gas
(from now on we will refer to these runs as S1 and S2, respectively). The
simulations have been evolved from an initial redshift of $z_{\rm ini}=30$ for
S1, and $z_{\rm ini}=50$ for S2, producing 25 outputs uniformly spaced in the
logarithm of the expansion factor. Additionally, we selected two other
clusters with final virial masses substantially smaller (g676) and bigger (g1)
than the S1/S2-cluster. These clusters have been extracted form a cosmological
$\Lambda$CDM simulation of box-size $479\, h^{-1}{\rm Mpc}$
\citep{Yoshida01,Jenkins01}, and again have been resimulated with the ZIC
technique at higher resolution \citep{Dolag2004}.

The simulation of the g1 galaxy cluster includes several other smaller
systems in the high-resolution region which we also included in our analysis.
Tables~\ref{tab_simpar} and~\ref{tab_GCpar} provide a summary of the
main properties of our set of simulated galaxy clusters. In all runs,
the cosmological parameters were that 
of a flat concordance $\Lambda$CDM model, with $\Omega_{m}=0.3$,
$\Omega_{\Lambda}=0.7$, $\Omega_b=0.04$, a normalisation of the power spectrum
given by $\sigma_8=0.9$, and a Hubble constant at the present epoch of $H=70 \ 
{\rm km} \ {\rm s}^{-1} \ {\rm Mpc}^{-1}$.

\begin{table*}
\bc
\begin{tabular}{crrccccc}
\hline
\hline
Simulation & $N_{\rm HR}$ & $N_{\rm gas}$ & $m_{\rm DM}$
[$\,h^{-1}{\rm M}_\odot\,$] & $m_{\rm gas}$  [$\,h^{-1}{\rm
    M}_\odot\,$] & $z_{\rm start}$ & $z_{\rm end}$ & $\epsilon$
[$\,h^{-1}{\rm kpc}\,$] \\
\hline
S1 & $450088$ & $450088$   & $5.96\times 10^{9}$ & $0.92\times 10^{9}$
& $30$ & $0$  & $14.5$  \\
S2 & $1999978$ & $1999978$ & $1.18\times 10^{9}$ & $0.18\times 10^{9}$
& $50$ & $0$ & $8.5$ \\
g676 & $314518$  & $314518$  & $1.13\times 10^9$ & $0.17\times 10^9$ &
$60$ & $0$ & $5.0$ \\
g1   & $4937886$ & $4937886$ & $1.13\times 10^9$ & $0.17\times 10^9$ &
$60$ & $0$ & $5.0$ \\

\hline
\hline
\end{tabular}
\caption{Numerical parameters of the cosmological galaxy cluster simulations used in this
  study. The values listed from the second to the fifth column refer
  to the number and to the mass of high resolution dark matter
  particles and of gas particles. Note that the actual values of
  $N_{\rm gas}$ and $m_{\rm gas}$ vary in time due to star formation.
  The last three columns give the initial and final redshifts of the
  runs, and the gravitational softening length $\epsilon$.
\label{tab_simpar}}
\ec
\end{table*}

\begin{table*}
\bc
\begin{tabular}{cccccc}
\hline
\hline
Cluster & $R_{\rm 200}$ [$\,h^{-1}{\rm kpc}\,$] & $M_{\rm 200}$
[$\,h^{-1}{\rm M}_\odot\,$] & $T_{\rm mw}$
[$K$] & $T_{\rm ew}$ [$K$] & $L_{\rm X}$ [$\,\rm ergs^{-1}\,$]    \\
\hline
S1     & $2427$ & $9.98\times10^{14}$ & $5.2\times10^7$ &
$8.7\times10^7$ & $9.8\times10^{44}$ \\
S2     & $2466$ & $1.05\times10^{15}$ & $5.1\times10^7$ &
$8.8\times10^7$ & $9.9\times10^{44}$ \\
g676   & $1176$ & $1.13\times10^{14}$ & $1.4\times10^7$ &
$2.6\times10^7$ & $1.6\times10^{43}$ \\
g1\_a  & $2857$ & $1.63\times10^{15}$ & $7.3\times10^7$ &
$1.3\times10^8$ & $1.0\times10^{45}$\\
g1\_b  & $1914$ & $4.89\times10^{14}$ & $3.1\times10^7$ &
$4.1\times10^7$ & $1.2\times10^{44}$\\
g1\_c  & $1448$ & $2.12\times10^{14}$ & $1.5\times10^7$ &
$2.5\times10^7$ & $3.0\times10^{43}$\\
g1\_d  & $1258$ & $1.39\times10^{14}$ & $1.6\times10^7$ &
$1.9\times10^7$ & $1.8\times10^{43}$\\
g1\_e  & $1085$ & $8.92\times10^{13}$ & $1.1\times10^7$ &
$1.6\times10^7$ & $7.0\times10^{42}$\\
\hline
\hline
\end{tabular}
\caption{Physical properties of our sample of simulated galaxy
  clusters at $z=0$ and at $200\rho_c$. For different galaxy clusters,
  labeled in the 
  first column, cluster radius, total mass, mass-- and
  emission--weighted gas temperature and X--ray luminosity are listed,
  respectively. Note that the values refer to the simulations with
  cooling and star formation, without bubble heating included.     
\label{tab_GCpar}}
\ec
\end{table*}

Unlike simulations of isolated clusters, cosmological simulations require a
special algorithmic method for placing bubbles, since the position of the
cluster centre and properties like virial mass are not known a priori, and
change with time.  To address this problem, we run for every AGN duty cycle a
fast parallel FOF group finder on the fly as a part of the simulation code,
obtaining a list of all halos with their basic properties. We then adopt two
different schemes for injecting bubbles.  We either consider only the most
massive halo found in the high-resolution zone, which can be identified with
the most massive progenitor of the final cluster, or we introduce AGN-driven
bubbles in all halos above a given fixed mass threshold value. The injection
of bubbles in all large halos is motivated by the observational indications
that probably most if not all of the spheroidal galaxies harbour a
supermassive black hole at their centres.  Note that the larger number of
bubbles in this second scenario can also cause additional effects during
merger events, where bubble material can be torn apart and mixed into outer
regions of the cluster.

\subsection{Global gas properties of simulated galaxy clusters}\label{Gas prop}

Before analysing the properties of simulated galaxy clusters with and
without AGN bubble heating, we briefly discuss issues of numerical
convergence. For this purpose we consider the S1 and S2 runs, and
compare their spherically averaged radial profiles at two epochs,
namely at $z=3$ and $z=0$\footnote{These two epochs delimit the time
interval during which the bubble heating is active, and hence the
period of time where our analysis is performed.}. The dark
matter and stellar density profiles of the S1 and S2 galaxy clusters are
in excellent agreement at both epochs, as well as the gas density
profiles, with the residual differences at early times being
consistent with what is expected from the increased noise.  Both the
emission-weighted and the mass-weighted temperature profiles do not
noticeably differ at low redshifts, while there is a hint a of
slightly higher gas temperature for the S2 cluster at early
times. Thus, we conclude that for radii larger than the gravitational
softening length the properties of our simulated galaxy clusters are
numerically robust and have converged quite well.

\begin{figure*}
\centerline{
\hbox{
\psfig{file=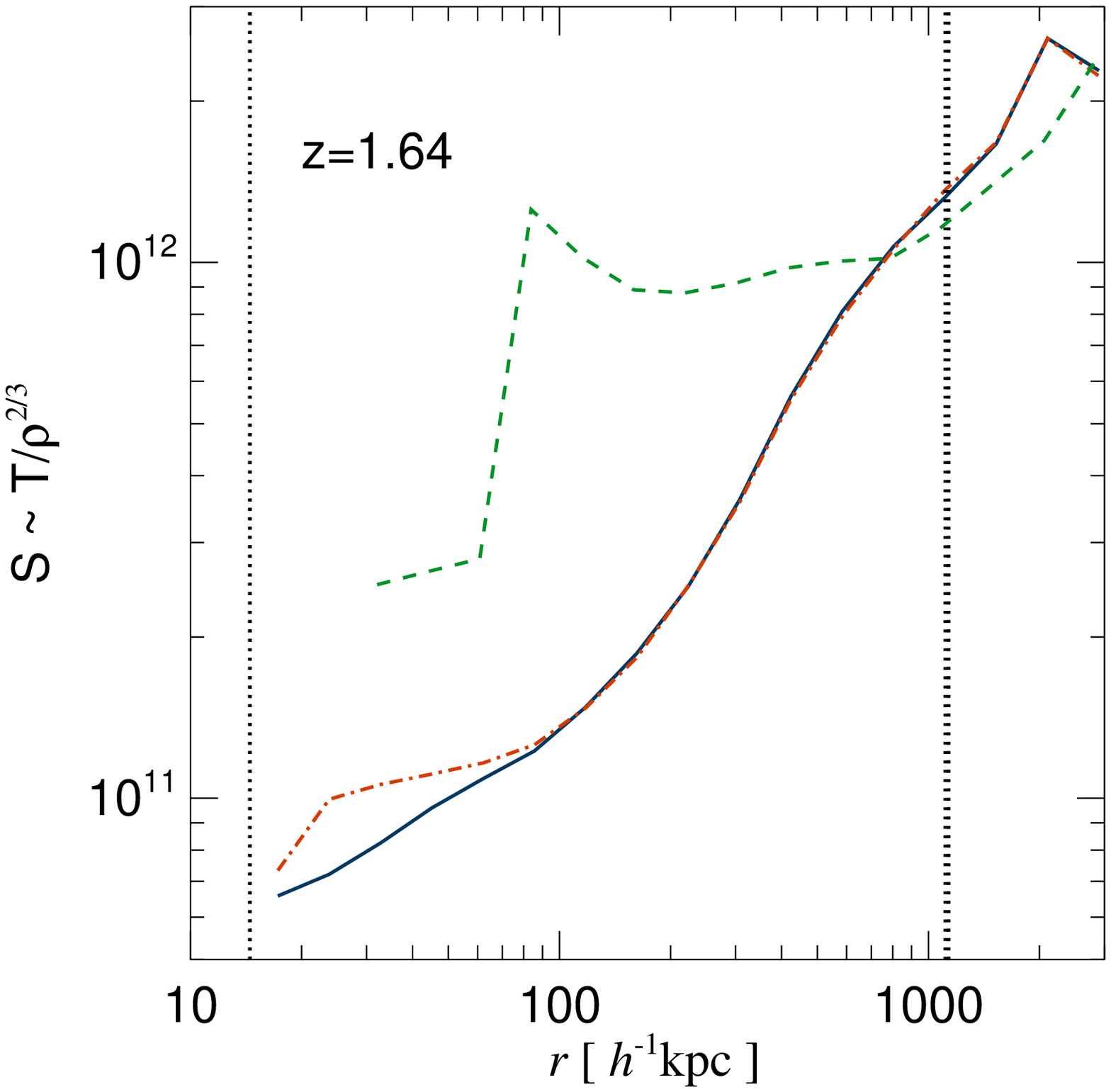,width=6.2truecm,height=6.2truecm}
\hspace{-0.3truecm}
\psfig{file=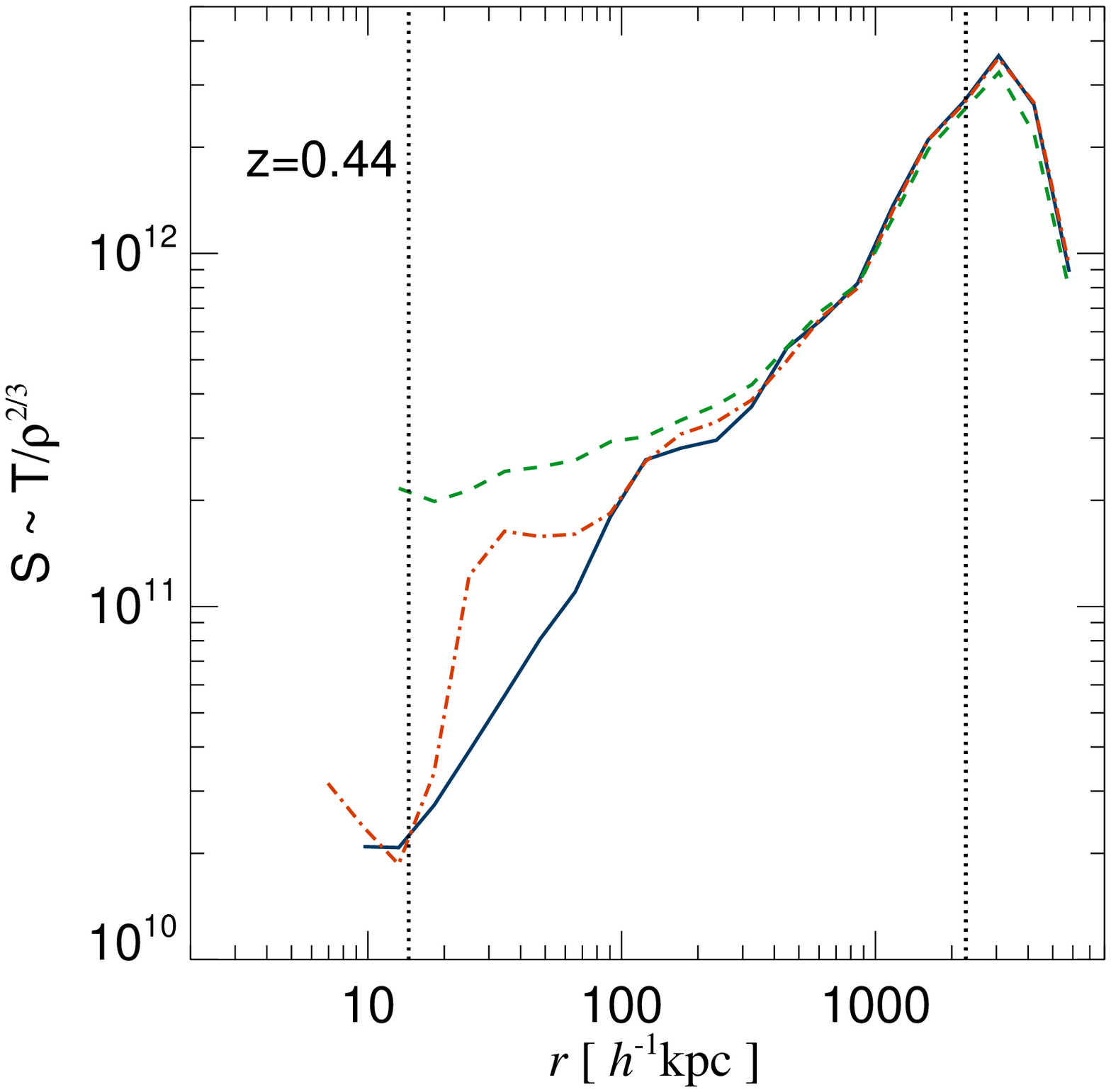,width=6.2truecm,height=6.2truecm}
\hspace{-0.3truecm}
\psfig{file=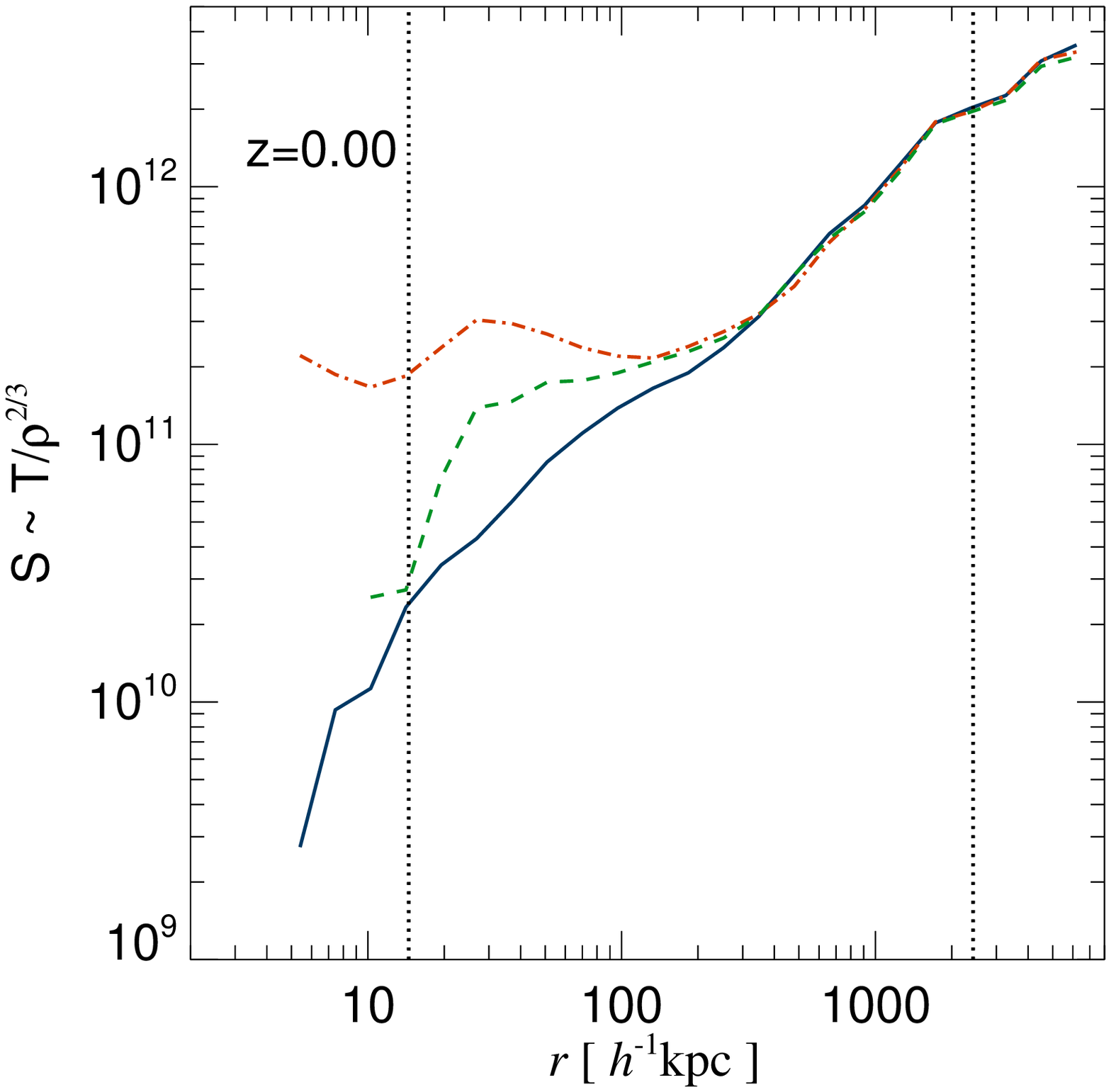,width=6.2truecm,height=6.2truecm}
}}
\caption{Radial profiles of gas entropy of the S1 galaxy cluster
  simulation. The blue continuous lines are for the run with cooling and star
  formation only, the red dot-dashed lines refer to the case when AGN heating
  based on our ``Magorrian-like scheme'' is included as well, while the green
  dashed lines are for the BHAR-based model. The dotted vertical lines
  denote the 
  gravitational softening and the virial radius at the given redshift; the
  latter is indicated in the upper-left corner. The profiles do not extent
  down to vanishingly small radii because they have been calculated
  exclusively from the hot gas component (basically the gas above $1\,{\rm
    keV}$), excluding the cold dense gas in the centre.  Note that both the
  spatial and the entropy scale vary between the three different panels.}
\label{Shot_S1_z}
\end{figure*}

\begin{figure*}
\centerline{\vbox{
\hbox{
\psfig{file=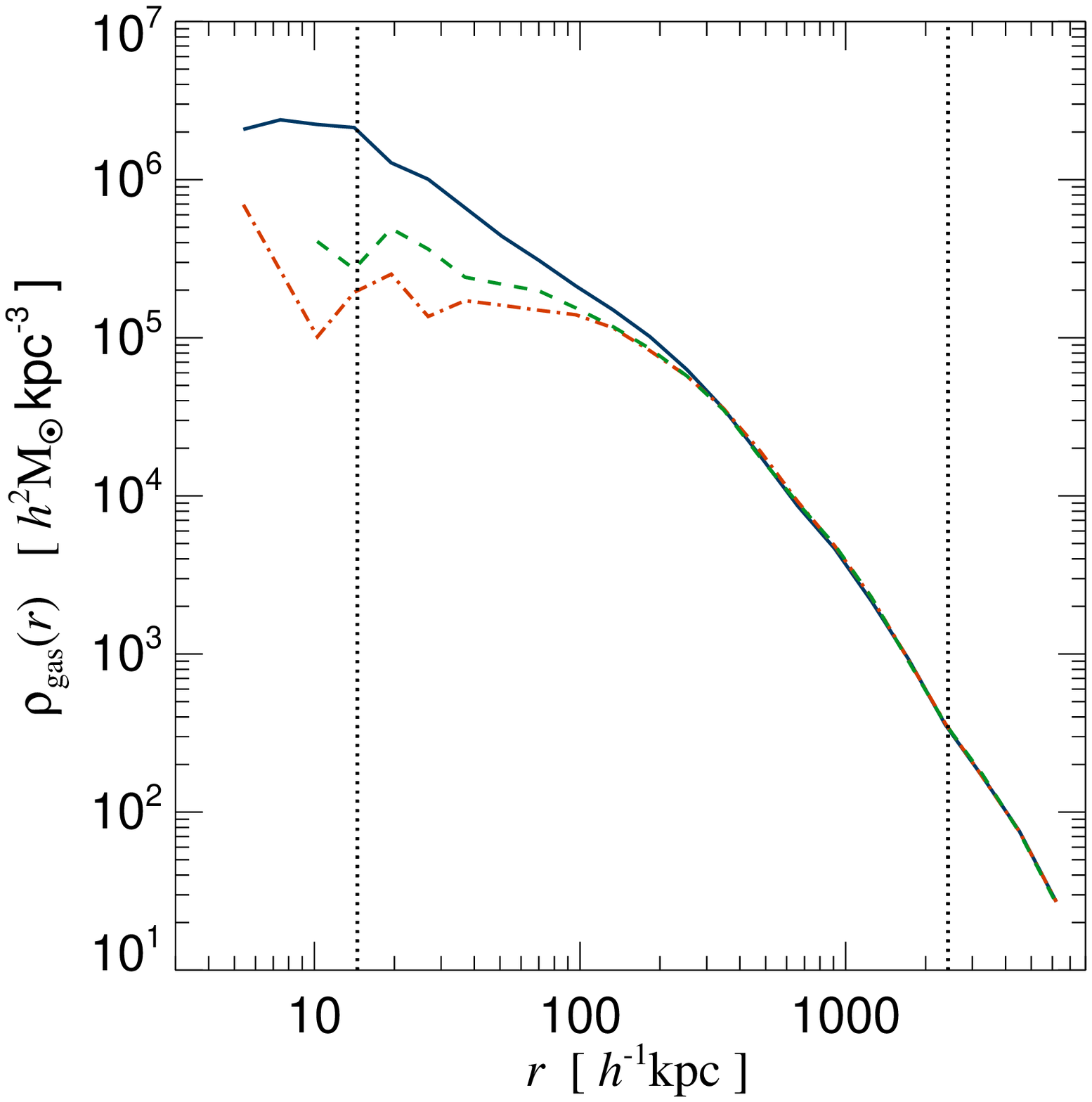,width=8.truecm,height=8truecm}
\hspace{0.truecm}
\psfig{file=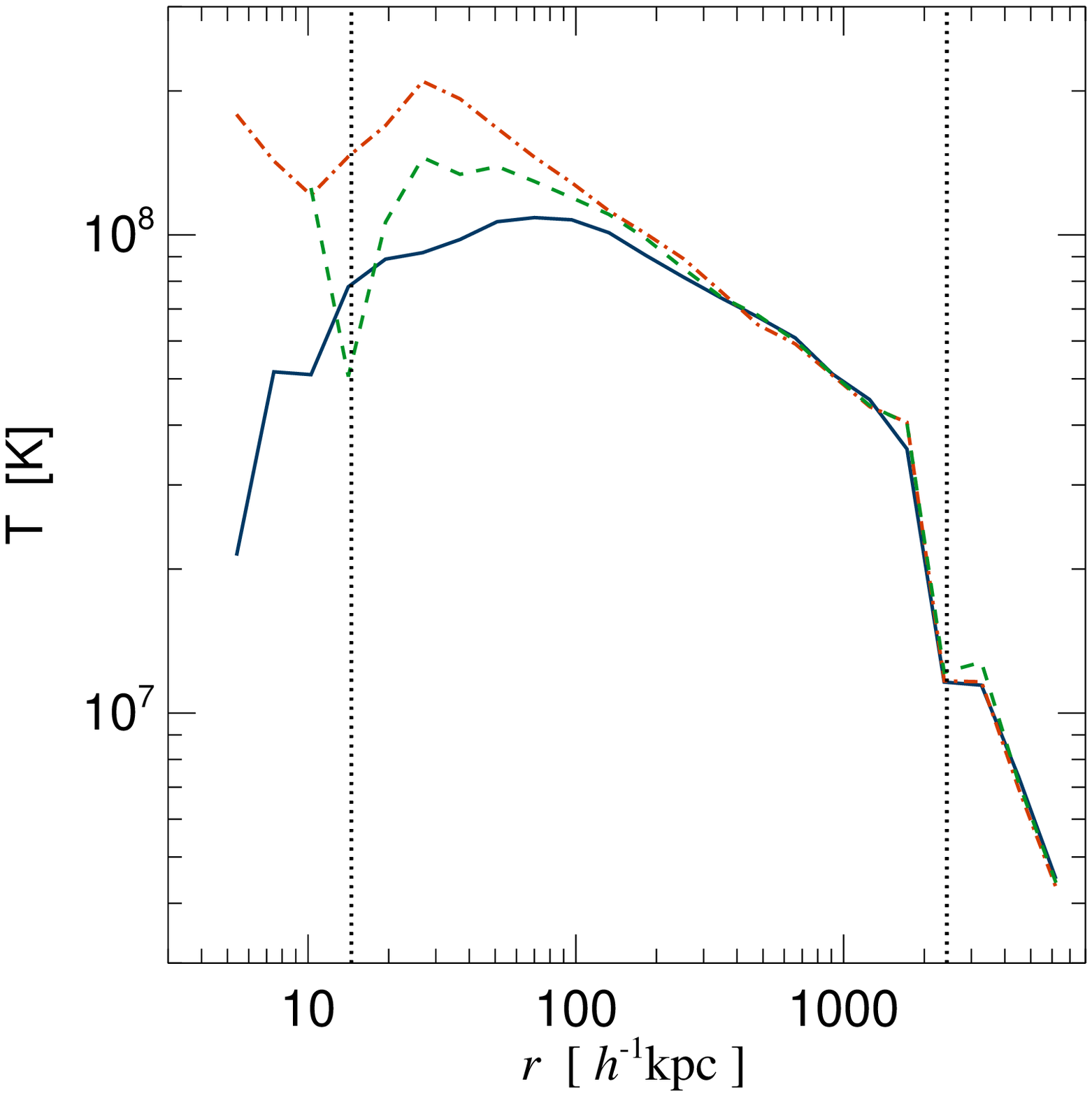,width=8.truecm,height=8truecm}
}
\hbox{
\psfig{file=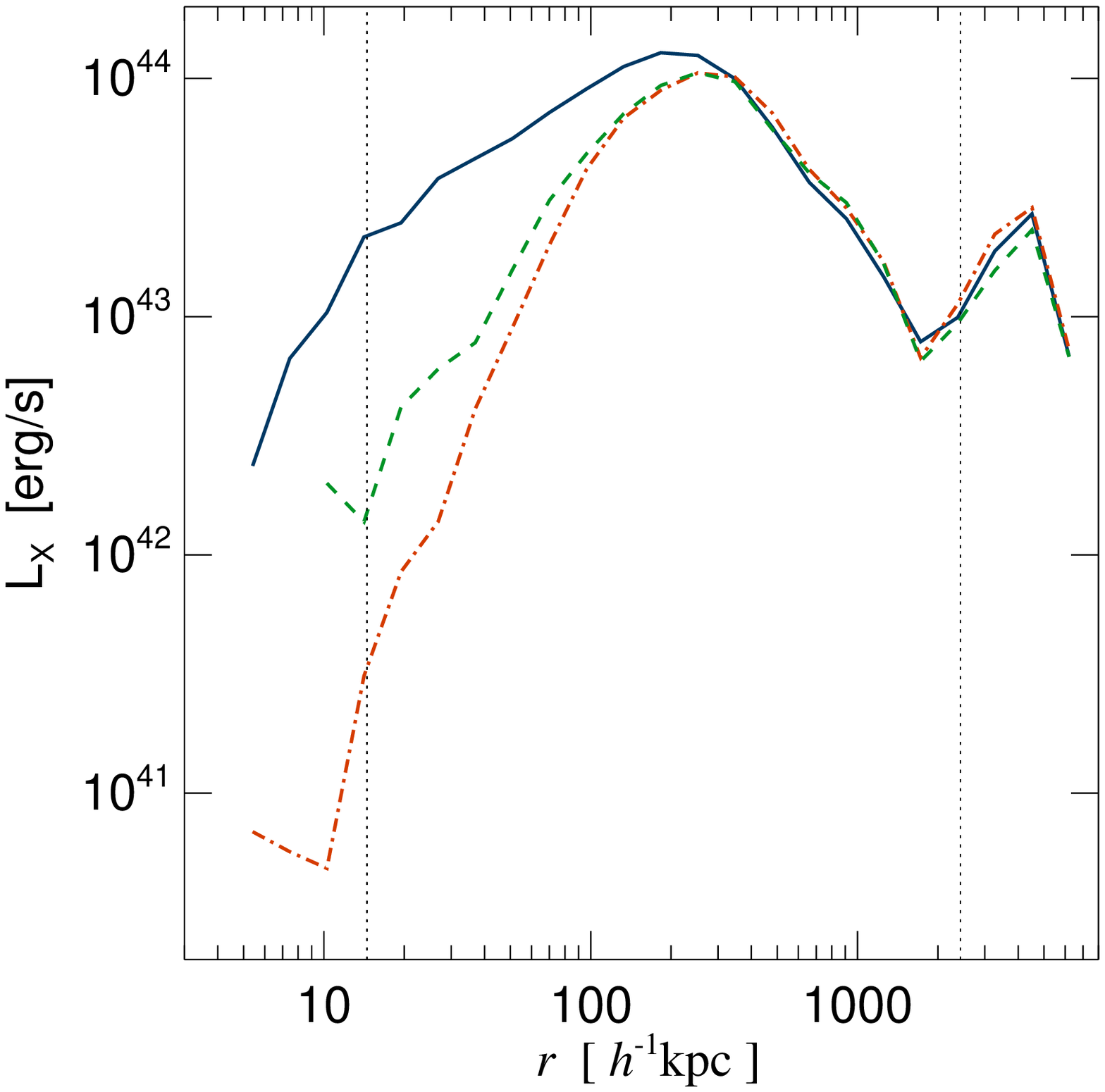,width=8.truecm,height=8truecm}
\hspace{0.truecm}
\psfig{file=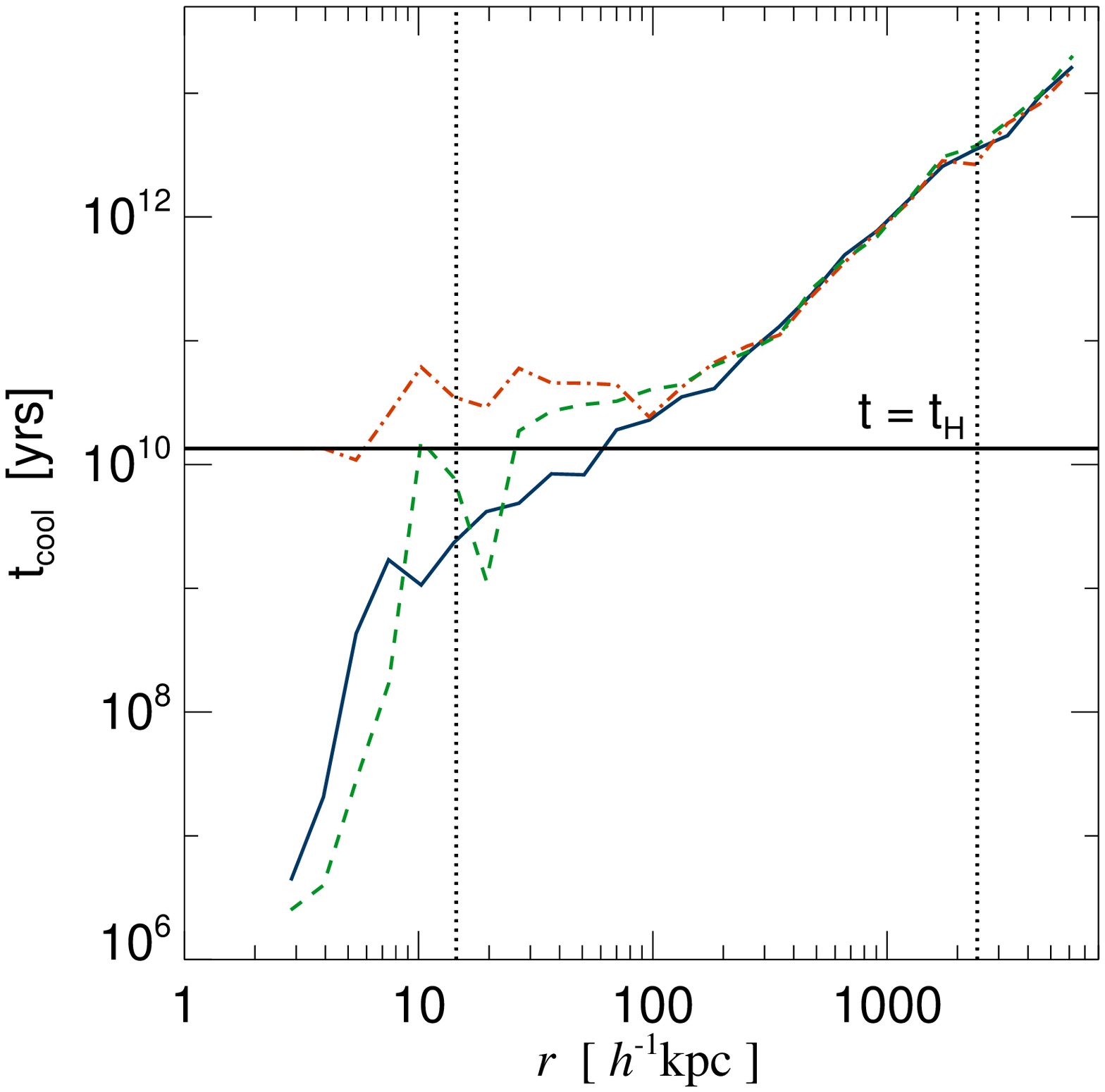,width=8.truecm,height=8truecm}
}
}}
\caption{Radial profiles at $z=0$ of gas density (upper-left panel),
  emission-weighted temperature (upper-right panel), and X--ray
  luminosity (lower-left panel) estimated with the bremsstrahlung
  approximation given by eq.~(\ref{Lx_eq}). The lower-right panel shows
  the cooling time of all gas particles as a function of radius,
  computed using eq.~(\ref{tcool_eq}). The continuous horizontal line
  indicates the Hubble time at $z=0$. The blue continuous lines are
  for the case without AGN feedback, the red dot-dashed lines are for
  the model where $E_{\rm bub} \propto M_{200}^{4/3}(z)$, while the
  green dashed lines are for the scenario where the bubble energy
  depends on the BHAR given by eq.~(\ref{Ebub_BHAR_eq}).}
\label{Prof_S1}
\end{figure*}

In Figure~\ref{Shot_S1_z}, we compare the gas entropy profiles with
and without bubble heating at three different epochs, $z=1.64$,
$z=0.44$ and $z=0$. For this comparison, we use both of our AGN
heating models, the one based on the Magorrian relationship and the
``BHAR model''. The entropy has been estimated by calculating the
ratio of the emission-weighted temperature to the gas density
elevated to the $2/3$ power, where the temperature is measured in
Kelvin and the gas density is given in $h^2 {\rm M}_{\odot}{\rm
kpc}^{-3}$.  We selected only the hot gas component to compute these
profiles, i.e.~we avoided the cold, star-forming gas by imposing a cut
in density and ionisation level. With this choice, the gas profiles
are smoother because they have no contributions from cool
substructures at various radii.  Nonetheless, it is also important to
investigate the fate of the cold gas in the central cluster region, an
issue we will address separately in Section~\ref{Stellar prop}. The
blue continuous lines are for the run without AGN feedback, the red
dot-dashed lines correspond to the ``Magorrian model'', while the
green dashed lines are for the ``BHAR model''. The vertical dotted lines
denote the softening length and the virial radius at the different
epochs, respectively. When $E_{\rm bub}$ is computed from the ``BHAR
model'', the effect of bubbles is less prominent at low redshifts than
in our other AGN heating scenario.  However, at early times the
situation is opposite, as expected. Here the ``BHAR model'' heats the
ICM gas more prominently, right from the initial injection epoch
($z=3$) until $z \sim 0.4$. It is interesting to note that at $z \sim
0.4$ the bubble energy content is already much lower than in the
``Magorrian model'', indicating that the efficient heating at early
times has a prolonged effect on the thermodynamic state of the galaxy
cluster. We find that the ``Magorrian model'' starts to affect the gas
entropy from $z \approx 1.6$ in a noticeable way and it becomes very
important at late times, where it suppresses the cooling flow
completely at $z=0$.
As a further comparison, we show in Figure~\ref{Prof_S1} radial
profiles of gas density, temperature, X--ray luminosity, and 
local cooling time for the S1 galaxy cluster at $z=0$. AGN heating
alters the gas properties out to a radius of $\approx 300\,h^{-1}{\rm
kpc}$, reducing the central gas density and increasing its
temperature. The X--ray emissivity, being more sensitive to the gas
density, is substantially lower when bubble feedback is active. In the
lower-right panel of Figure~\ref{Prof_S1}, we show the cooling time of
all ICM gas, estimated isobarically as \citep{Sarazin} 
\be \label{tcool_eq} t_{\rm cool} = 8.5
\times 10^{10} \bigg( \frac{n_p}{10^{-3}\,{\rm cm}^{-3}} \bigg)^{-1} \bigg(
\frac{T}{10^8\,{\rm K}} \bigg)^{1/2}\, [{\rm yrs}], 
\ee 
where $n_p$ is the number density of hydrogen. When AGN heating is
not included, the cooling radius, i.e. the radius where $t_{\rm cool}
= t_{\rm H}$, lies at $\sim 60\,h^{-1}{\rm kpc}$, while it gets
reduced to $\sim 25\,h^{-1}{\rm kpc}$ in our ``BHAR model''. However,
the cooling radius vanishes for the ``Magorrian model'', where the
bubbles injection heats the gas above $1\,{\rm keV}$.

Even though the spatial extent of the bubble particles reaches out to the
virial radius of the cluster, they are not capable of heating the gas in outer
regions, simply because their entropy content becomes comparable to the
entropy of the surrounding ICM gas at intermediate radii. This finding is
analogous to the previously discussed case of isolated halos, but the
dynamical evolution of the cluster with its associated merger processes makes
spreading of bubble material towards the outskirts more efficient.

\begin{figure*}
\centerline{\vbox{
\hbox{
\psfig{file=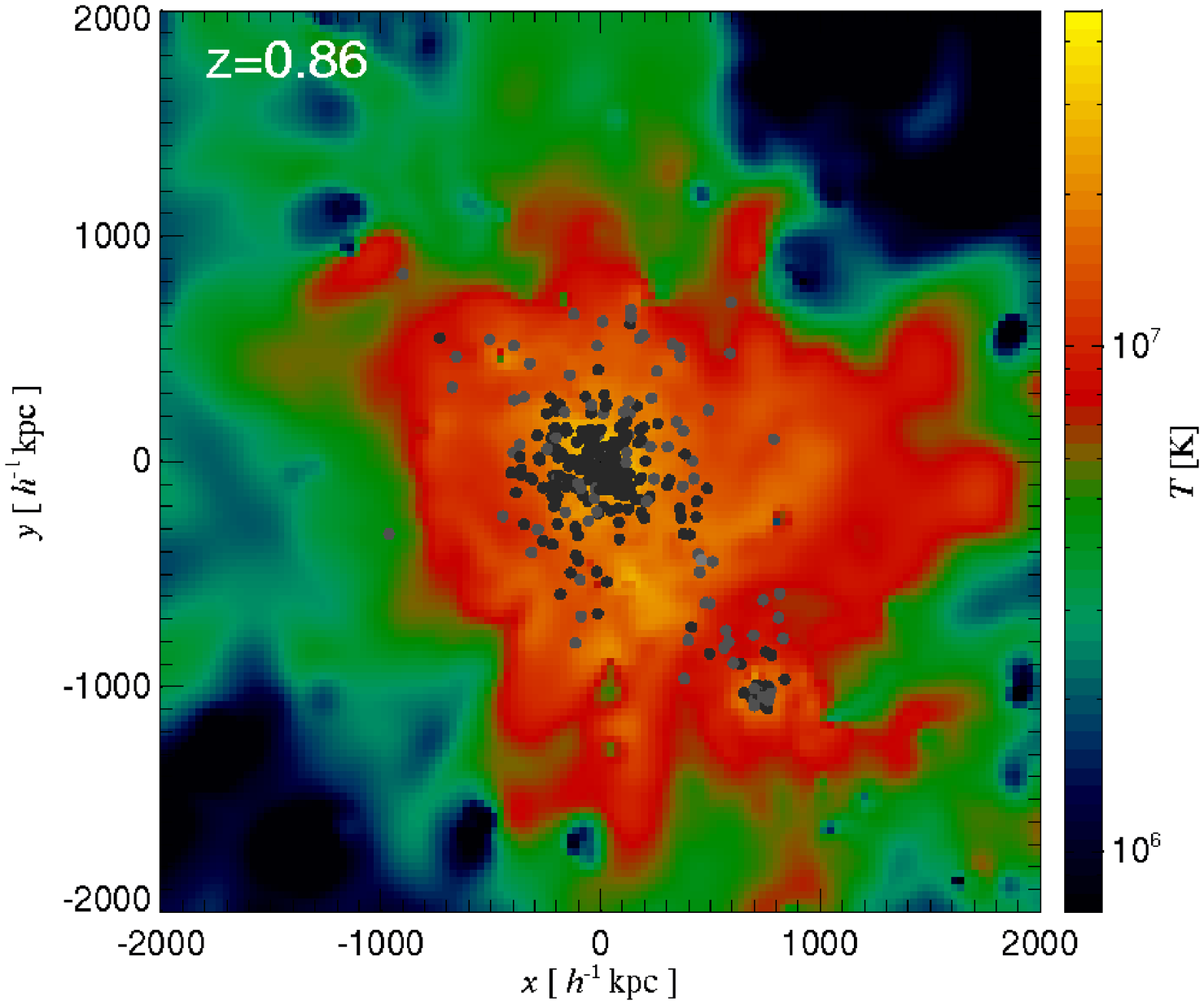,width=8.5truecm,height=8.truecm}
\hspace{-0.truecm}
\psfig{file=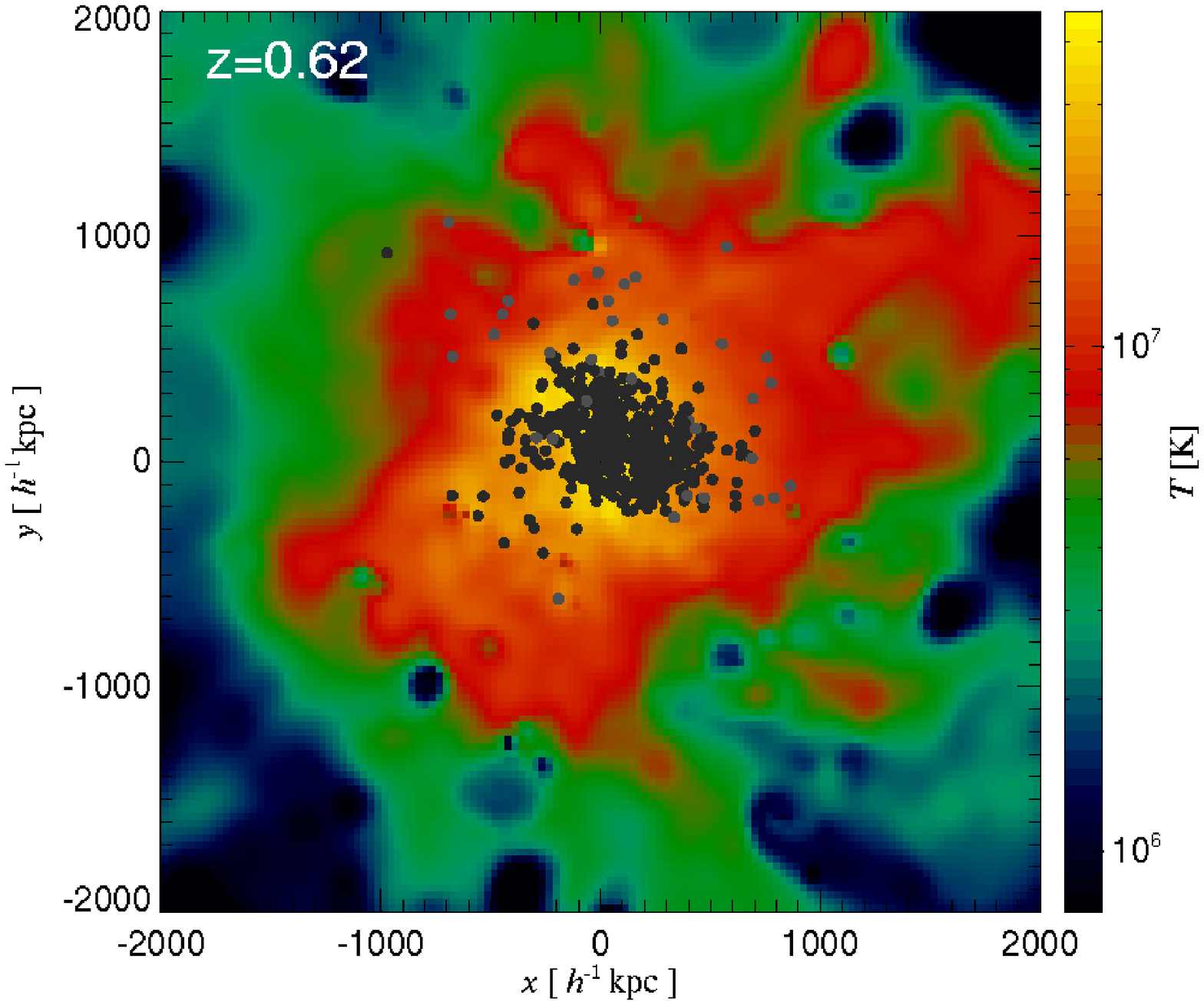,width=8.5truecm,height=8.truecm}
}
\hbox{
\psfig{file=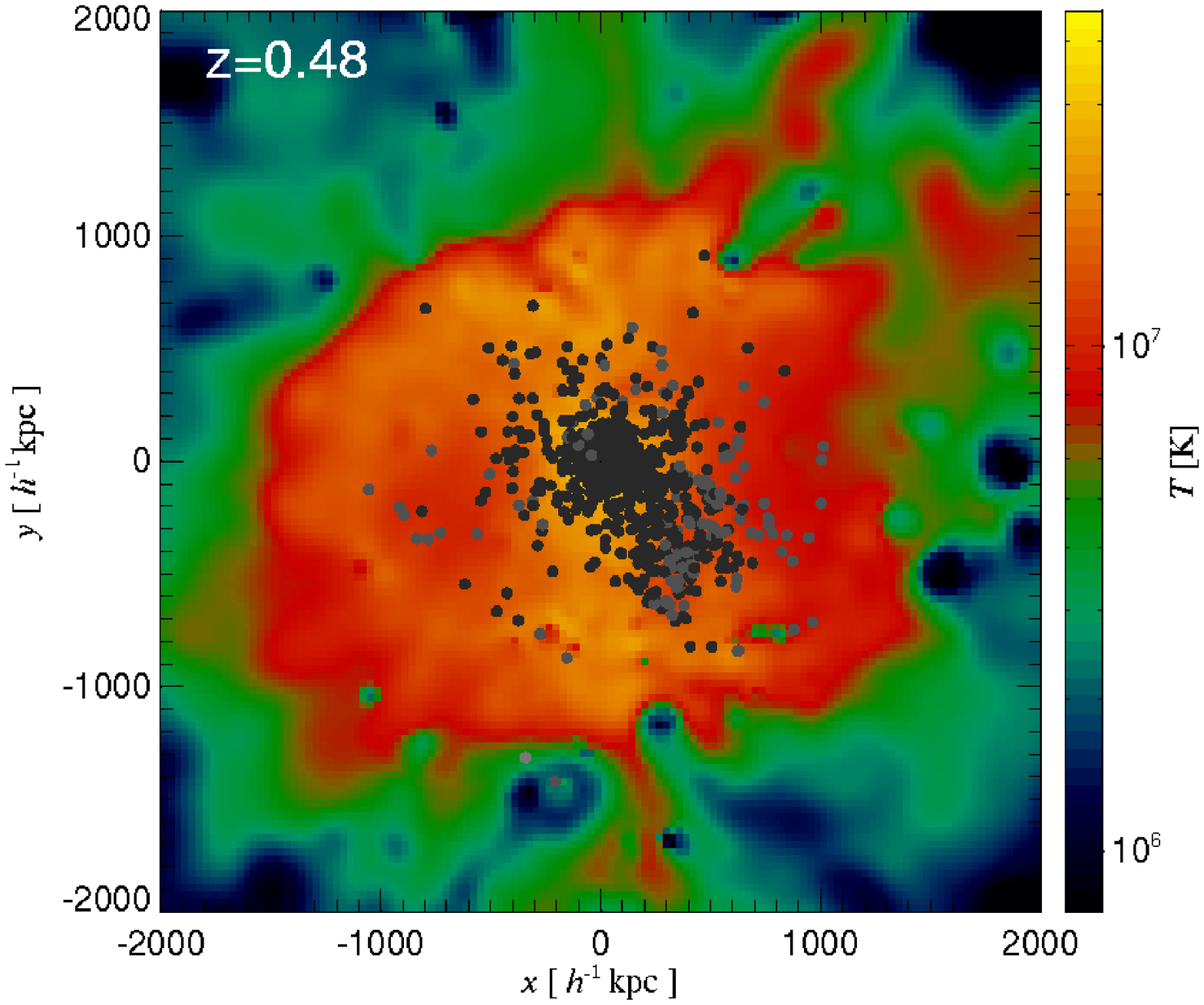,width=8.5truecm,height=8.truecm}
\hspace{-0.truecm}
\psfig{file=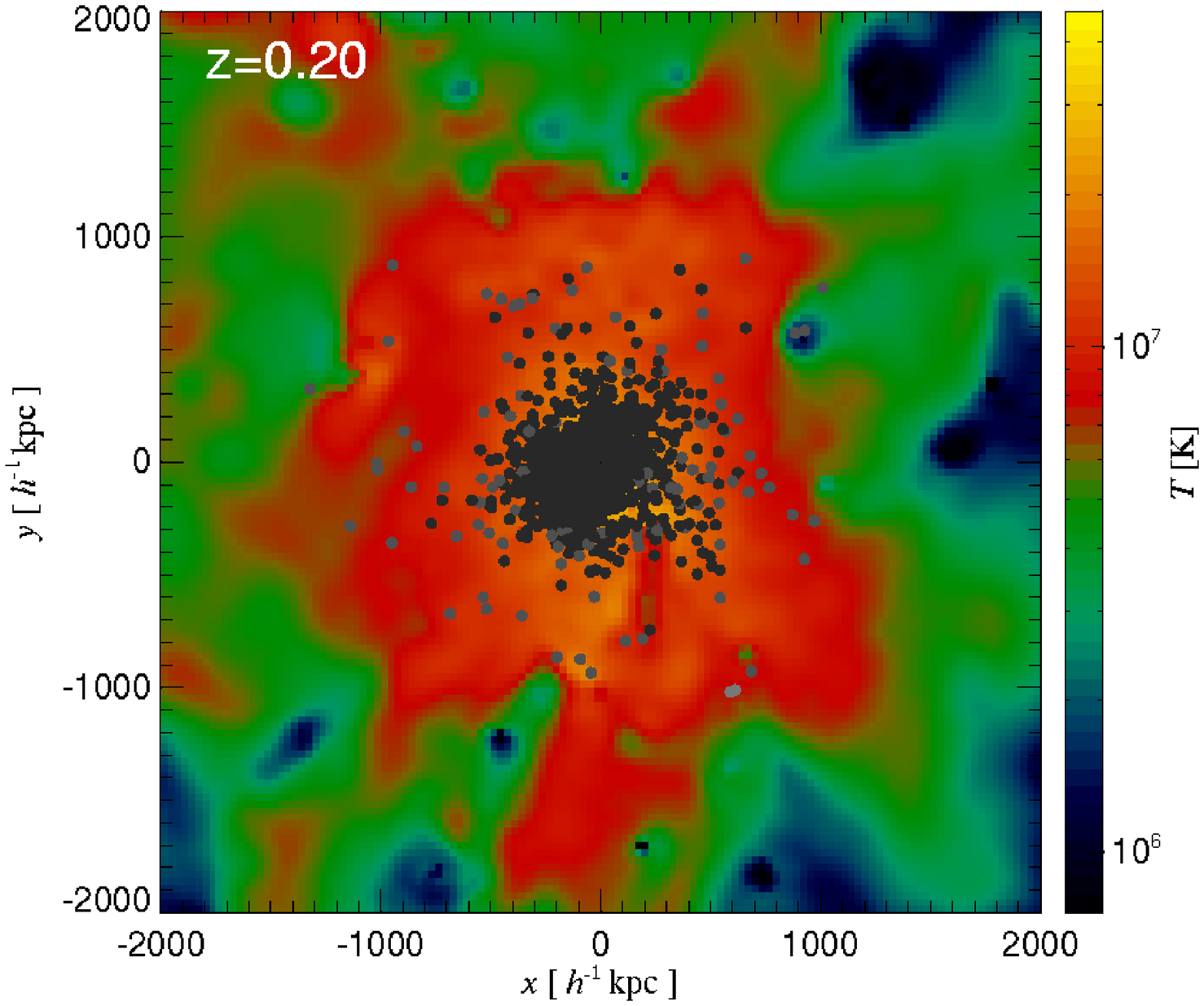,width=8.5truecm,height=8.truecm}
}
}}
\caption{Emission-weighted temperature maps of the g676 galaxy cluster
simulation during a major merger event at $z=0.86$, $0.62$, $0.48$ and
$0.20$, respectively. Over-plotted dots represent gas particles that
have been at least once part of a bubble, and they are colour-coded
according to their temperature, the darkest ones being the hottest. It
can be noticed that both the spatial distribution of bubbles and their
energy content are drastically influenced by the passage of the
smaller halo through the central region of the massive cluster.}
\label{MM_g676}
\end{figure*}

In Figure~\ref{MM_g676}, we show emission-weighted temperature maps of
the g676 galaxy cluster at four different epochs, in order to more
closely discuss the spatial distribution of bubbles during merger
events.  The over-plotted dots represent the particles that at least
once belonged to a bubble, and they are colour-coded according to their
temperature, the darkest ones are particles with $T > 10^8\,{\rm K}$,
while the lightest have $T < 10^4\,{\rm K}$. For this analysis, we
have introduced AGN feedback in all halos above $5 \times 10^{10}
h^{-1}{\rm M}_\odot$ and we scaled the energy content of the bubbles
with $M_{200}^{4/3}(z)$ of the host galaxy cluster.

In the first panel at redshift $z=0.86$, there is a smaller halo on
the lower right corner which enters the most massive cluster
progenitor at that epoch (which roughly has three times larger mass),
which is located at the centre of the panel. Both the massive halo and
the smaller one are AGN heated, but the bubbles are less energetic for
the infalling halo due to our assumed mass dependence. Moreover, it
can be noticed that the bubbles occupying the central regions are
hotter, both because they are more recent and thus have had less time
to lose their energy content, and also because at later times bubbles
are intrinsically more energetic in our ``Magorrian scenario''. The
second panel of Figure~\ref{MM_g676} (at $z=0.62$) illustrates what
happens to the bubble distribution when the smaller halo is crossing
the central region of the massive cluster. The bubbles are literally
pushed out of the way, upwards and to the right, and they are also
heated. The next panel (at $z=0.48$) shows that the bubble
distribution is still quite asymmetric, but at the same time, bubbles
have spread efficiently into outer regions and also have
cooled. Finally, the last panel shows how the cluster appears at
$z=0.20$, where it starts to be fairly relaxed with a quite symmetric
distribution of bubbles.

\subsection{Stellar properties of galaxy clusters}\label{Stellar prop}

In this section, we analyse the effects of AGN heating on the
properties of the stellar components of galaxy clusters. We
concentrate on the properties of the central cluster galaxy, which is
the one affected most by the bubble heating. From the initial
injection epoch ($z=3$) until $z=0$, we compute for this purpose the
stellar and gaseous mass, star formation rates, stellar ages, and
colours of the cD galaxy that sits in the main progenitor of our final
halo.

\begin{figure}
\bc
\centerline{\includegraphics[width=8.5truecm,height=8.5truecm]{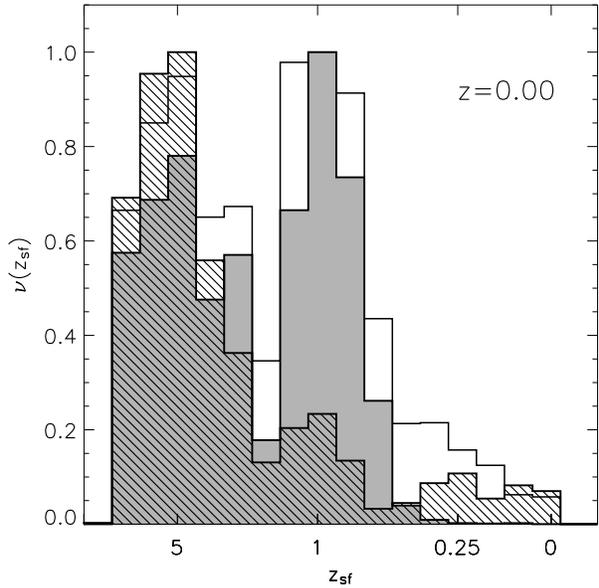}}
\caption{Distribution of formation redshifts of the stars belonging to
  the cD galaxy of the S1 galaxy cluster at $z=0$.  The white histogram
  corresponds to the case with cooling and star formation only; here it can be
  seen that stars continue to be formed until $z=0$. The grey coloured
  histogram shows how the stellar formation times change when our ``Magorrian
  model'' of bubble heating is switched on, which essentially suppresses any
  central star formation for $z < 0.25$ completely. Finally, the result for
  the ``BHAR model'' of AGN feedback is illustrated with the hatched
  histogram, which shows a considerable suppression of central star formation
  at intermediate redshifts, $0.3<z<2.0$.}
\label{CD_zf}
\ec
\end{figure}

In Figure \ref{CD_zf}, we show a histogram of the formation times of stars
belonging to the cD galaxy at $z=0$ of our S1 galaxy cluster. The histograms
have been computed by binning the expansion factors that correspond to each
stellar formation time, and they have been normalised to the maximum bin. The
white histogram is for the run without AGN heating, the grey coloured one for
the ``Magorrian model'', while the hatched histogram gives the result for
``BHAR model''. When AGN heating is absent, the histogram of stellar formation
redshifts clearly shows an extended tail at low redshifts; together with the
considerable SFR of order of $100\,{\rm M}_{\odot}{\rm yr}^{-1}$, this implies
that stars are formed until $z=0$ in situ. Nevertheless, there is a
possibility that some of these stars have been formed elsewhere, e.g.~in
merging substructures, and that they only ended up later in the cD galaxy by
merging. For redshifts less than $0.3$ this is certainly not an important
mechanism, because in our ``Magorrian model'', the central SFR is completely
suppressed for $z < 0.3$ and at the same time the $z_{\rm sf}$ histogram is
truncated. Thus, the difference of $ \sim 6 \times
10^{11} h^{-1}{\rm M}_{\odot}$ in the stellar mass of the final cD galaxy in
these two cases gives an indication on how many stars have formed in the cD
galaxy from $z=0.3$ until today, when AGN feedback is not present. At the same
time, also the mass of the cold gas (below $1\,{\rm keV}$) is reduced in the
``Magorrian model'', from $\sim 2 \times 10^{10}\,h^{-1}{\rm M}_{\odot}$ to
zero.  Moreover, the suppression of star formation at late times has an
immediate effect on colours of the cD galaxy, which becomes redder. To estimate
the colours we used Bruzual \& Charlot's stellar population synthesis models
\citep{Charlot}, computing rest-frame magnitudes in the SDSS bands, assuming
Solar metallicity and a Chabrier initial mass function. The $u$-band magnitude
is changed from $-23.8$ to $-22.8$ and the $u-r$ colour is increased from
$2.0$ to $2.6$.

In contrast, the star formation of the ``BHAR model'' proceeds in a quite
different manner, as can been seen from the hatched histogram, which is
substantially lower for $0.4 < z < 2.0$, but quite similar to the case without
AGN feedback at very low redshifts, $z < 0.3$. The first feature arises due to
two different processes happening at the same time. The second peak present in
the histograms is due to a major merger event that happens roughly at $z=1$.
One consequence of this merger event is a central burst of star formation,
which happens to be absent in the ``BHAR model'' due to its very efficient
bubble heating at this epoch. Nonetheless, there are still some stars becoming
part of the cD galaxy at $z=0$ that have formation times in the corresponding
time interval. Tracing back these stars in time and considering that the SFR
of the cD galaxy in the ``BHAR model'' during this epoch is practically
zero, we see that these stars have been produced in other small galaxies and
were indeed accreted onto the central cD galaxy at later times.  Assuming that a
similar amount of stars accreted onto the cD galaxy also in the case without
AGN feedback, a total mass of $\sim 10^{12} \, h^{-1}{\rm M}_{\odot}$ in stars
has formed in situ between $z=2$ and $z=0.4$, with a considerable part created
as part of the central starburst induced by the major merger event.

The tail at low redshifts present in the hatched histogram can be explained by
the reduced energy content of the bubbles, which are not efficient any more in
suppressing the cooling flow, and consequently the star formation in the central
cD galaxy. The total SFR within $R_{200}$ follows a similar trend as the SFR
of the cD galaxy, with comparable systematic differences for the different runs,
implying that the bubble heating mainly affects the central stellar properties
and not the residual star formation in the cluster volume.

\subsection{The metallicity distribution in the simulated clusters}

In this section, we analyse the metal distribution in our simulated galaxy
clusters. It is a well known problem that numerical simulations which include
cooling and star formation processes in general fail to reproduce the observed
shallow metallicity gradients, especially so if efficient feedback mechanisms
that help spreading the metals are absent. In particular, metals produced by
stars remain locked in the dense, star forming regions, even though supernovae
feedback is included which regulates the star formation process itself. As a
result, the metallicity distribution remains lumpy and exhibits a rather step
gradient. Moreover, most of the metals are produced in the central cD galaxy
due to its excessive star formation, which is a manifestation of the cooling
flow problem. 

This motivates the search for physical feedback processes that can spread and
mix metals more efficiently, acting both in the central region and on the
scale of the whole galaxy cluster. While the galactic wind model suggested by
\cite{SH03} helps in reducing this discrepancy and also diminishes the total
SFR over cosmological time, the model fails to qualitatively change the gas
and stellar properties of galaxy clusters discussed in Sections~\ref{Gas prop}
and \ref{Stellar prop}. Especially at late times, mixing of metals due to
winds from the cD galaxy proves inefficient; here the cluster potential well
is simply to deep and the ram pressure of the ICM too high to allow winds to
travel far. Bubble heating may fare considerably better in this respect. In
the following we therefore analyse the effect of AGN on the metal
distribution in our simulations, and we also compare simulations
with or without galactic winds of velocity  $\sim 480\,{\rm km\,s^{-1}}$.
    
\begin{figure}
\bc
\centerline{\includegraphics[width=8.5truecm,height=8.5truecm]{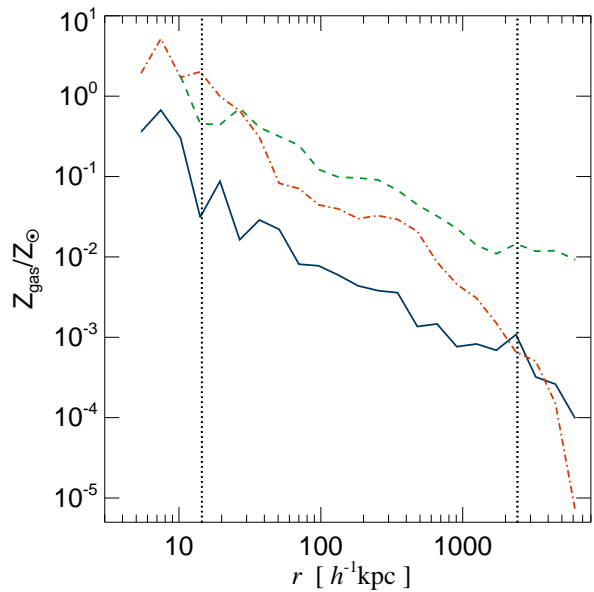}}
\caption{Radial profiles of the gas metallicity of the S1 galaxy
  cluster. Only the diffuse hot gas component has been used to estimate the
  metallicity, which is given in Solar units. The blue solid line is for the
  run without AGN heating, the red dot-dashed line is for bubble injection
  based on our ``Magorrian model'', while the green dashed line is also for
  the ``Magorrian model'' but with additional inclusion of feedback by
  galactic winds. We can see that the feedback processes manage to better
  spread the metals into the diffuse ICM gas, and the increased mixing leads
  to a less clumpy metallicity distribution.}
\label{Zhot_S1}
\ec
\end{figure}

\begin{figure*}
\centerline{
\psfig{file=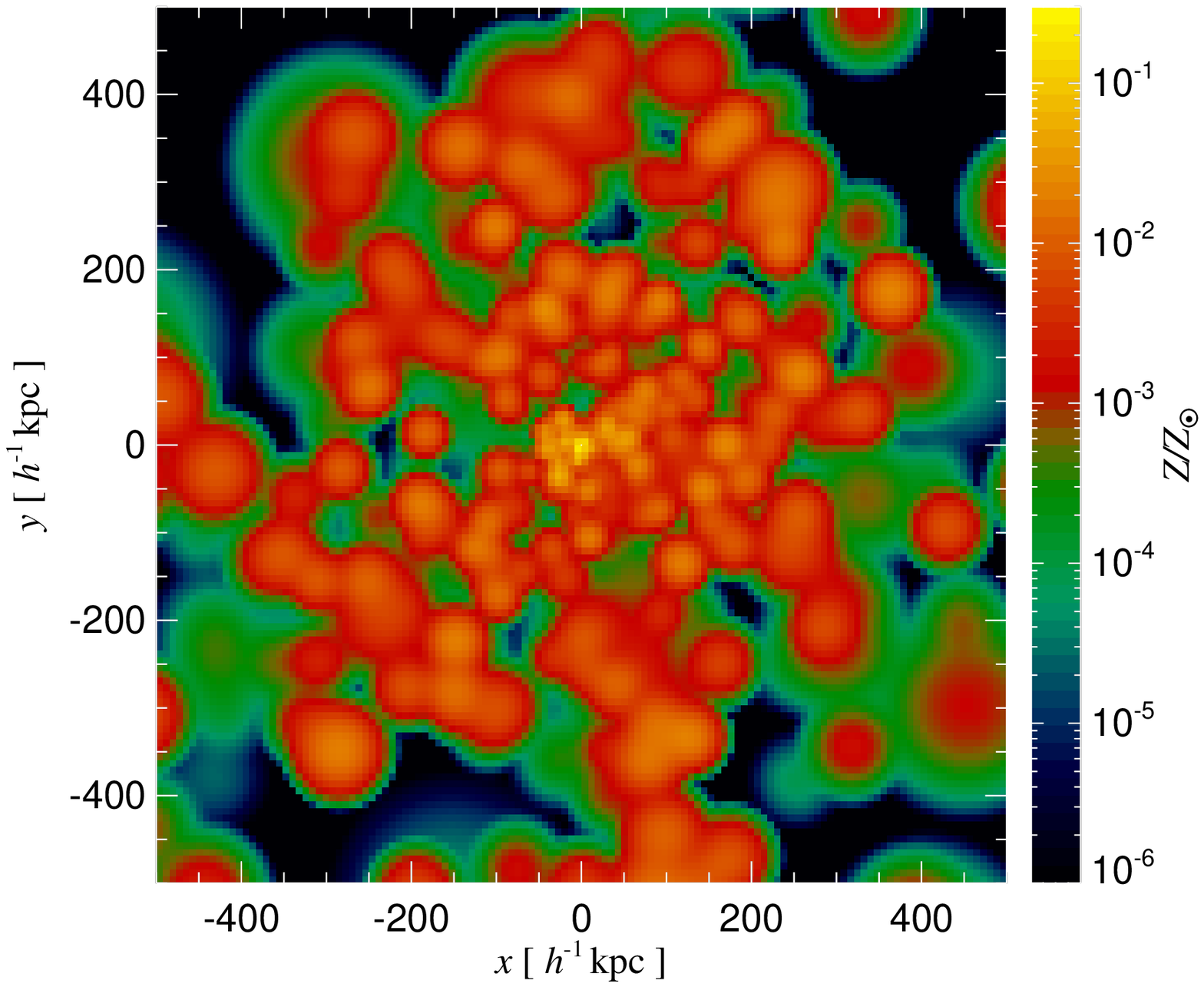,width=6.1truecm,height=5.5truecm}
\hspace{-0.truecm}
\psfig{file=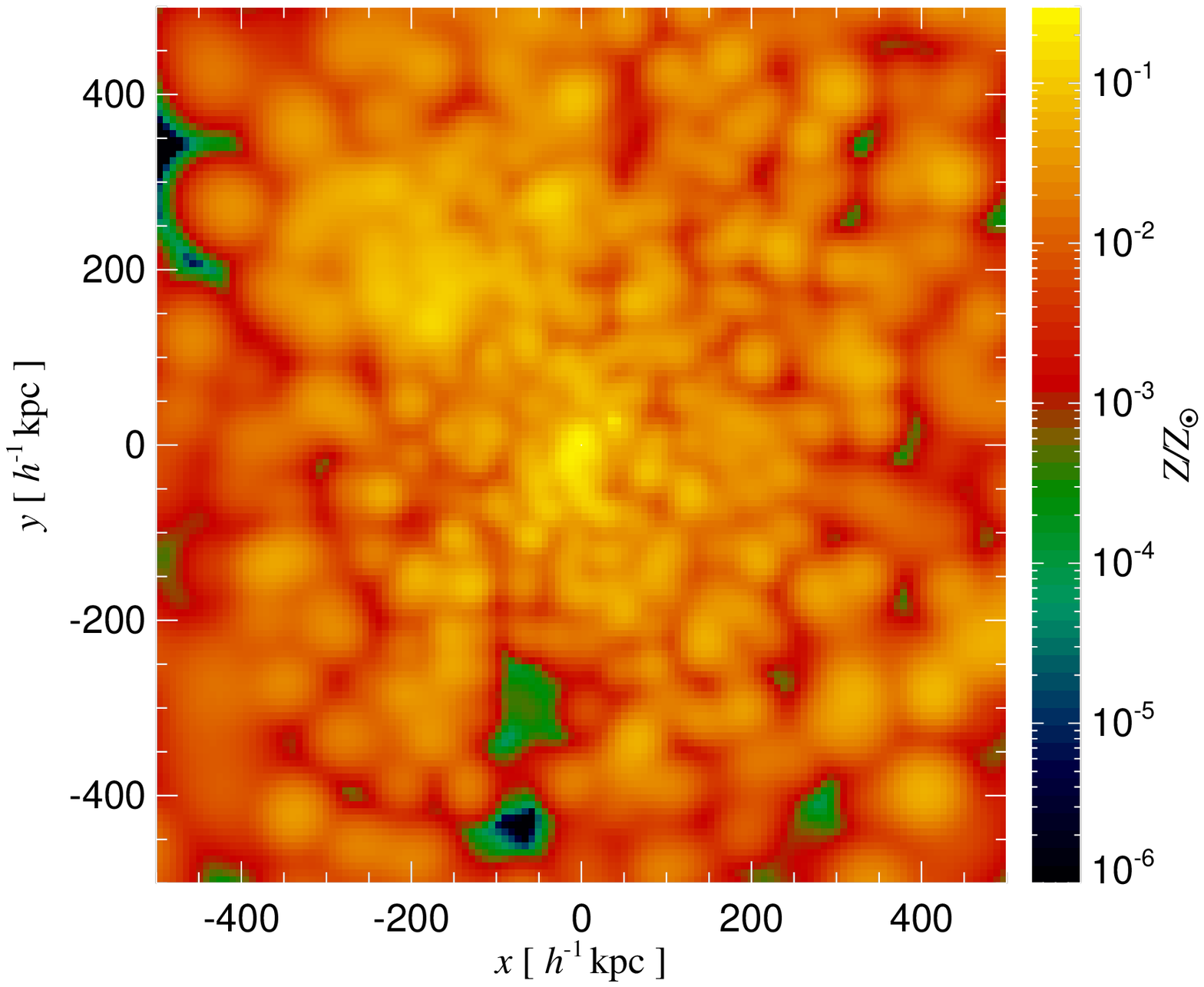,width=6.1truecm,height=5.5truecm} 
\hspace{-0.truecm}
\psfig{file=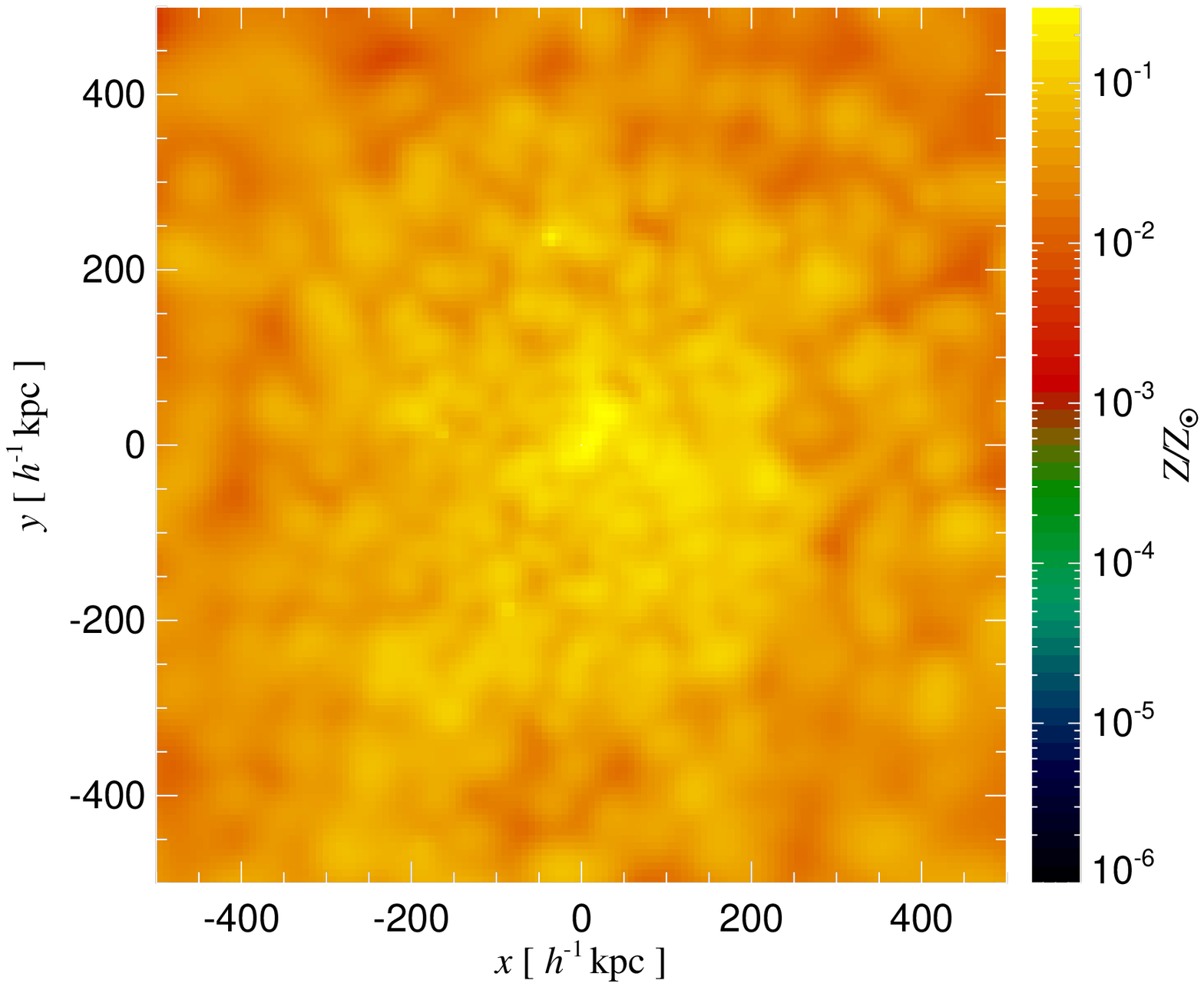,width=5.9truecm,height=5.5truecm}
}
\caption{Mass-weighted gas metallicity maps of the S1 galaxy cluster at
  $z=0$. The left panel corresponds to the case with cooling and star
  formation only. The middle panel shows how the metallicity of the hot gas
  component changes when AGN heating (here the ``Magorrian model'' was
  assumed) is included, while the right panel illustrates the case where in
  addition galactic winds with a velocity of $ \sim 480\,{\rm km \, s^{-1}}$
  were included, making the metallicity distribution more homogeneous.}
\label{Zmaps}
\end{figure*}

In Figure~\ref{Zhot_S1}, we show radial profiles of the gas
metallicity of the S1 galaxy cluster, and in Figure~\ref{Zmaps} we
illustrate the corresponding mass-weighted gas metallicity maps.  When
additional feedback mechanisms are taken into account, the amount of
metals in the hot gas component is increased with respect to runs with
cooling and star formation only (continuous blue line on
Figure~\ref{Zhot_S1} and left panel of Figure~\ref{Zmaps}). Also, the
metallicity distribution becomes less lumpy because the metals are
more efficiently transported out of the dense regions, increasing the
fraction of enriched gas. It can be noticed that already bubble
heating without winds (red dot-dashed line in Figure~\ref{Zhot_S1},
and middle panel of Figure~\ref{Zmaps}) is capable of producing a more
homogeneous metallicity distribution, while the AGN heating coupled
with the galactic winds (green dashed line in Figure~\ref{Zhot_S1} and
right panel of Figure~\ref{Zmaps}) slightly decreases the radial
metallicity gradient and makes the spreading of the metals even more
efficient. 

Nevertheless, the total amount of metals in the three different
ICM phases -- hot ICM gas, cold star-forming gas, and stars --
remains very similar in all runs, implying that there are no
substantial metal-enriched gas outflows from the galaxy cluster
itself. The situation can be rather different if winds with similar
velocities are present in less massive systems, as demonstrated by
\cite{SH03}. When the galactic wind velocities become comparable to
the escape velocity from the system in consideration, then the winds
may lead to gaseous outflows, eventually polluting the surrounding
medium with metals produced by cluster stars. The heating provided by
bubbles might help the metal spreading even more. Considering our
``BHAR model'' at early times where it is very efficient and
when the corresponding halo mass is considerably smaller, there is
evidence that not only many metals are transfered into the ``hot
phase'', but also that metal enriched gas is pushed towards the
cluster outskirts, and in part beyond the virial radius. However, this
effect is transitory, because $E_{\rm bub}$ decreases with time,
and more importantly, because the forming cluster is so massive that
it behaves effectively like a closed box at late times. Consequently,
at $z=0$, the distribution of metals looks quite similar to the
``Magorrian model'' discussed above.  Finally, note
that even though bubble heating improves metal spreading in the
ICM, at the same time it does not disrupt the central metallicity
gradient of the galaxy clusters, as can be clearly seen from
Figure~\ref{Zhot_S1}. Therefore, our AGN heating prescription is
qualitatively in good agreement with the metallicity gradients observed in cool
core clusters \citep[e.g.][]{Boehringer2002,DeGrandi2004,Boehringer04}.   

\subsection{Sound waves or merger induced weak shocks?} \label{S2_24}

While the unsharp mask technique is very useful in detecting X-ray
cavities and associated sound waves, it is potentially easy to confuse
these ripples with shock waves stemming from merger events. To
demonstrate this danger in interpreting observations, we here analyse
a specific case where the presence of smaller systems passing trough
the cluster might induce such a misleading conclusion.  

\begin{figure*}
\centerline{\vbox{
\hbox{
\psfig{file=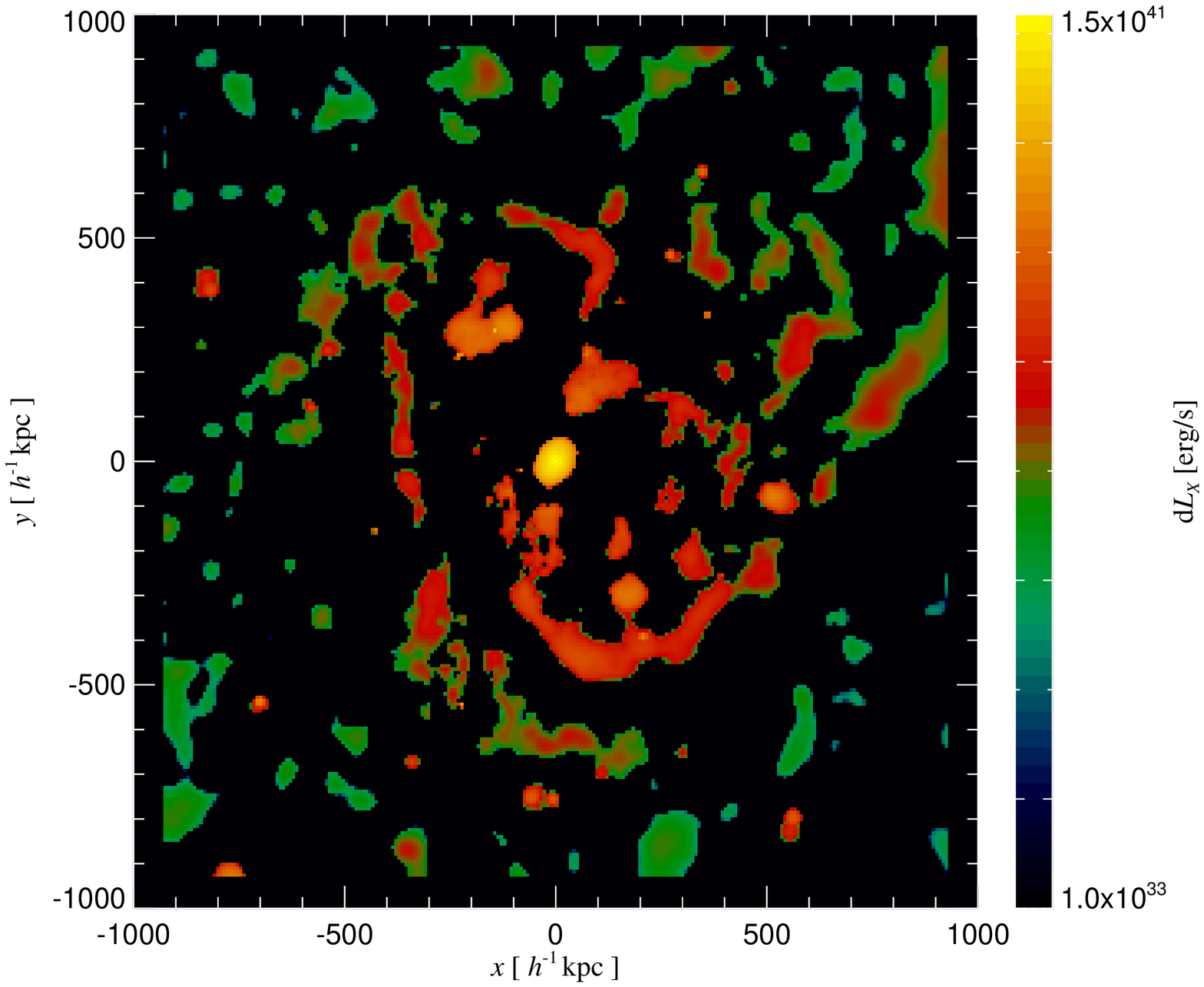,width=9.truecm,height=7.6truecm}
\hspace{-0.truecm}
\psfig{file=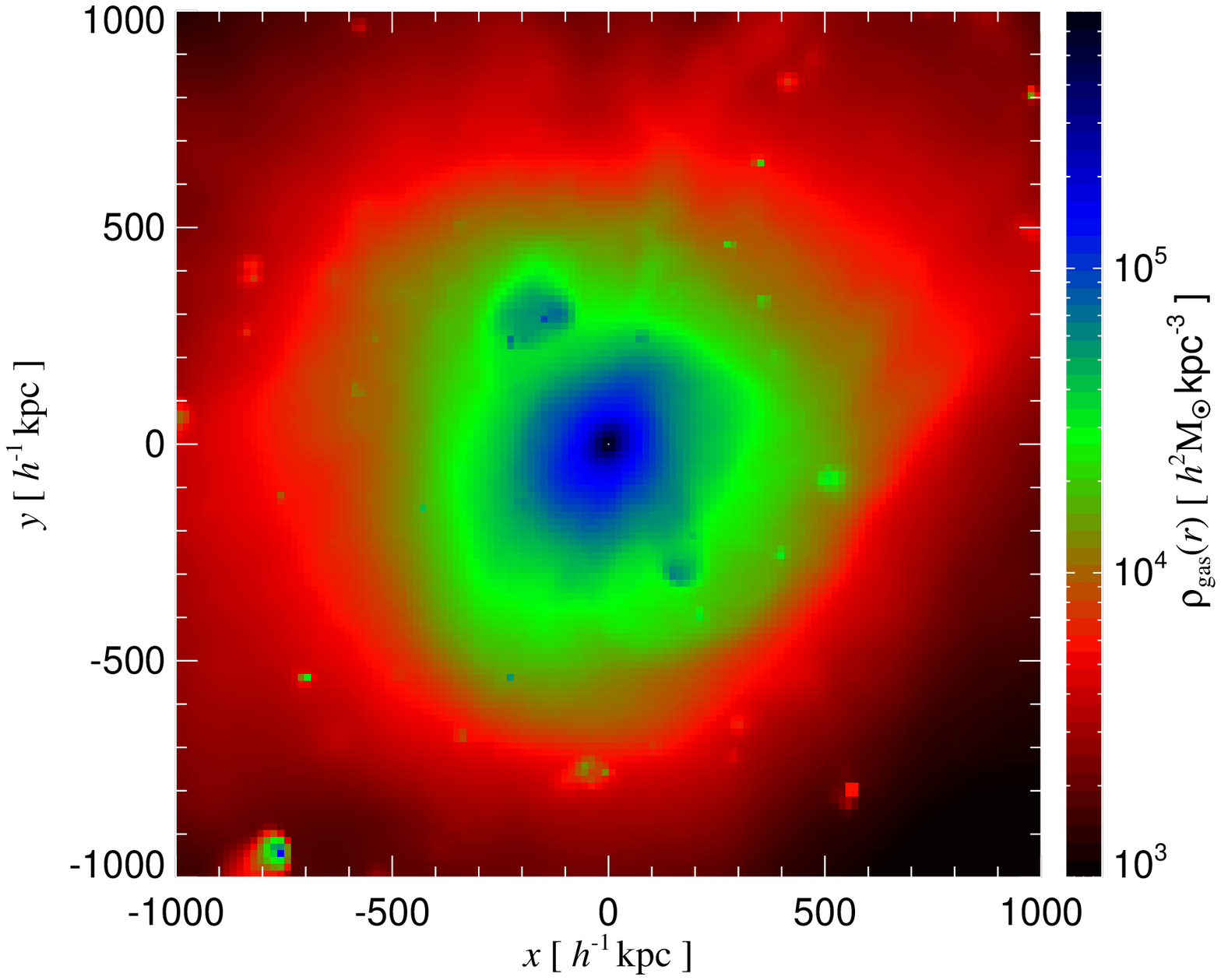,width=9.truecm,height=7.6truecm}
}
\hbox{
\psfig{file=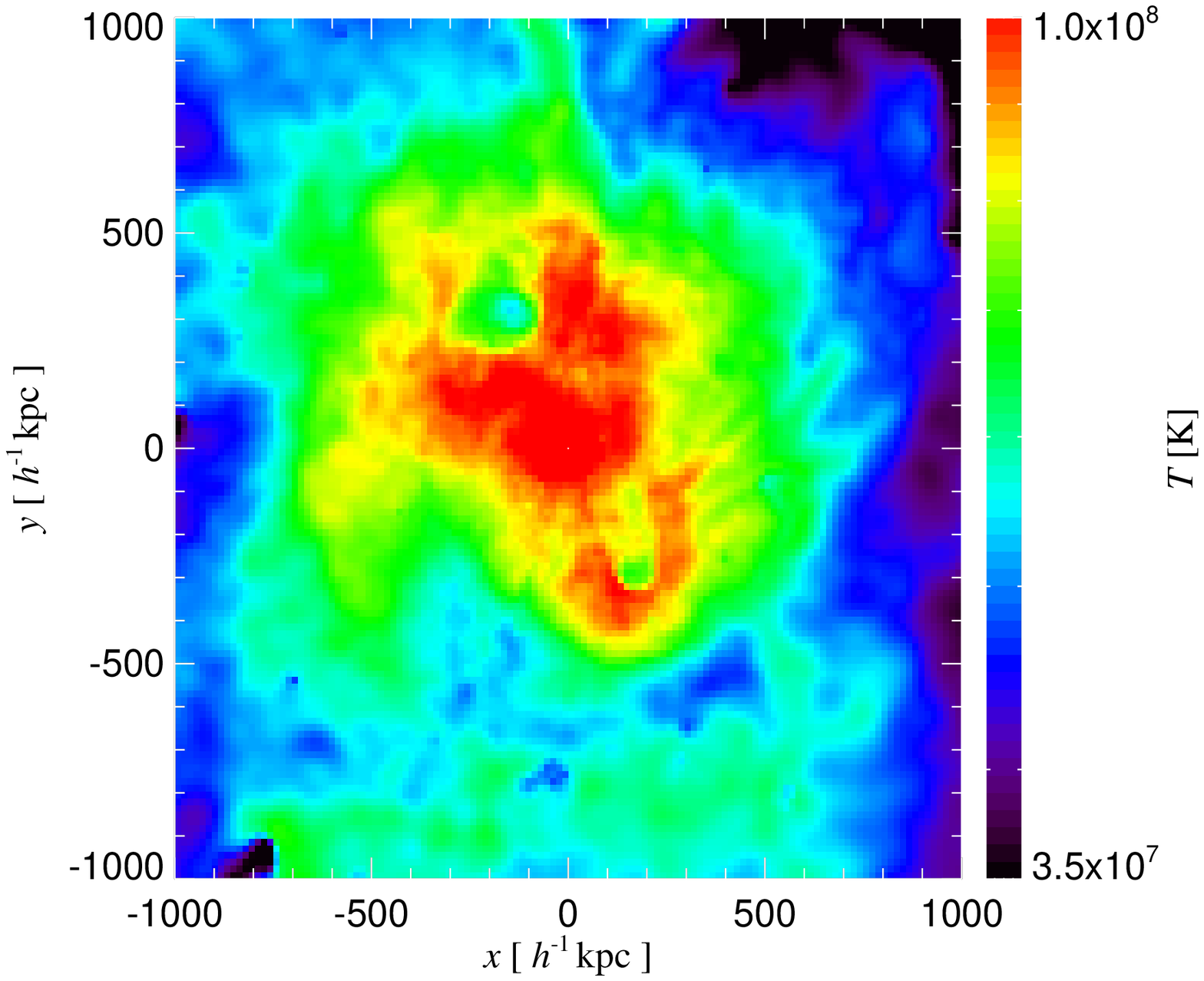,width=9.truecm,height=7.6truecm}
\hspace{-0.truecm}
\psfig{file=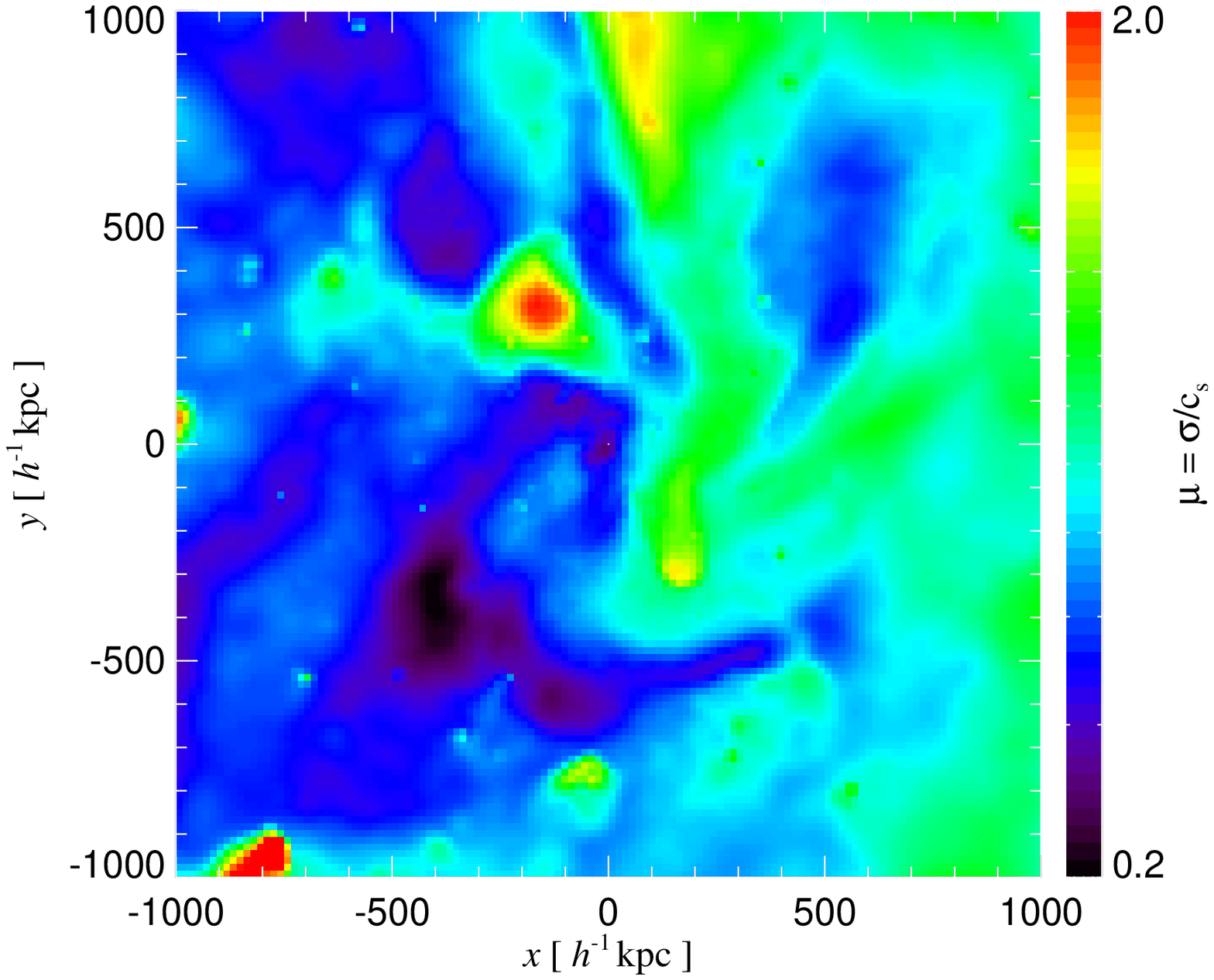,width=9.truecm,height=7.6truecm}
}
}}
\caption{Projected maps of different gas properties of the 
  S2 galaxy cluster at $z=0.13$, where a merger with two smaller subclumps is
  in progress. AGN heating has not been included in this simulation.  The
  upper left panel shows the unsharped masked image of the X--ray luminosity
  on a scale of $160\,h^{-1}{\rm kpc}$, and the upper right panel illustrates
  the gas density map. On the lower left panel we show a map of mass-weighted
  cluster gas temperature, while the lower right panel gives a crude estimate
  of the mass-weighted Mach number.}
\label{S2_24maps}
\end{figure*}

To this end, we computed projected maps of the S2 galaxy cluster
without AGN heating at $z=0.13$. At this time, two substructures at
radial distances of about half the virial radius are moving towards
the centre.  These substructures are still visible in the gas density
map (upper right panel of Figure~\ref{S2_24maps}), but almost
completely vanish when the X--ray luminosity map is
computed. Nevertheless, when the unsharp masking procedure is applied
to the $L_{\rm X}$ map, a two-lobed feature is clearly visible (upper left
panel of Figure~\ref{S2_24maps}). Even though this feature at 
first glance looks strikingly similar to the ripples produced by
bubble heating events, it is exclusively a product of the specific
spatial geometry of the substructures in the cluster, and is also
enhanced due to projection effects.

In the mass-weighted temperature map (lower left panel of
Figure~\ref{S2_24maps}), two spherical regions can be noticed that are
slightly cooler than their surroundings. They correspond to the two
substructures. Especially around the substructure in the lower right
part of the map, hotter surrounding gas can be seen. To distinguish
between a cold front or a shock, we can analyse the pressure of the
surrounding gas and compute the Mach number map\footnote{We made a
crude estimate of the Mach number by calculating for every gas
particle its velocity in the galaxy cluster reference frame and
dividing it by the local sound speed.} (lower right panel of
Figure~\ref{S2_24maps}). A clear jump of pressure in the region
adjacent to the hot gas around the substructure together with the
mildly supersonic motion of the substructure indicates the presence of
weak shocks.  Note that these shocks cannot be easily identified
observationally as being caused by infalling substructures. Thus, if a
careful analysis is not performed, these features could be
associated mistakingly with sound waves caused by AGN bubbles, especially if at
the same time some fossil but unrelated bubble is detected in the
cluster by its radio emission.

\subsection{AGN heating in galaxy clusters of different mass}

Here we discuss the effect of AGN heating in clusters spanning a wide range in
mass, following their cosmological evolution up to $z=0$.  Their main
properties are summarized in Tables~\ref{tab_simpar} and~\ref{tab_GCpar}. We
analyze how the gas properties of clusters of different mass and at various
radii, normalized to $R_{200}$, change due to the AGN-driven bubbles.

\begin{figure}
\bc
\centerline{\includegraphics[width=8.5truecm,height=8.5truecm]{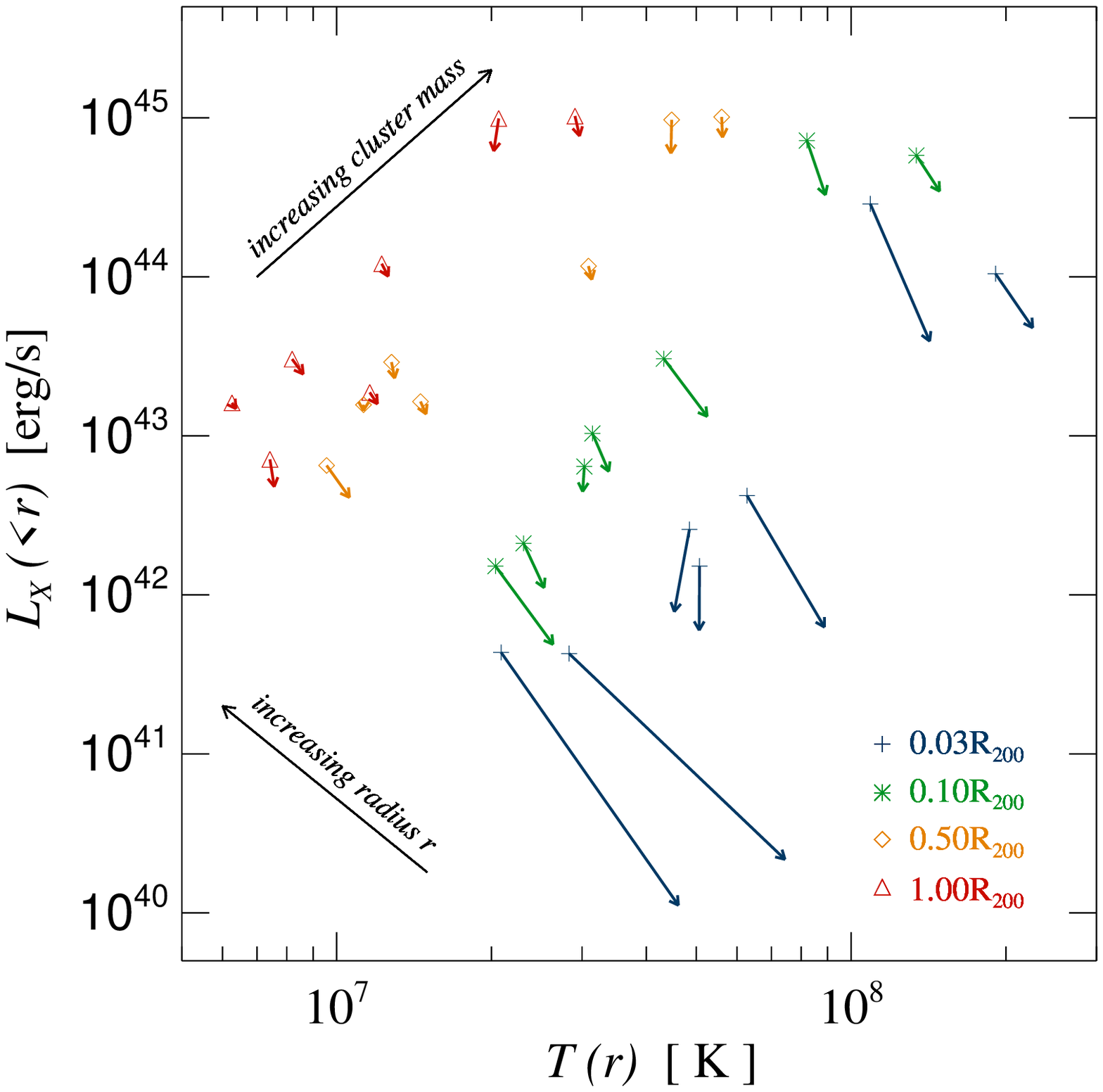}}
\caption{Cumulative X--ray luminosity of seven galaxy clusters as
  a function of their mass-weighted gas temperature. We give results for
  redshift $z=0$, and at four different radii. Different symbols and colours
  denote estimates of $L_{\rm X}$ and $T$ at different radii normalized to
  $R_{200}$: blue crosses correspond to $0.03\,R_{200}$, green stars to
  $0.1\,R_{200}$, yellow diamonds to $0.5\,R_{200}$, and red triangles to
  $R_{200}$. The arrows indicate how the cluster luminosity and temperature
  change when AGN heating is included, i.e.~systematically decreasing $L_{\rm
    X}$ for all the considered radii. More massive clusters are located in the
  upper-right part of the panel, while less massive systems are found in the
  lower-left corner of the figure, as indicated.}
\label{L_T}
\ec
\end{figure}

In Figure~\ref{L_T}, we show for all clusters in our sample at $z=0$ the
cumulative X--ray luminosity and mean gas temperature at four different radii,
$r\,=\,0.03\,R_{200}$, $0.1\,R_{200}$, $0.5\,R_{200}$ and $R_{200}$, marked
with different colours and symbols.  In this $L_{\rm X}-T$ plane, clusters are
found at different places according to their mass, with less massive systems
being located in the lower-left part of the panel and more massive ones in the
upper-right part, as indicated in the figure. Also, cluster luminosity and
temperature vary systematically with increasing radius $r$, with values at the
virial radius occupying the upper-left part of the plot. The arrows attached
to each cluster model show how $L_{\rm X}$ and $T$ change when bubble heating
is present.  Clearly, there is a systematic trend of decreasing X--ray
luminosity for all clusters and at all considered radii.  This effect is more
pronounced in the most inner regions and it is important both for massive
clusters and smaller systems. Thus, in the cosmological simulations the bubble
heating efficiency is not as clearly related to the mass of the host cluster
as is the case for the isolated halos. From Figure~\ref{L_T}, it can be seen
that the two low mass clusters exhibit prominent imprints caused by AGN
activity, highlighting that the bubble heating is more complex in cosmological
simulations due to the hierarchical merging histories of clusters.

The gas temperature does not follow an equally clear trend as $L_{\rm
X}$ for all the clusters in our sample, but in most cases, and
especially for central clusters regions, it is boosted towards higher
values, implying that bubble injection leads to en effective heating
of the ICM. These trends in $L_{\rm X}$ and $T$ confirm our previous
findings for the S1/S2 clusters (presented in Figure~\ref{Prof_S1}),
and show that they are general features of our AGN heating model.
Interestingly, recent observational work by \cite{Croston05} indicates
that radio-loud elliptical-dominated groups have $L_{\rm X}-T$ scaling
relations systematically different from those of elliptical-dominated
radio-quiet groups. Their observed trends in the $L_{\rm X}-T$ plane
are qualitatively similar with what we find, showing that for a given
X--ray luminosity radio-loud groups have systematically higher gas
temperature values.  Nevertheless, since they are probing smaller mass
systems, simulated galaxy groups with a matching range in mass are
needed in order to make a more detailed comparison.  Recent
observational works \citep{McNamara05, Nulsen05} revealed the presence
of very powerful outbursts due to the AGN activity, which are also
accompanied with large-scale shocks. These clusters are located above
the mean $L_{\rm X}-T$ relation, probably because we are witnessing
the very early stages of bubble feedback, where pressure
equilibrium with the surrounding ICM has not yet been established.

\section{Discussion and Conclusions} \label{DIS} 

In this work, we discussed a simulation model for AGN heating in the
form of hot, buoyant bubbles, which are inflated by active phases of
supermassive black holes at the centres of massive halos. The
motivation for such a mode of feedback stems from the rich
phenomenology of X-ray cavities and radio bubbles observed in clusters
of galaxies, and the suggestion that this AGN activity may represent
the solution of the `cooling flow puzzle' posed by clusters of
galaxies.

Several previous studies in the literature \citep[e.g.][]{Churazov01,Quilis01,
Brueggen03,Hoeft04,DVecchia04} have analysed bubble feedback in
isolated galaxy clusters using hydrodynamical mesh codes. We here
present the first implementation of this feedback in a SPH code, so an
 important goal was to test whether our results are consistent
with these previous studies based on very different hydrodynamical
techniques.  Reassuringly, this is the case, both qualitatively and
quantitatively. In particular, for galaxy clusters of
similar masses as considered by \citet{Quilis01, DVecchia04}, and with
bubbles parametrised in an analogous way, we find changes induced by
AGN-heating in gas properties, e.g. density and temperature radial
profiles, central mass deposition rates, that are in excellent
agreement. Also, the morphology of the bubbles and their time
evolution are very similar. This is important because it implies that
the SPH technique, which is more easily applicable to full blown
cosmological simulations of cluster formation, can be reliably used to
study bubble feedback.

In our simulations, we considered both, isolated halos of different mass, and
cosmological simulations of the $\Lambda$CDM model that follow galaxy cluster
assembly from high redshift to the present.  The isolated simulations served
as a laboratory to study the dynamics of bubbles in detail. By considering
halos with masses ranging from $10^{12}\,h^{-1}{\rm M}_{\odot}$ to
$10^{15}\,h^{-1}{\rm M}_{\odot}$, they also allowed us to gain some insight in
how the coupling of radiative cooling to the AGN heating varies with cluster
mass. An important conclusion from these experiments is that for systems of
mass lower than $\sim 10^{13}\,h^{-1}{\rm M}_{\odot}$ bubbles with reasonable
energy content are not capable of preventing excessive gas cooling. If one
nevertheless allows for very large energy in these systems, a stable
suppression of the cooling requires a delicate fine tuning of the bubbles.
However, the inefficiency of bubble heating in lower mass systems is also
caused in part by our injection prescription, which is based on global
properties of the host galaxy cluster without accounting for the actual amount
of cooling gas present in the very centre.

In our cosmological simulations, we find that bubble injection can
substantially affect galaxy cluster properties, especially in massive,
relaxed clusters and at late cosmological times. Central cluster gas
is efficiently heated, and thus both the amount of cold baryons and
the star formation in the central cD galaxy is reduced. Also, an
excessive mass deposition rate by cooling flows is
prevented. AGN-driven bubbles not only modify the properties
of the most central cluster parts, but alter the whole
inner region of massive clusters, out to radii of order $\sim
300\,h^{-1}{\rm kpc}$, where the gas density is reduced and the
temperature is increased.  As a result, the X--ray luminosity is
considerably reduced, while the gas entropy exhibits a flat core in
the central region. These trends are all in the direction required to
reconcile hydrodynamical simulations of cluster formation in the
$\Lambda$CDM model with observations of real galaxy clusters.

We tried several variants of our bubble model in order to explore the
dependence of obtained results on the detailed assumptions made about
how the energy is released in the bubbles. In particular, we compared
an instantaneous injection of energy into the bubbles with a scheme
where the energy is released over a certain period of time (from
$5\times10^7$yrs to $5\times10^8$yrs), we imposed that the bubble
particles should not cool during a given time interval (e.g. $\sim
10^8$ yrs), we tried different spatial patterns for the bubble
placement, and we also varied the initial epoch where our AGN heating
started (from $z=6$ to $z=3$). None of these changes was really
capable of modifying our results considerably. However, it appears
that our findings are much more sensitive to the adopted model for the
time evolution of the bubble energy. We explicitly demonstrated this
by changing the rate at which the energy is released with time, under
the constraint that the total energy injected from $z=3$ to $z=0$ was
kept constant.  When we linked the bubble energy content in this way
to a model for the BH accretion rate, the fraction of cold gas and the
star formation rate can be noticeably reduced even at early times, but
this is compensated by a reduced efficiency of bubble heating at late
times, such that cooling flows are not suppressed sufficiently at
$z=0$. The ``Magorrian model'', where the feedback occurs primarily at
low redshift fares better in this respect, and gives therefore a
better match to the properties of observed rich clusters of
galaxies.  In addition, we explored yet another scaling between
the energy of the bubbles and the mass of the host galaxy cluster,
namely $E_{\rm bub} \propto M_{200}^{5/3}$, which can be motivated by the
observed $M_{\rm BH}-\sigma$ relation as well.  Also, \cite{Ferrarese2005}
pointed out a relationship between the mass of the black hole and that
of the hosting dark matter halo, in the form of $M_{\rm BH} \propto
M_{\rm DM}^{1.65}$. Thus, if the energy content of the bubbles is a
linear function of the black hole mass, the above scaling is
obtained. One also arrives at $E_{\rm bub} \propto M_{200}^{5/3}$ if
one assumes that the energy in the bubbles is some small fraction of
the total thermal cluster energy, which roughly scales as
$M_{200}^{5/3}$. In performing the analysis with the modified slope,
we fixed the normalization of the $E_{\rm bub} \propto M_{200}^{5/3}$
relation at redshift $z=0$ to be equal to the value in our ordinary
``Magorrian model''. The results of this analysis showed that the
change in the slope from $4/3$ to $5/3$ produces qualitatively very
similar results, and hence it follows that the properties of our
simulated galaxy clusters are not very sensitive to such a modest
change of the AGN heating prescription.

In the newly emerging picture for the joint evolution of galaxies and
supermassive black holes, the interplay between AGN and their host galaxies
may be composed of two modes. One mode is caused by the quiescent accretion of
intracluster gas onto the central BH, provoking periodic AGN activity which
manifests itself in jets and radio bubbles. This mode plays a more important
role at late cosmological epochs, in massive and more relaxed systems, and it
is often referred to as a ``radio-mode'' \citep[e.g.][]{Croton05}. The other
mode occurs in merging pairs of galaxies, where strong tidal forces
efficiently funnel large amounts of {\em cold} gas towards the nuclei of the
merging galaxies, where it becomes available for fueling the embedded
supermassive BHs~\citep{DiMatteo05,Springel04}. The associated intense
accretion triggers quasar activity, which is more frequent at higher redshift
due to the larger merger rates there. If a small fraction of the bolometric
luminosity of the quasar couples thermally to the surrounding gas, a prominent
gas outflow can eventually be created during the formation of ellipticals,
which then shuts off further accretion and star formation~\citep{Springel2005}
and establishes the $M_{\rm BH}-\sigma$ relationship~\citep{DiMatteo05}.

The latter process, the `quasar mode', has already been explored in direct
simulation models of galaxy mergers, but not yet in cosmological simulations.
It would therefore be extremely interesting to couple these two modes of AGN
feedback in a unified simulation model for supermassive black hole growth, and
to carry out cosmological simulations with it.  In such a model, the
energetics of the bubbles and the periods of AGN activity can then be made
directly dependent on the current BH mass and on the local physics of
accreting gas, removing much of the freedom in our present bubble models. We
will present such a model in forthcoming work. In addition, our work suggests
that for a more complete picture of the ICM dynamics, additional physical
processes should be incorporated as well. This includes thermal conduction,
even though it is probably relevant only in the most massive systems. We also
suggest that the physical viscosity expected for the ICM gas should be
considered as well, since the efficiency of non-local AGN heating by viscous
dissipation of sound waves will depend crucially on this input. Finally, radio
observations strongly suggest that bubbles are prevalently filled with
relativistic particles, which appear as radio lobes, and in many cases are
coincident with X--ray cavities. Therefore, it would be important to address,
using fully cosmological simulation of cluster formation, the role of
non-thermal radio plasma in bubbles for heating of the ICM.  It appears that
for some time to come clusters of galaxies will remain one of the most
interesting places to study complex hydrodynamic phenomena in the Universe.

\section*{Acknowledgements}

We thank Martin Jubelgas for providing a code to set-up isolated halos in
equilibrium, Elena Rasia for help in producing mock observations with {\small
  X-MAS}, and Klaus Dolag for providing cluster initial conditions. We are
indebted to Simon White and Eugene Churazov for very useful comments on the
manuscript. DS acknowledges the PhD fellowship of the International Max Planck
Research School in Astrophysics, and received support from Marie Curie Host
Fellowship for Early Stage Research Training.

\bibliographystyle{mnras}

\bibliography{paper}

\begin{thebibliography}{107}
\expandafter\ifx\csname natexlab\endcsname\relax\def\natexlab#1{#1}\fi

\bibitem[{Allen}(2000)]{Allen00}
{Allen} S.~W., 2000, \mnras, 315, 269

\bibitem[{Allen} et~al.(1995){Allen}, {Fabian}, {Edge}, {Bohringer} \&
  {White}]{Allen95}
{Allen} S.~W., {Fabian} A.~C., {Edge} A.~C., {Bohringer} H., {White} D.~A.,
  1995, \mnras, 275, 741

\bibitem[{Allen} et~al.(2001{\natexlab{a}}){Allen}, {Fabian}, {Johnstone},
  {Arnaud} \& {Nulsen}]{Allen2001}
{Allen} S.~W., {Fabian} A.~C., {Johnstone} R.~M., {Arnaud} K.~A., {Nulsen}
  P.~E.~J., 2001{\natexlab{a}}, \mnras, 322, 589

\bibitem[{Allen} et~al.(2001{\natexlab{b}}){Allen}, {Schmidt} \&
  {Fabian}]{Allen01}
{Allen} S.~W., {Schmidt} R.~W., {Fabian} A.~C., 2001{\natexlab{b}}, \mnras,
  328, L37

\bibitem[{B{\^ i}rzan} et~al.(2004){B{\^ i}rzan}, {Rafferty}, {McNamara},
  {Wise} \& {Nulsen}]{Birzan04}
{B{\^ i}rzan} L., {Rafferty} D.~A., {McNamara} B.~R., {Wise} M.~W., {Nulsen}
  P.~E.~J., 2004, \apj, 607, 800

\bibitem[{B{\" o}hringer} et~al.(2004){B{\" o}hringer}, {Matsushita},
  {Churazov}, {Finoguenov} \& {Ikebe}]{Boehringer04}
{B{\" o}hringer} H., {Matsushita} K., {Churazov} E., {Finoguenov} A., {Ikebe}
  Y., 2004, \aap, 416, L21

\bibitem[{B{\" o}hringer} et~al.(2002){B{\" o}hringer}, {Matsushita},
  {Churazov}, {Ikebe} \& {Chen}]{Boehringer2002}
{B{\" o}hringer} H., {Matsushita} K., {Churazov} E., {Ikebe} Y., {Chen} Y.,
  2002, \aap, 382, 804

\bibitem[{Babul} et~al.(2002){Babul}, {Balogh}, {Lewis} \& {Poole}]{Babul02}
{Babul} A., {Balogh} M.~L., {Lewis} G.~F., {Poole} G.~B., 2002, \mnras, 330,
  329

\bibitem[{Balogh} et~al.(2001){Balogh}, {Pearce}, {Bower} \& {Kay}]{Balogh01}
{Balogh} M.~L., {Pearce} F.~R., {Bower} R.~G., {Kay} S.~T., 2001, \mnras, 326,
  1228

\bibitem[{Blanton} et~al.(2001){Blanton}, {Sarazin}, {McNamara} \&
  {Wise}]{Blanton01}
{Blanton} E.~L., {Sarazin} C.~L., {McNamara} B.~R., {Wise} M.~W., 2001, \apjl,
  558, L15

\bibitem[{Borgani} et~al.(2001){Borgani}, {Governato}, {Wadsley}
  et~al.]{Borgani01}
{Borgani} S., {Governato} F., {Wadsley} J., et~al., 2001, \apjl, 559, L71

\bibitem[{Borgani} et~al.(2004){Borgani}, {Murante}, {Springel}
  et~al.]{Borgani04}
{Borgani} S., {Murante} G., {Springel} V., et~al., 2004, \mnras, 348, 1078

\bibitem[{Bower}(1997)]{Bower97}
{Bower} R.~G., 1997, \mnras, 288, 355

\bibitem[{Br{\" u}ggen}(2003)]{Brueggen03}
{Br{\" u}ggen} M., 2003, \apj, 592, 839

\bibitem[{Br{\" u}ggen} \& {Kaiser}(2002)]{Brueggen02b}
{Br{\" u}ggen} M., {Kaiser} C.~R., 2002, \nat, 418, 301

\bibitem[{Br{\" u}ggen} et~al.(2002){Br{\" u}ggen}, {Kaiser}, {Churazov} \&
  {En{\ss}lin}]{Brueggen02}
{Br{\" u}ggen} M., {Kaiser} C.~R., {Churazov} E., {En{\ss}lin} T.~A., 2002,
  \mnras, 331, 545

\bibitem[{Br{\"u}ggen} et~al.(2005){Br{\"u}ggen}, {Ruszkowski} \&
  {Hallman}]{Brueggen05}
{Br{\"u}ggen} M., {Ruszkowski} M., {Hallman} E., 2005, \apj, 630, 740

\bibitem[{Bruzual} \& {Charlot}(2003)]{Charlot}
{Bruzual} G., {Charlot} S., 2003, \mnras, 344, 1000

\bibitem[{Churazov} et~al.(2001){Churazov}, {Br{\" u}ggen}, {Kaiser}, {B{\"
  o}hringer} \& {Forman}]{Churazov01}
{Churazov} E., {Br{\" u}ggen} M., {Kaiser} C.~R., {B{\" o}hringer} H., {Forman}
  W., 2001, \apj, 554, 261

\bibitem[{Churazov} et~al.(2005){Churazov}, {Sazonov}, {Sunyaev}, {Forman},
  {Jones} \& {B{\"o}hringer}]{Churazov2005}
{Churazov} E., {Sazonov} S., {Sunyaev} R., {Forman} W., {Jones} C.,
  {B{\"o}hringer} H., 2005, \mnras, 363, L91

\bibitem[{Churazov} et~al.(2002){Churazov}, {Sunyaev}, {Forman} \& {B{\"
  o}hringer}]{Churazov02}
{Churazov} E., {Sunyaev} R., {Forman} W., {B{\" o}hringer} H., 2002, \mnras,
  332, 729

\bibitem[{Cowie} \& {Binney}(1977)]{Cowie77}
{Cowie} L.~L., {Binney} J., 1977, \apj, 215, 723

\bibitem[{Croston} et~al.(2005){Croston}, {Hardcastle} \&
  {Birkinshaw}]{Croston05}
{Croston} J.~H., {Hardcastle} M.~J., {Birkinshaw} M., 2005, \mnras, 357, 279

\bibitem[{Croton} et~al.(2005){Croton}, {Springel}, {White} et~al.]{Croton05}
{Croton} D.~J., {Springel} V., {White} S.~D.~M., et~al., 2005, \mnras,  1055--+

\bibitem[{Dalla Vecchia} et~al.(2004){Dalla Vecchia}, {Bower}, {Theuns},
  {Balogh}, {Mazzotta} \& {Frenk}]{DVecchia04}
{Dalla Vecchia} C., {Bower} R.~G., {Theuns} T., {Balogh} M.~L., {Mazzotta} P.,
  {Frenk} C.~S., 2004, mnras, submitted (preprint astro-ph/0402441)

\bibitem[{De Grandi} et~al.(2004){De Grandi}, {Ettori}, {Longhetti} \&
  {Molendi}]{DeGrandi2004}
{De Grandi} S., {Ettori} S., {Longhetti} M., {Molendi} S., 2004, \aap, 419, 7

\bibitem[{De Grandi} \& {Molendi}(2002)]{DeGrandiM02}
{De Grandi} S., {Molendi} S., 2002, \apj, 567, 163

\bibitem[{Di Matteo} et~al.(2003){Di Matteo}, {Croft}, {Springel} \&
  {Hernquist}]{DiMatteo03}
{Di Matteo} T., {Croft} R.~A.~C., {Springel} V., {Hernquist} L., 2003, \apj,
  593, 56

\bibitem[{Di Matteo} et~al.(2005){Di Matteo}, {Springel} \&
  {Hernquist}]{DiMatteo05}
{Di Matteo} T., {Springel} V., {Hernquist} L., 2005, \nat, 433, 604

\bibitem[{Dolag}(2004)]{Dolag2004}
{Dolag} K., 2004, in { The Riddle of Cooling Flows in Galaxies and Clusters of
  galaxies\/}

\bibitem[{Dolag} et~al.(2004){Dolag}, {Jubelgas}, {Springel}, {Borgani} \&
  {Rasia}]{Dolag04}
{Dolag} K., {Jubelgas} M., {Springel} V., {Borgani} S., {Rasia} E., 2004,
  \apjl, 606, L97

\bibitem[{Donahue} et~al.(2005){Donahue}, {Voit}, {O'Dea}, {Baum} \&
  {Sparks}]{Donahue2005}
{Donahue} M., {Voit} G.~M., {O'Dea} C.~P., {Baum} S.~A., {Sparks} W.~B., 2005,
  \apjl, 630, L13

\bibitem[{Edge}(2001)]{Edge01}
{Edge} A.~C., 2001, \mnras, 328, 762

\bibitem[{Edge} \& {Frayer}(2003)]{Edge03}
{Edge} A.~C., {Frayer} D.~T., 2003, \apjl, 594, L13

\bibitem[{Edge} et~al.(2002){Edge}, {Wilman}, {Johnstone}, {Crawford}, {Fabian}
  \& {Allen}]{Edge02}
{Edge} A.~C., {Wilman} R.~J., {Johnstone} R.~M., {Crawford} C.~S., {Fabian}
  A.~C., {Allen} S.~W., 2002, \mnras, 337, 49

\bibitem[{Ensslin} \& {Vogt}(2005)]{Ensslin2005}
{Ensslin} T.~A., {Vogt} C., 2005, ArXiv Astrophysics e-prints, astro-ph/0505517

\bibitem[{Fabian}(1994)]{Fabian94}
{Fabian} A.~C., 1994, \araa, 32, 277

\bibitem[{Fabian} et~al.(2001){Fabian}, {Mushotzky}, {Nulsen} \&
  {Peterson}]{Fabian01}
{Fabian} A.~C., {Mushotzky} R.~F., {Nulsen} P.~E.~J., {Peterson} J.~R., 2001,
  \mnras, 321, L20

\bibitem[{Fabian} \& {Nulsen}(1977)]{FabianNulsen77}
{Fabian} A.~C., {Nulsen} P.~E.~J., 1977, \mnras, 180, 479

\bibitem[{Fabian} et~al.(1984){Fabian}, {Nulsen} \& {Canizares}]{Fabian84}
{Fabian} A.~C., {Nulsen} P.~E.~J., {Canizares} C.~R., 1984, \nat, 310, 733

\bibitem[{Fabian} et~al.(2003{\natexlab{a}}){Fabian}, {Sanders}, {Allen}
  et~al.]{Fabian2003}
{Fabian} A.~C., {Sanders} J.~S., {Allen} S.~W., et~al., 2003{\natexlab{a}},
  \mnras, 344, L43

\bibitem[{Fabian} et~al.(2003{\natexlab{b}}){Fabian}, {Sanders}, {Crawford},
  {Conselice}, {Gallagher} \& {Wyse}]{Fabian03}
{Fabian} A.~C., {Sanders} J.~S., {Crawford} C.~S., {Conselice} C.~J.,
  {Gallagher} J.~S., {Wyse} R.~F.~G., 2003{\natexlab{b}}, \mnras, 344, L48

\bibitem[{Fabian} et~al.(2000){Fabian}, {Sanders}, {Ettori} et~al.]{Fabian00}
{Fabian} A.~C., {Sanders} J.~S., {Ettori} S., et~al., 2000, \mnras, 318, L65

\bibitem[{Ferrarese} \& {Ford}(2005)]{Ferrarese2005}
{Ferrarese} L., {Ford} H., 2005, Space Science Reviews, 116, 523

\bibitem[{Gardini} et~al.(2004){Gardini}, {Rasia}, {Mazzotta}, {Tormen}, {De
  Grandi} \& {Moscardini}]{Gardini04}
{Gardini} A., {Rasia} E., {Mazzotta} P., {Tormen} G., {De Grandi} S.,
  {Moscardini} L., 2004, \mnras, 351, 505

\bibitem[{Hoeft} \& {Br{\" u}ggen}(2004)]{Hoeft04}
{Hoeft} M., {Br{\" u}ggen} M., 2004, \apj, 617, 896

\bibitem[{Jenkins} et~al.(2001){Jenkins}, {Frenk}, {White} et~al.]{Jenkins01}
{Jenkins} A., {Frenk} C.~S., {White} S.~D.~M., et~al., 2001, \mnras, 321, 372

\bibitem[{Johnstone} et~al.(1987){Johnstone}, {Fabian} \&
  {Nulsen}]{Johnstone87}
{Johnstone} R.~M., {Fabian} A.~C., {Nulsen} P.~E.~J., 1987, \mnras, 224, 75

\bibitem[{Jubelgas} et~al.(2004){Jubelgas}, {Springel} \& {Dolag}]{Jubelgas04}
{Jubelgas} M., {Springel} V., {Dolag} K., 2004, \mnras, 351, 423

\bibitem[{Kaastra} et~al.(2001){Kaastra}, {Ferrigno}, {Tamura}, {Paerels},
  {Peterson} \& {Mittaz}]{Kaastra01}
{Kaastra} J.~S., {Ferrigno} C., {Tamura} T., {Paerels} F.~B.~S., {Peterson}
  J.~R., {Mittaz} J.~P.~D., 2001, \aap, 365, L99

\bibitem[{Kaiser}(1986)]{Kaiser86}
{Kaiser} N., 1986, \mnras, 222, 323

\bibitem[{Kaiser}(1991)]{Kaiser91}
{Kaiser} N., 1991, \apj, 383, 104

\bibitem[{Katz} et~al.(1996){Katz}, {Weinberg} \& {Hernquist}]{KWH96}
{Katz} N., {Weinberg} D.~H., {Hernquist} L., 1996, \apjs, 105, 19

\bibitem[{Kauffmann} et~al.(1999){Kauffmann}, {Colberg}, {Diaferio} \&
  {White}]{K99}
{Kauffmann} G., {Colberg} J.~M., {Diaferio} A., {White} S.~D.~M., 1999, \mnras,
  303, 188

\bibitem[{Lewis} et~al.(2000){Lewis}, {Babul}, {Katz}, {Quinn}, {Hernquist} \&
  {Weinberg}]{Lewis00}
{Lewis} G.~F., {Babul} A., {Katz} N., {Quinn} T., {Hernquist} L., {Weinberg}
  D.~H., 2000, \apj, 536, 623

\bibitem[{Markevitch} et~al.(2001){Markevitch}, {Vikhlinin} \&
  {Mazzotta}]{Markevitch01}
{Markevitch} M., {Vikhlinin} A., {Mazzotta} P., 2001, \apjl, 562, L153

\bibitem[{Mazzotta} et~al.(2002){Mazzotta}, {Kaastra}, {Paerels}
  et~al.]{Mazzotta02}
{Mazzotta} P., {Kaastra} J.~S., {Paerels} F.~B., et~al., 2002, \apjl, 567, L37

\bibitem[{McCarthy} et~al.(2004){McCarthy}, {Balogh}, {Babul}, {Poole} \&
  {Horner}]{McCarthy04}
{McCarthy} I.~G., {Balogh} M.~L., {Babul} A., {Poole} G.~B., {Horner} D.~J.,
  2004, \apj, accepted (preprint astro-ph/0406329)

\bibitem[{McNamara} et~al.(2005){McNamara}, {Nulsen}, {Wise}
  et~al.]{McNamara05}
{McNamara} B.~R., {Nulsen} P.~E.~J., {Wise} M.~W., et~al., 2005, \nat, 433, 45

\bibitem[{McNamara} et~al.(2000){McNamara}, {Wise}, {Nulsen}
  et~al.]{McNamara00}
{McNamara} B.~R., {Wise} M., {Nulsen} P.~E.~J., et~al., 2000, \apjl, 534, L135

\bibitem[{Muanwong} et~al.(2001){Muanwong}, {Thomas}, {Kay}, {Pearce} \&
  {Couchman}]{Muanwong01}
{Muanwong} O., {Thomas} P.~A., {Kay} S.~T., {Pearce} F.~R., {Couchman}
  H.~M.~P., 2001, \apjl, 552, L27

\bibitem[{Narayan} \& {Medvedev}(2001)]{Narayan01}
{Narayan} R., {Medvedev} M.~V., 2001, \apjl, 562, L129

\bibitem[{Navarro} et~al.(1995){Navarro}, {Frenk} \& {White}]{Navarro95}
{Navarro} J.~F., {Frenk} C.~S., {White} S.~D.~M., 1995, \mnras, 275, 720

\bibitem[{Navarro} et~al.(1996){Navarro}, {Frenk} \& {White}]{NFW96}
{Navarro} J.~F., {Frenk} C.~S., {White} S.~D.~M., 1996, \apj, 462, 563

\bibitem[{Navarro} et~al.(1997){Navarro}, {Frenk} \& {White}]{NFW97}
{Navarro} J.~F., {Frenk} C.~S., {White} S.~D.~M., 1997, \apj, 490, 493

\bibitem[{Nulsen} et~al.(2003){Nulsen}, {McNamara}, {David} \&
  {Wise}]{Nulsen03}
{Nulsen} P.~E., {McNamara} B.~R., {David} L.~P., {Wise} M., 2003, The Cosmic
  Cauldron, 25th meeting of the IAU, Joint Discussion 10, 18 July 2003, Sydney,
  Australia, 10

\bibitem[{Nulsen} et~al.(2005{\natexlab{a}}){Nulsen}, {Hambrick}, {McNamara}
  et~al.]{Nulsen05}
{Nulsen} P.~E.~J., {Hambrick} D.~C., {McNamara} B.~R., et~al.,
  2005{\natexlab{a}}, \apjl, 625, L9

\bibitem[{Nulsen} et~al.(2005{\natexlab{b}}){Nulsen}, {McNamara}, {Wise} \&
  {David}]{Nulsen2005}
{Nulsen} P.~E.~J., {McNamara} B.~R., {Wise} M.~W., {David} L.~P.,
  2005{\natexlab{b}}, \apj, 628, 629

\bibitem[{O'Connell} \& {McNamara}(1989)]{O'Connell89}
{O'Connell} R.~W., {McNamara} B.~R., 1989, \aj, 98, 180

\bibitem[{Oh} \& {Benson}(2003)]{OhBenson03}
{Oh} S.~P., {Benson} A.~J., 2003, \mnras, 342, 664

\bibitem[{Owen} et~al.(2000){Owen}, {Eilek} \& {Kassim}]{Owen2000}
{Owen} F.~N., {Eilek} J.~A., {Kassim} N.~E., 2000, \apj, 543, 611

\bibitem[{Peterson} et~al.(2003){Peterson}, {Kahn}, {Paerels}
  et~al.]{Peterson03}
{Peterson} J.~R., {Kahn} S.~M., {Paerels} F.~B.~S., et~al., 2003, \apj, 590,
  207

\bibitem[{Peterson} et~al.(2001){Peterson}, {Paerels}, {Kaastra}
  et~al.]{Peterson01}
{Peterson} J.~R., {Paerels} F.~B.~S., {Kaastra} J.~S., et~al., 2001, \aap, 365,
  L104

\bibitem[{Quilis} et~al.(2001){Quilis}, {Bower} \& {Balogh}]{Quilis01}
{Quilis} V., {Bower} R.~G., {Balogh} M.~L., 2001, \mnras, 328, 1091

\bibitem[{Reynolds} et~al.(2005){Reynolds}, {McKernan}, {Fabian}, {Stone} \&
  {Vernaleo}]{Reynolds05}
{Reynolds} C.~S., {McKernan} B., {Fabian} A.~C., {Stone} J.~M., {Vernaleo}
  J.~C., 2005, \mnras, 357, 242

\bibitem[{Ruszkowski} \& {Begelman}(2002)]{Ruszkowski02}
{Ruszkowski} M., {Begelman} M.~C., 2002, \apj, 581, 223

\bibitem[{Ruszkowski} et~al.(2004){Ruszkowski}, {Br{\" u}ggen} \&
  {Begelman}]{Ruszkowski04}
{Ruszkowski} M., {Br{\" u}ggen} M., {Begelman} M.~C., 2004, \apj, 615, 675

\bibitem[{Salom{\'e}} \& {Combes}(2004)]{Salome2004}
{Salom{\'e}} P., {Combes} F., 2004, \aap, 415, L1

\bibitem[{Sanders} \& {Fabian}(2002)]{Sanders02}
{Sanders} J.~S., {Fabian} A.~C., 2002, \mnras, 331, 273

\bibitem[{Sanderson} et~al.(2005){Sanderson}, {Finoguenov} \&
  {Mohr}]{Sanderson2005}
{Sanderson} A.~J.~R., {Finoguenov} A., {Mohr} J.~J., 2005, \apj, 630, 191

\bibitem[{Sarazin}(1988)]{Sarazin}
{Sarazin} C.~L., 1988, {X-ray emission from clusters of galaxies}, Cambridge
  Astrophysics Series, Cambridge: Cambridge University Press, 1988

\bibitem[{Springel}(2005)]{Gadget2}
{Springel} V., 2005, \mnras, submitted (preprint astro-ph/0505010)

\bibitem[{Springel} et~al.(2005{\natexlab{a}}){Springel}, {Di Matteo} \&
  {Hernquist}]{Springel2005}
{Springel} V., {Di Matteo} T., {Hernquist} L., 2005{\natexlab{a}}, \apjl, 620,
  L79

\bibitem[{Springel} et~al.(2005{\natexlab{b}}){Springel}, {Di Matteo} \&
  {Hernquist}]{Springel04}
{Springel} V., {Di Matteo} T., {Hernquist} L., 2005{\natexlab{b}}, \mnras, 361,
  776

\bibitem[{Springel} \& {Hernquist}(2002)]{SH02}
{Springel} V., {Hernquist} L., 2002, \mnras, 333, 649

\bibitem[{Springel} \& {Hernquist}(2003)]{SH03}
{Springel} V., {Hernquist} L., 2003, \mnras, 339, 289

\bibitem[{Springel} et~al.(2001{\natexlab{a}}){Springel}, {White}, {Tormen} \&
  {Kauffmann}]{Springel01}
{Springel} V., {White} S.~D.~M., {Tormen} G., {Kauffmann} G.,
  2001{\natexlab{a}}, \mnras, 328, 726

\bibitem[{Springel} et~al.(2001{\natexlab{b}}){Springel}, {Yoshida} \&
  {White}]{SYW01b}
{Springel} V., {Yoshida} N., {White} S.~D.~M., 2001{\natexlab{b}}, New
  Astronomy, 6, 79

\bibitem[{Tamura} et~al.(2001){Tamura}, {Kaastra}, {Peterson} et~al.]{Tamura01}
{Tamura} T., {Kaastra} J.~S., {Peterson} J.~R., et~al., 2001, \aap, 365, L87

\bibitem[{Tormen} et~al.(1997){Tormen}, {Bouchet} \& {White}]{T97}
{Tormen} G., {Bouchet} F.~R., {White} S.~D.~M., 1997, \mnras, 286, 865

\bibitem[{Tornatore} et~al.(2003){Tornatore}, {Borgani}, {Springel},
  {Matteucci}, {Menci} \& {Murante}]{Tornatore03}
{Tornatore} L., {Borgani} S., {Springel} V., {Matteucci} F., {Menci} N.,
  {Murante} G., 2003, \mnras, 342, 1025

\bibitem[{Tozzi} \& {Norman}(2001)]{TozziN01}
{Tozzi} P., {Norman} C., 2001, \apj, 546, 63

\bibitem[{Tremaine} et~al.(2002){Tremaine}, {Gebhardt}, {Bender}
  et~al.]{Tremaine02}
{Tremaine} S., {Gebhardt} K., {Bender} R., et~al., 2002, \apj, 574, 740

\bibitem[{Voigt} \& {Fabian}(2003)]{Voigt03}
{Voigt} L.~M., {Fabian} A.~C., 2003, \mnras, submitted (preprint
  astro-ph/0308352)

\bibitem[{Voigt} et~al.(2002){Voigt}, {Schmidt}, {Fabian}, {Allen} \&
  {Johnstone}]{Voigt02}
{Voigt} L.~M., {Schmidt} R.~W., {Fabian} A.~C., {Allen} S.~W., {Johnstone}
  R.~M., 2002, \mnras, 335, L7

\bibitem[{Voit}(2004)]{Voit2004}
{Voit} G.~M., 2004, ArXiv Astrophysics e-prints, astro-ph/0410173

\bibitem[{Voit} et~al.(2003){Voit}, {Balogh}, {Bower}, {Lacey} \&
  {Bryan}]{Voit03}
{Voit} G.~M., {Balogh} M.~L., {Bower} R.~G., {Lacey} C.~G., {Bryan} G.~L.,
  2003, \apj, 593, 272

\bibitem[{Voit} \& {Bryan}(2001)]{VoitBryan01}
{Voit} G.~M., {Bryan} G.~L., 2001, \nat, 414, 425

\bibitem[{Voit} et~al.(2002){Voit}, {Bryan}, {Balogh} \& {Bower}]{Voit02}
{Voit} G.~M., {Bryan} G.~L., {Balogh} M.~L., {Bower} R.~G., 2002, \apj, 576,
  601

\bibitem[{Voit} \& {Donahue}(2005)]{Voit2005}
{Voit} G.~M., {Donahue} M., 2005, ArXiv Astrophysics e-prints, astro-ph/0509176

\bibitem[{White} et~al.(1994){White}, {Fabian}, {Allen} et~al.]{White94}
{White} D.~A., {Fabian} A.~C., {Allen} S.~W., et~al., 1994, \mnras, 269, 589

\bibitem[{White} et~al.(1997){White}, {Jones} \& {Forman}]{WJF97}
{White} D.~A., {Jones} C., {Forman} W., 1997, \mnras, 292, 419

\bibitem[{White} et~al.(1993){White}, {Navarro}, {Evrard} \& {Frenk}]{White93}
{White} S.~D.~M., {Navarro} J.~F., {Evrard} A.~E., {Frenk} C.~S., 1993, \nat,
  366, 429

\bibitem[{Wu} \& {Xue}(2002)]{WuXue02}
{Wu} X., {Xue} Y., 2002, \apjl, 572, L19

\bibitem[{Yoshida} et~al.(2001){Yoshida}, {Sheth} \& {Diaferio}]{Yoshida01}
{Yoshida} N., {Sheth} R.~K., {Diaferio} A., 2001, \mnras, 328, 669

\bibitem[{Yoshida} et~al.(2002){Yoshida}, {Stoehr}, {Springel} \&
  {White}]{Yoshida02}
{Yoshida} N., {Stoehr} F., {Springel} V., {White} S.~D.~M., 2002, \mnras, 335,
  762

\bibitem[{Zakamska} \& {Narayan}(2003)]{Zakamska03}
{Zakamska} N.~L., {Narayan} R., 2003, \apj, 582, 162

\end{thebibliography}

\end{document}